\documentclass[twocolumn]{aastex631}
\begin{document}
\title{Investigating Differences in the Palomar-Green Blazar Population Using Polarization}
\correspondingauthor{Janhavi Baghel}
\email{jbaghel@ncra.tifr.res.in}
\author[0000-0002-0367-812X]{Janhavi Baghel}
\affiliation{National Centre for Radio Astrophysics (NCRA) - Tata Institute of Fundamental Research (TIFR)\\ 
S. P. Pune University Campus, Ganeshkhind, Pune 411007, India}
\author{P. Kharb}
\affiliation{National Centre for Radio Astrophysics (NCRA) - Tata Institute of Fundamental Research (TIFR)\\ 
S. P. Pune University Campus, Ganeshkhind, Pune 411007, India}
\author{T. Hovatta}
\affiliation{Finnish Centre for Astronomy with ESO, FINCA, University of Turku, Turku, Finland}
\affiliation{Aalto University Metsähovi Radio Observatory, Metsähovintie 114, 02540 Kylmälä, Finland}
\author{Luis C. Ho}
\affiliation{Kavli Institute for Astronomy and Astrophysics, Peking University, Beijing 100871, China}
\affiliation{Department of Astronomy, School of Physics, Peking University, Beijing 100871, China}
\author{C. Harrison}
\affiliation{School of Mathematics, Statistics and Physics, Newcastle University, Newcastle upon Tyne NE1 7RU, UK}
\author{E. Lindfors}
\affiliation{Finnish Centre for Astronomy with ESO, FINCA, University of Turku, Turku, Finland}
\author{Silpa S.}
\affiliation{Departamento de Astronom{\`i}a, Universidad de Concepci{\`o}n, Barrio Universitario s/n, Concepci{\`o}n, Chile}
\author{S. Gulati}
\affiliation{National Centre for Radio Astrophysics (NCRA) - Tata Institute of Fundamental Research (TIFR)\\ 
S. P. Pune University Campus, Ganeshkhind, Pune 411007, India}

\begin{abstract}
We present polarization images with the Karl G. Jansky Very Large Array (VLA) in A and B-array configurations at 6~GHz of 7 radio-loud (RL) quasars and 8 BL~Lac objects belonging to the Palomar-Green (PG) `blazar' sample. This completes our arcsecond-scale polarization study of an optically-selected volume-limited blazar sample comprising 16 radio-loud quasars and 8 BL Lac objects. Using the VLA, we identify kpc-scale polarization in the cores and jets/lobes of all the blazars, with fractional polarization varying from around $0.8 \pm 0.3$\% to $37\pm6$\%. The kpc-scale jets in PG RL quasars are typically aligned along their parsec-scale jets and show apparent magnetic fields parallel to jet directions in their jets/cores and magnetic field compression in their hotspots. The quasars show evidence of interaction with their environment as well as restarted AGN activity through morphology, polarization and spectral indices. These quasi-periodic jet modulations and restarted activity may be indicative of an unstable accretion disk undergoing transition. We find that the polarization characteristics of the BL~Lacs are consistent with their jets being reoriented multiple times, with no correlation between their core apparent magnetic field orientations and pc-scale jet directions. We find that the low synchrotron peaked BL Lacs show polarization and radio morphology features typical of `strong' jet sources as defined by \citet{Meyer2011} for the `blazar envelope scenario', which posits a division based on jet profiles and velocity gradients rather than total jet power. 
\end{abstract}


\section{Introduction} \label{sec:intro}
Active galactic nuclei (AGN) are highly energetic phenomena produced by the central supermassive black hole of a galaxy devouring the inflowing gas and dust \citep{Rees84,Rawlings1991}. This spiraling debris forms a rotating accretion disk that emits intense radiation due to frictional heating \citep{LyndenBell1969, Bardeen1970}. A fraction of these AGN also expel relativistic jets of synchrotron-emitting plasma orthogonal to the accretion disks and are bright at radio frequencies \citep{Blandford1974}.  Radio-loud (RL) AGN have strong, large-scale jets and are { defined} as having their 5~GHz radio flux density at least 10 times greater than their optical B-band flux densities \citep{Kellermann1989}. These jets have the potential to significantly influence their host galaxy and surrounding environment, as they heat the intracluster medium \citep{McNamara2007,Chang2012} and modulate star formation rates \citep{Cattaneo2009, Shin2019}. 

There are many observational sub-classifications of RL AGN and most can be explained through differences in viewing angles and beaming \citep{Rees1966,Readhead1978}. However, some delineations are independent of orientation. One such dichotomy is based on the appearance of two distinctive radio jet morphologies - Fanaroff and Riley type I and II \citep{Fanaroff1974}. The FRI jets are fainter, diffuse plumes whereas the FRII jets are brighter, highly collimated, and produce terminal hot spots when they impact the ambient medium \citep{Blandford1974, Mingo2019}. 

The underlying cause of the FR dichotomy continues to be a subject of much debate as is its initial correlation with the extended radio luminosity and subsequently jet power \citep{Blackman2022}. Potential explanations can be broadly categorized into intrinsic factors relating to differences in jet kinetic power arising from differences in black hole spins \citep{Baum1995, Meier1999}, accretion modes \citep{Best2012, Hardcastle2018}, jet formation mechanisms \citep{BZ1977,BP1982}, or jet profile and particle compositions \citep{Scheuer1974, Laing2002, Croston2018} and extrinsic factors relating to jet-medium interaction such as jet decollimation due to turbulence-induced plasma flow instabilities \citep{Bicknell1984, Laing2002} and stellar mass loading \citep{Komissarov1994, Wykes2015} or the fast deceleration of an initially supersonic jet due to interaction with a denser ambient medium \citep{Biretta1995, Hardcastle2003}. 

As per the radio-loud unification scheme \citep{Urry95}, RL quasars are the pole-on counterparts of FRII radio galaxies, and BL~Lac objects are the pole-on counterparts of FRI radio galaxies. Collectively referred to as blazars, their jets are at small angles to our line-of-sight. Blazars were originally differentiated based on their optical spectra. BL Lacs have weak optical line emission compared to quasars, which are characterized by stronger optical line emission \citep[equivalent widths $>5$\AA;][]{Stickel1991,Stocke1991}. Given their high beaming and strong variability and no evidence of differences in beaming among the two classes, the BL Lacs are thought to be at similar angles to the flat spectrum radio quasars (FSRQs) \citep{Ghisellini1993, Gabuzda2000}. {  The steep spectrum radio quasars (SSRQs) and FSRQs both display broad (velocity widths $\geq 1000$~km~s$^{-1}$) emission lines in their optical/UV spectra. Given that the half-opening angle of the dusty obscuring tori is $\leq 50\degr$ \citep{Simpson96}, their jets are assumed to be inclined at angles $\leq 50\degr$ to our lines of sight. The SSRQ and FSRQ are therefore progressively more aligned versions of FR II galaxies \citep{Barthel1989}.} BL Lacs are also characterized by systematically lower apparent jet speeds compared to FSRQs at parsec scales \citep{Ghisellini1993, Gabuzda2000, Kellermann2004} which given their similar orientations suggests a difference in the actual intrinsic velocities of relativistic bulk motion.

With polarimetric Very Long Baseline Interferometry (VLBI) observations, differences in the apparent magnetic (B-) fields among the parsec-scale jets of the two blazar classes have been found \citep{Gabuzda1992, Cawthrone1993}. BL~Lacs tend to have their parsec-scale { electric vector position angle (EVPA)} parallel to the jet direction whereas RL quasars tend to show a perpendicular relative orientation \citep{Lister2005, Lister2013}. The inferred B-field structures are perpendicular to the EVPAs for optically thin \footnote{The optical thickness required to change the inference of B-field direction is considered to be $\tau\sim7$ \citep{Cawthrone2013, Wardle2018}}\label{footnote1} emission \citep{Pacholczyk70}, so this would mean that the BL~Lacs show a predominance of jet B fields orthogonal to the jet direction while { quasars primarily show jet B fields along the jet direction. One explanation for such a B-field structure could be the tightness or looseness of the helical magnetic fields in BL Lacs and quasars respectively \citep[e.g.,][]{Gabuzda2015}.}

Blazars are also subcategorized based on their spectral energy distributions (SEDs) into high-, intermediate-, and low-spectrally peaked based on the position of their synchrotron peak frequency \citep{Abdo2010,Padovani2015}. Contradicting the notion of the FR dichotomy, several studies have found a broad continuum in parameters such as synchrotron peak frequencies and spectral indices \citep{Giommi1994,Sambruna1996}. Notably, \citet{Fossati1998} found an anti-correlation between the luminosity and frequency of the synchrotron peak. This anti-correlation, now referred to as the `blazar sequence', implies that the total jet power is the only parameter informing the spectral type and broadband SED characteristics of an AGN. Later studies by \citet{Ghisellini2017} found the sequence to exist for the BL Lac objects but not for the quasars, which was interpreted as a sequence of radiative cooling due to the presence of the broad line region (BLR) and the dusty torus in the case of quasars. 

More recent work by \citet{Meyer2011} and \citet{Keenan2021} have given rise to the `blazar envelope' picture. Under this scenario, a range of progressively misaligned but intrinsically different ‘weak’ and ‘strong’ jets are hosted by sources with inefficient accretion and low-excitation spectral lines or efficient accretion and high-excitation spectral lines, respectively. The FRII radio galaxies and quasars have 'strong' jets and so do low-synchrotron-peaked (LSP) BL Lacs whereas FRI radio galaxies and high-synchrotron-peaked (HSP) BL Lacs have 'weak' jets with intermediate-synchrotron-peaked (ISP) BL Lacs coming from a mixed population. Low to moderately high jet power is generated by either type of jet but extreme high-power jets are exclusively 'strong' jet sources. 

Given that high- and low-excitation radio galaxies both exhibit FRI and FRII morphologies \citep{Best2012, Mingo2022} and the fact that hybrid morphology sources exist, it implies that the environment may play a larger role in determining the radio morphology. Beaming and light travel time differences between the two lobes may also cause an intrinsically FRII source to look like a hybrid source \citep{Kharb2015, Gopal-Krishna1996, Hardwood2020, Ghosh2023}.

Given the complexity of these classifications and their interrelated nature, we aimed to disentangle the base causes of these differences in jet profiles. Since FRI jets decelerate and de-collimate on kpc-scales \citep{Bicknell1994, Laing2002} and magnetic fields play a critical role in bulk acceleration and jet propagation \citep{Meier2001,Hawley2015}, kpc-scale radio polarimetric observations form an important means of investigating the differences in jet characteristics. Blazars with their higher total and polarized flux densities, form ideal candidates to investigate both the FR dichotomy as well as the blazar divide.

With this in mind, we set out to investigate the differences in blazars using kpc-scale polarization with the upgraded Giant Metrewave Radio Telescope (uGMRT) and the Karl G. Jansky Very Large Array (VLA). Our objective was to investigate whether the differences in polarization among the blazar subclasses at the parsec-scale persist till the kiloparsec scale. We also aim to increase the number of high-sensitivity kpc-scale images of BL Lacs, which are currently in short supply compared to quasars \citep{Giroletti2006}. 

This paper is a continuation of our study of the Palomar-Green (PG) sample \citep{Green1986} of blazars \citep{Baghel2023, Baghel2024}. Results from 
uGMRT Band-4 (650 MHz) data for the PG BL Lac objects were presented in \citet{Baghel2024}, while VLA C-band (6 GHz) B-array data for nine out of 16 quasars was presented in \citet{Baghel2023}. In this paper, we present the VLA polarization images for the remaining RL quasars and all BL Lac objects { (see Table \ref{tab:PG})} to study the kpc-scale B-field structures of the entire PG `blazar' sample. 
 
The sample selection is discussed briefly in Section \ref{sec:PGSample}. The radio data reduction and calibration and imaging details are presented in Section \ref{sec:radioanalysis}. Our observational results are discussed in Section \ref{sec:results} along with a { brief} discussion on existing literature on the sources. In Section \ref{sec:corr} we present the global correlations for the entire PG `blazar' sample. Section \ref{sec:disc} discusses our findings, and Section \ref{sec: Conclusions} provides our conclusions. 

Throughout this paper we have adopted a $\Lambda$CDM cosmology with $\mathrm{H_0 = 73~km~s^{-1} Mpc^{-1}}$, $\mathrm{\Omega_m = 0.3}$ and $\mathrm{\Omega_v = 0.7}$ and used flat $\Lambda$CDM subroutine of {\tt{astropy.cosmology}} subpackage \citep{astropy:2013,astropy:2018}. The spectral index $\alpha$ is defined such that flux density at frequency $\nu$, $S_\nu \propto \nu^\alpha$.

\section{The PG `blazar' Sample}\label{sec:PGSample}
With the intention of selecting a radio unbiased blazar sample, we chose the Palomar-Green UV-excess survey { \citep{Green1986}} as the parent sample. It remains the largest complete optically selected survey for unobscured AGNs at low redshift and is one of the most well-studied samples of AGNs having extensive supplementary multiband data available in the literature \citep{Kellermann1989, Boroson1992, Miller1993}.

The sample selection is described in \citet{Baghel2023}. The key points for the sample 
presented here are as follows:

\begin{enumerate}
    \item We selected the 16 RL quasars and 8 BL~Lac objects to form the PG `blazar' sample based on (i) a redshift cut-off ($z < 0.5$), as well as (ii) having core-to-lobe extents greater than $\geq15$ arcsec. This was to ensure that these blazars could be well resolved with both the VLA $1\arcsec$ and the uGMRT $5\arcsec$ resolution observations.
    \item The sample includes steep spectrum radio quasars (SSRQs) along with the flat spectrum radio quasars (FSRQs) to create a statistically { representative} sample. The kpc-scale polarization properties would be negligibly affected by jet orientation between these two quasar sub-classes.
    \item This paper completes the VLA observations of the PG sample with the remaining 7 quasars and 8 BL Lac objects with new observations (See Table \ref{tab:PG} and \ref{tab:PGObs}).
\end{enumerate}

\begin{table*}
\centering
\caption{\label{tab:PG}The subset of PG `blazar' sample presented in this paper}
\begin{tabular}{cccccccccc}
\hline\hline
S.No. & Name  
& Other Name  & RA & DEC & Redshift & Extent ($\arcsec$) & Extent (kpc) & Type \\
\hline

1 &
{PG 0851+203} &
OJ 287 &
08h54m48.87s &
+20$\degr$06$\arcmin$30.64$\arcsec$ &
0.306501 &
20 & 87 & 
BL Lac \\

2 &
{PG 1101+384} &
Mrk 421 &
11h04m27.31s &
+38$\degr$12$\arcmin$31.79$\arcsec$ &
0.030893 &
30 & 18 & 
BL Lac \\

3 &
{PG 1218+304} &
RBS 1100 &
12h21m21.94s &
+30$\degr$10$\arcmin$37.16$\arcsec$ &
0.184537 &
45 & 134 &
BL Lac \\

4 &
{PG 1418+546} &
OQ +530 &
14h19m46.59s &
+54$\degr$23$\arcmin$14.78$\arcsec$ &
0.152845 &
45 & 115 & 
BL Lac\\

5 &
{PG 1424+240} &
OQ +240&
14h27m00.39s &
+23$\degr$48$\arcmin$00.03$\arcsec$ &
0.160680$^\ast$ &
45 & 290 & 
BL Lac\\

6 &
{PG 1437+398} &
RBS 1414 &
14h39m17.47s &
+39$\degr$32$\arcmin$42.80$\arcsec$ &
0.344153 &
 25 & 117  & 
BL Lac\\

7 &
{PG 1553+113} &
RBS 1538 &
15h55m43.04s &
+11$\degr$11$\arcmin$24.36$\arcsec$ &
0.360365 &
60 & 290 & 
BL Lac\\

8 &
{PG 2254+075} &
OY +091 &
22h57m17.30s &
+07$\degr$43$\arcmin$12.30$\arcsec$ &
0.188765 &
60 & 181 & 
BL Lac\\

9 & 
{PG 1302$-$102} &
RBS 1212&
13h05m33.01s &
$-$10$\degr$33$\arcmin$19.43$\arcsec$ &
0.27949 &
15 & 61 & 
FSRQ \\

10 & 
{PG 2209+184} &
II Zw 171 &
22h11m53.88s &
+18$\degr$41$\arcmin$49.85$\arcsec$ &
0.06873 &
 12 & 15 &
FSRQ\\

11 & 
{PG 1425+267} &
Ton 202 &
14h27m35.60s &
+26$\degr$32$\arcmin$14.54$\arcsec$ &
0.364262  &
218&1060 &
SSRQ
\\

12 & 
{PG 1512+370} &
4C +37.43 &
15h14m43.06s &
+36$\degr$50$\arcmin$50.35$\arcsec$ &
0.370922 &
58&287 &
SSRQ
\\

13 & 
{PG 1545+210} &
3C 323.1 &
15h47m43.53s &
+20$\degr$52$\arcmin$16.61$\arcsec$ &
0.264659  &
70&275  &
SSRQ 
\\

14 &
{PG 2251+113} &
4C +11.72 &
22h54m10.42s &
+11$\degr$36$\arcmin$38.74$\arcsec$ &
0.32427 &
15 & 68 &
SSRQ 
\\

15 & 
{PG 2308+098} &
4C +09.72 &
23h11m17.75s &
+10$\degr$08$\arcmin$15.75$\arcsec$ &
0.432064 &
83&447 &
SSRQ
\\
\hline
\multicolumn{9}{l}{Note. Column (1): Serial Number. Column (2): PG names of sources. Column (3): Other common names of sources. }\\
\multicolumn{9}{l}{Column (4): Right Ascension. Column (5): Declination. Column (6): Redshift$^\dagger$. Column (7): Radio extents in arcsec { $^{\dagger\dagger}$.} }\\
\multicolumn{9}{l}{Column (8): Radio extents in kpc. Column (9): Type of blazar.}\\
\multicolumn{9}{l}{$^\ast$ $z=0.60468$ has been suggested by \citet{Paiano17}. We have used this updated value { of $z=0.60468$} in our calculations{.}}\\
\multicolumn{9}{l}{$^\dagger$ All redshift values reported are {from NASA NED} corrected to the reference frame defined by the 3K CMB on NASA NED\footnote{The NASA/IPAC Extragalactic Database (NED) is operated by the Jet Propulsion Laboratory, California Institute of Technology, under contract with the National Aeronautics and Space Administration.}.}\\ 
\multicolumn{9}{l}{{ $^{\dagger\dagger}$} Radio extents derived from \citet{Miller1993} at 5~GHz or from 1.4~GHz VLA FIRST / NVSS images for sources}\\
\multicolumn{9}{l}{unresolved at 5~GHz.}
\end{tabular}
\end{table*}

\begin{table*}
\centering
\caption{Observational Parameters from the VLA images of PG BL Lacs and 7 PG Quasars}\label{tab:BQMSMT}
\begin{tabular}{ccccccccc}
\hline \hline
Source &$I_{rms}$($\mu$Jy/beam) &$P_{rms}$($\mu$Jy/beam) &Region & &P~(mJy) &I~(mJy) &FP~(\%) &$\alpha$ \\ 
\hline
\multicolumn{9}{c}{BL Lacs}\\
\hline
PG 0851+203 &920 &585 &Core & &209$\pm$19 &3841.4 &13$\pm$2 &0.50$\pm$0.09 \\
PG 1101+385 &56.2 &188 &Core & &17$\pm$1 &394 &16$\pm$2 &$-$0.22$\pm$0.06 \\
PG 1218+304 &12.6 &8.7 &Core & &1.8$\pm$0.1 &487 &17$\pm$1 &$-$0.12$\pm$0.01 \\
PG 1418+546 &157 &296 &Core & &35$\pm$2 &638 &3.59$\pm$0.05 &$-$0.14$\pm$0.03 \\
PG 1424+240 &99 &110 &Core & &16.5$\pm$0.7 &360 &18.6$\pm$0.9 &$-$0.16$\pm$0.04 \\
& & &Lobes &N &16$\pm$3 &30 &24$\pm$6 & \nodata$^*$ \\
& & & &S &17$\pm$2 &21 &33$\pm$6 &\nodata$^*$\\ 
PG 1437+398 &24.8 &16.6 &Core & &2.39$\pm$0.07 &48.3 &17$\pm$2 &$-$0.23$\pm$0.04 \\
& & &Lobes &S &5.3$\pm$0.5 &3.15 &\nodata$^*$&\nodata$^*$\\
PG 1553+113 & 157 & 15.8 &Core & &17.6$\pm$0.3 &321 &14$\pm$2 &$-$0.07$\pm$0.02 \\
PG 2254+075 & 174 & 131 &Core & &14$\pm$4 &601 &1.1$\pm$0.7 &0.20$\pm$0.09 \\ 
\hline
\multicolumn{9}{c}{Quasars: FSRQs}\\
\hline
PG 1302$-$102 &116 &42.9 &Core & &21.2$\pm$0.6 & 682 &9.8$\pm$0.7 &$-$0.3$\pm$0.1 \\
& & &Lobes &NE (up) &6.9$\pm$0.6 & 21.8 &\nodata$^*$&\nodata$^*$\\ 
& & & &SE (up) &3.8$\pm$0.2 & 3.46 & \nodata$^*$&\nodata$^*$\\ 
PG 2209+184 &85.8 &32.3 &Core & &1.0$\pm$0.1 & 108 &1.50$\pm$1.1 &$-$0.7$\pm$0.1\\
\hline
\multicolumn{9}{c}{Quasars: SSRQs}\\
\hline
PG 1425+267 &7.45 &2.77&Core & &0.87$\pm$0.06 & 26 &9.5$\pm$0.9 & $-$0.4$\pm$0.1\\ 
& & &Lobes &NE (up) &0.09$\pm$0.03 & 5.68 &\nodata$^*$&\nodata$^*$\\
& & & &SW (dn) &4.4$\pm$0.7 & 30.2 & 30$\pm$5  & $-$1.2$\pm$0.1 \\ 
& & &Jet &SW &0.5$\pm$0.1 & 1.58 &\nodata$^*$&\nodata$^*$\\
PG 1512+370 &8.88 &3.77 &Core & &0.96$\pm$0.07 & 43.1 & 10$\pm$2  &$-$0.33$\pm$0.01\\
& & &Hotspot &W &7.7$\pm$0.2 & 37.4 & 21$\pm$2  &$-$0.65$\pm$0.04 \\
& & & &E &8.55$\pm$0.09 & 50.5 &16.3$\pm$0.4 &$-$0.91$\pm$0.03 \\
& & &Lobes &W &18$\pm$2 & 86.4 &25$\pm$4  &$-$1.09$\pm$0.09 \\
& & & &E &20$\pm$1& 105 &22$\pm$3  &$-$0.88$\pm$0.08 \\
PG 1545+210 &20.2 &13.3 &Core & &1.4$\pm$0.2 & 46.4 &5.1$\pm$0.5 &$-$0.1$\pm$0.1 \\
& & &Hotspot &N &33.7$\pm$0.5 & 354 &20.8$\pm$0.9 &$-$0.9$\pm$0.03 \\
& & & &S &16.5$\pm$0.3 & 162 &16$\pm$1  &$-$0.91$\pm$0.01 \\
& & &Lobes &N &23$\pm$3 & 60.3 &37$\pm$6  & \nodata$^*$\\ 
& & & &S &11.8$\pm$0.3 & 65.1 &26$\pm$5  & \nodata$^*$\\ 
& & &Jet &S &1.5$\pm$2 & 10.5 &20$\pm$4  & \nodata$^*$\\ 
PG 2251+113 &29 &7.9 &Core & &0.6$\pm$0.1 & 31.5 &3.6$\pm$0.7 &$-$0.13$\pm$0.01 \\
& & &Lobes &NW &6.2$\pm$0.6 &165 &7$\pm$1  &$-$0.70$\pm$0.06 \\
& & & &SE &15.5$\pm$0.8 & 315 &7.3$\pm$0.9 &$-$0.9$\pm$0.1 \\
PG 2308+098 &49 &10.6 &Core & &2.6$\pm$0.1 & 88.7 &4.2$\pm$0.6 &0.08$\pm$0.01 \\
& & &Lobes &SE &4$\pm$1 & 37.8 &30$\pm$8  &$-$1.1$\pm$0.3 \\
& & &Hotspot &SE &3.3$\pm$0.3 & 45.6 &10$\pm$1  &$-$1.0$\pm$0.2 \\
& & & &NW T1 &0.38$\pm$0.06 & 7.24 &9$\pm$2  &$-$1.2$\pm$0.1 \\
& & &Jet knots &NW T2 &1.7$\pm$0.1 & 20.7 &10$\pm$2  &$-$0.86$\pm$0.05 \\
& & & &NW T3 &0.20$\pm$0.06 & 7.25 &11$\pm$4  &$-$1.2$\pm$0.2\\
& & & &NW T4 &0.07$\pm$0.02 & 0.648 &19$\pm$5  &$-$1.5$\pm$0.3\\
\hline
 \multicolumn{9}{l}{Note. Column (1): PG source name. Column (2): R.M.S noise in Stokes I (total intensity) image. Column (3):  R.M.S noise in the}\\
  \multicolumn{9}{l}{ polarized intensity image. Column (4): Region of the source. Column (5): Polarized flux density. Column (6): Total flux density.} \\
\multicolumn{9}{l}{Column (7): Fractional Polarization. Column (8): Spectral index. }\\
\multicolumn{9}{l}{$*$: High error values }\\
\end{tabular}
\end{table*}

\section{Radio Data Analysis}\label{sec:radioanalysis}
The radio data for these blazars were obtained with the VLA in C-band ($6~$GHz) in B-array to BnA$\rightarrow$A array configurations from 2023 January 15 to 2023 June 22 (Project ID: 23A-038) with a resolution of 1.1$\arcsec$ as noted in Table \ref{tab:PGObs}. The frequency range of 4.5–6.6~GHz was spanned by 16 spectral windows with 64 channels each. The average time on source was around 40 min. Polarization calibrators 3C 286 and 3C 138 were used. For the initial calibration and flagging, we employed the CASA calibration pipeline for VLA data reduction. We followed this with manual polarization calibration.

Initially, the CASA task \texttt{SETJY} was used to manually configure the model of a polarized calibrator. As the model parameters, we used the reference frequency, Stokes I flux density at the reference frequency, and the spectral index. The inputs also included the coefficients of the Taylor expansion of fractional polarization and polarization angle as a function of frequency, centered on the reference frequency, which was estimated by fitting a first-order polynomial to the values obtained from the NRAO VLA observing guide\footnote{ \url{https://science.nrao.edu/facilities/vla/docs/manuals/obsguide/modes/pol}}. The \citet{Perley2017} scale was used to estimate the Stokes I flux density values, the spectral index (alpha), and also the curvature (beta) about the reference frequency.

The process of polarization calibration involves three stages: 
(i) The task \texttt{GAINCAL} with gaintype = \texttt{KCROSS} in CASA was used to solve for the cross-hand (RL, LR) delays, using a polarized calibrator (either 3C 138 or 3C 286). (ii) the task \texttt{POLCAL} in CASA was used to solve for instrumental polarization (i.e., the leakage terms or ‘D-terms’). The task \texttt{POLCAL} with poltype = Df + QU  was used when using the polarized calibrators (either 3C 138 or 3C 286) and poltype = Df was used when using the unpolarized calibrator (3C 84). The average value of the D-term amplitude turned out to be $\approx$ 7\% (iii) The task \texttt{POLCAL} in CASA with poltype = Xf was used to solve for the frequency-dependent polarization angle using a polarized calibrator (either 3C 138 or 3C 286) with a known { EVPA}. These calibration solutions were
subsequently applied to the dataset.

The calibrated visibility data for the sources was extracted using the CASA task \texttt{SPLIT}, which also averaged the spectral channels such that bandwidth (BW) smearing effects were negligible.
Multiterm-multifrequency synthesis \citep[MT-MFS;][]{MTMFS2011} algorithm of \texttt{TCLEAN} task in CASA was used to then create the total intensity or the Stokes I image of the sources. {Three rounds of phase-only self-calibration followed by one round of amplitude and phase self-calibration were carried out for almost all the datasets, except those for PG 1545+210 and PG 2308+098, for which only two rounds of phase-only self-calibration were carried out. Images were created using natural weighting with {\tt robust=+0.5} in {\tt CASA}. The final self-calibrated visibility data was used to create the Stokes Q and U images.

The Stokes Q and U images were combined to produce the linear polarized intensity $\mathrm{P = \sqrt{Q^2 + U^2}}$ and EVPA or $\chi = 0.5~\mathrm{{tan}^{-1} (U/Q)}$ images using the AIPS\footnote{Astronomical Image Processing System; \citet{Wells1985}} task \texttt{COMB} with opcode = \texttt{POLC} (which corrects for Ricean bias) and \texttt{POLA}, respectively.  \texttt{COMB} blanked regions with polarized intensity values $< 3~ \times$ the rms noise in the $\mathrm{P}$ image and regions with values greater than $10^{\degr}$ error in the $\chi$ image. The \texttt{COMB} task with opcode = \texttt{DIV} was then used to get the fractional polarization $\mathrm{FP=P/I}$ images with regions with $\gtrsim$10\% fractional polarization errors blanked. 

The in-band spectral index images and spectral index noise images are produced while imaging with the MT-MFS algorithm of the \texttt{TCLEAN} task in CASA with two Taylor terms to model the frequency dependence of the sky emission by setting the parameter nterms = 2. We used AIPS to blank the regions with spectral index noise greater than about 0.3. This varied by image and was chosen such that regions with total intensity $> 3 \times$ r.m.s. noise are included. In cases of high noise, we also blanked the spectral index noise image using the same.

The average r.m.s. noise in the Stokes I images is $1.28 \times 10^{-4}$~mJy~beam$^{-1}$. The values obtained from the images are presented in Table~\ref{tab:BQMSMT}. Flux density values reported in the paper were obtained using the Gaussian-fitting AIPS task \texttt{JMFIT} for compact components like the core, and AIPS verb \texttt{TVSTAT} for extended emission. Care was taken to exclude the compact regions from selection for obtaining values only for the extended regions using \texttt{TVSTAT}. The FP and spectral index values noted are the mean values over the noted region. The errors in these values are noted from the mean values over the same region for the corresponding noise image. We note here that since these values are the average over a region, they may include blanked regions in the noise image especially in regions of diffuse and patchy emission leading to higher reported noise values in such regions. In Table \ref{tab:BQMSMT} we have not reported values for regions where noise was exceptionally high. The r.m.s. noise values were obtained using AIPS tasks \texttt{TVWIN} and \texttt{IMSTAT}. The AIPS procedure \texttt{TVDIST} was used to obtain spatial extents.

We note that Faraday rotation effects are not expected to be significant at 5~GHz in the kpc-scale observations of jets and lobes \citep[e.g.,][]{Saikia1987,Pudritz2012}. 
Integrated RM $\lesssim 50$~rad~m$^{-2}$ result in rotation of $\lesssim 10\degr$ at 5~GHz. RM values can indeed be higher than the integrated RM values in smaller local regions, which can influence the local inferred B-field directions \citep[e.g.,][]{McKean2016, Silpa2021B}. However, the absence of strong Faraday effects is supported by the observed polarization in the jets in these sources which tend to lie either parallel or perpendicular to the local jet direction, with the inferred B field directions being perpendicular to the direction of the EVPA vectors for optically-thin regions of emission; such orientations could be indicative of organized B-field structures \citep[see][]{Pudritz2012}. 

\section{Results from new data} \label{sec:results}
The VLA 6 GHz polarization images and in-band spectral index images of the eight BL Lac objects and the 7 RL quasars are presented here in Figures \ref{fig1} - \ref{figq7}. { The results have also been listed in Table \ref{tab:BQMSMT}.} The fractional polarization ranges from $1.1\pm0.7$\% to $37\pm6$\% in the radio cores, lobes, and hotspots of these blazars. The radio cores are mostly flat or inverted in the spectral index images with a few slightly steeper values occurring where it is hard to discern the boundary of the core with portions of the inner jet. The hotspots in several of the quasars are steeper than typical values of about $-0.5$ to $-0.7$ \citep{Hardwood2013}, ranging from $-0.7 \pm 0.1$ up to $-1.0 \pm 0.20$, as also seen in \citet{Baghel2023} indicating bow-shock-like structures that could arise due to episodic AGN activity \citep[][]{Clarke1992, Silpa2021, Ghosh2023}. Below we discuss the PG BL Lac objects collectively \citep[see individual notes in][]{Baghel2024} and the PG quasars individually.

\subsection{PG BL Lac objects}\label{subsec:notesB}
In this section, we describe the VLA results for the PG BL Lacs. The uGMRT data on these BL Lacs have been presented in \citet{Baghel2024}. We find that six out of eight PG BL Lacs show compact core-halo structures with the exception of PG 1424+240 which shows diffuse lobes to the north and south, and PG 1437+398 whose diffuse lobe extends to the southwest. 

{ Table \ref{tab4rev} lists the VLBI jet directions in the PG blazars with the parsec-scale mean jet PA obtained from multi-frequency ($1.4-86$ GHz), multi-epoch VLBA observations of \citet{Plavin2022} and 5~GHz VLBI imaging of \citet{Wang2023}.} Our VLA 6~GHz image shows core EVPA perpendicular to the source's VLBI jet direction for PG 0851+203 (Figure \ref{fig1}) and PG 1418+546 (Figure \ref{fig4}). PG 1437+398 (Figure \ref{fig6}) also displays core EVPA perpendicular to this source’s indicated diffuse lobe and possible jet direction. The diffuse lobe is also polarized although the errors in fractional polarization are high with no uniformity in the EVPA patterns in the lobe. { The EVPA structures in the uGMRT 650~MHz images of \citet{Baghel2024} of PG 0851+203, PG 1418+546, and PG 1437+398 are consistent with that of the new VLA images and are} perpendicular to the jet direction in the case of PG 0851+203 and PG 1437+398, but are parallel in the case of PG 1418+546. PG 1418+546 also shows a hotspot to the east in its uGMRT image, suggesting multiple jet reorientations in this source. 

The VLA 6~GHz core EVPA for PG 1101+385 (Figure \ref{fig2}), PG 1218+304 (Figure \ref{fig3}), PG 1553+113 (Figure \ref{fig7}), and PG 2254+075 (Figure \ref{fig8}) are complex in structure. A similar complex core is observed in the uGMRT observations of PG 1101+385 and PG 2254+075 whereas PG 1218+304 uGMRT core EVPA was perpendicular to its VLBI jet direction and in PG 1553+113 its uGMRT core EVPA was oblique to its VLBI jet direction. In PG 1553+113 our VLA core shows two components, which both have EVPA oblique to the VLBI jet direction, which lies parallel along the direction of the EVPA of the second component. Our EVPA of our uGMRT image \citep{Baghel2024} matches one of these components. This could again suggest a structured jet. For PG 2254+075, the VLBI jet direction lies to the southwest and the VLA image shows some indication of EVPA parallel to the jet direction there.

Finally, PG 1424+240 (Figure \ref{fig5}) shows its VLA 6~GHz core EVPA parallel to its VLBI jet direction whereas its uGMRT core EVPA was perpendicular. The double lobes seen in \citet{Baghel2024} are also resolved and show polarization with EVPA parallel along the outflowing lobes. This is in contrast to the uGMRT image where the lobes had EVPA perpendicular along the outflowing lobes, possibly suggesting a spine-sheath structured jet. However, we note that the diffuse lobe emission was patchy in both total intensity and polarized intensity, and as such the values of fractional polarization and spectral index are not reliable for these regions.

Overall, three out of eight of the BL Lacs have their EVPA perpendicular to the jet direction, one shows a spine-sheath-like structure and the remaining four show complex EVPA structures { suggesting} jet reorientations {or turbulent magnetic fields}. Of the three that have EVPA perpendicular to the jet direction, two are LSP BL Lacs with indications of terminal hotspots in their uGMRT images similar to FR type II radio galaxies. \citet{Plavin2022} have provided a mean jet PA of these BL Lacs with multi-frequency, multi-epoch GHz VLBA observations that have been denoted in the figures and mentioned in Table \ref{tab4rev}. These differ from the multi-epoch 15~GHz VLBA mean jet PA provided by \citet{Lister2021} and noted in our \citet{Baghel2024} paper by an average of $\sim 5\degr$ except for PG 1553+113 and PG 2254+075 where it differs by $34\degr$ and $13\degr$ respectively.

\begin{figure*}
\centering
\includegraphics[width=8.7cm,trim=30 220 30 220]{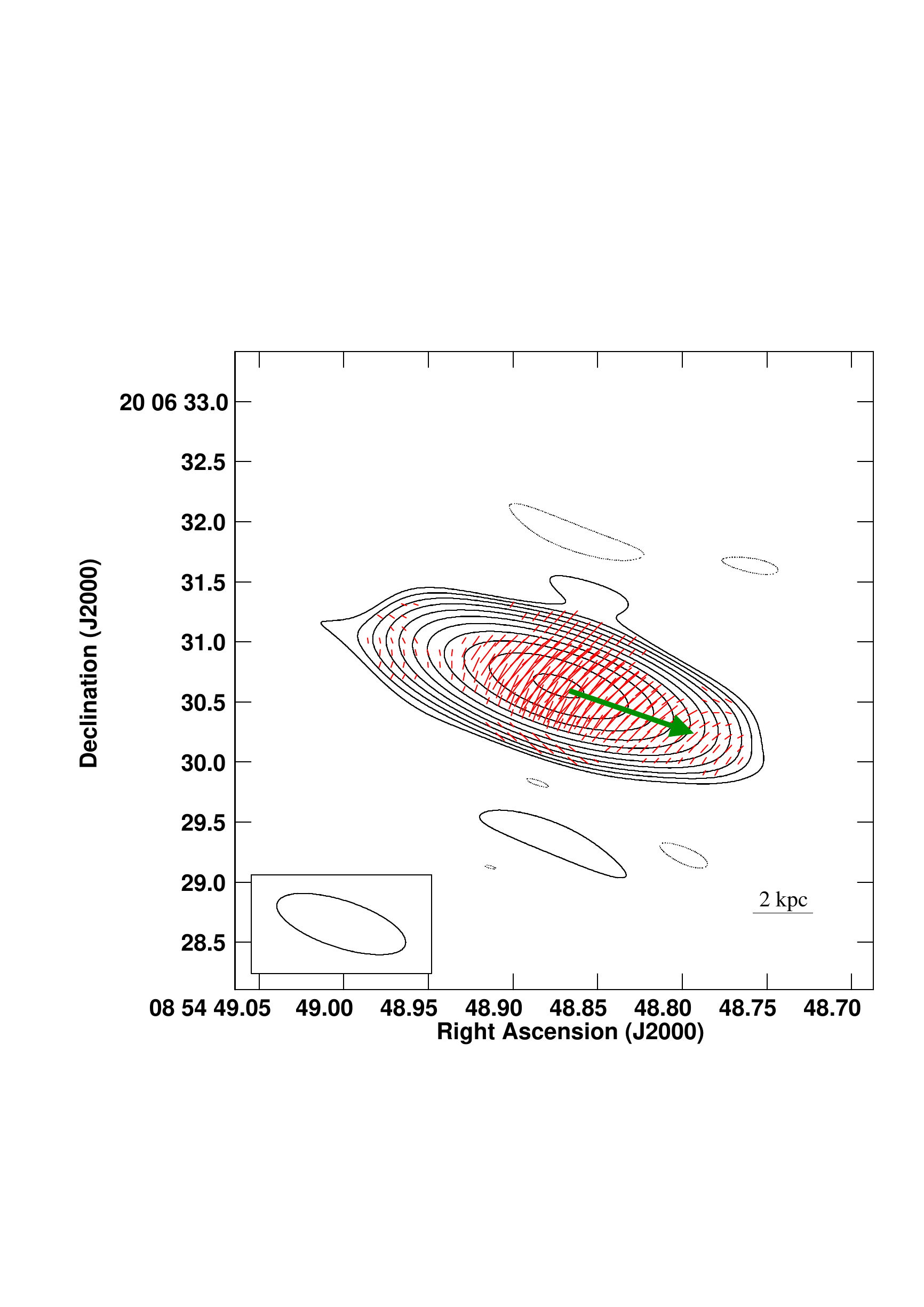}
\includegraphics[width=8.7cm,trim=30 150 30 220]{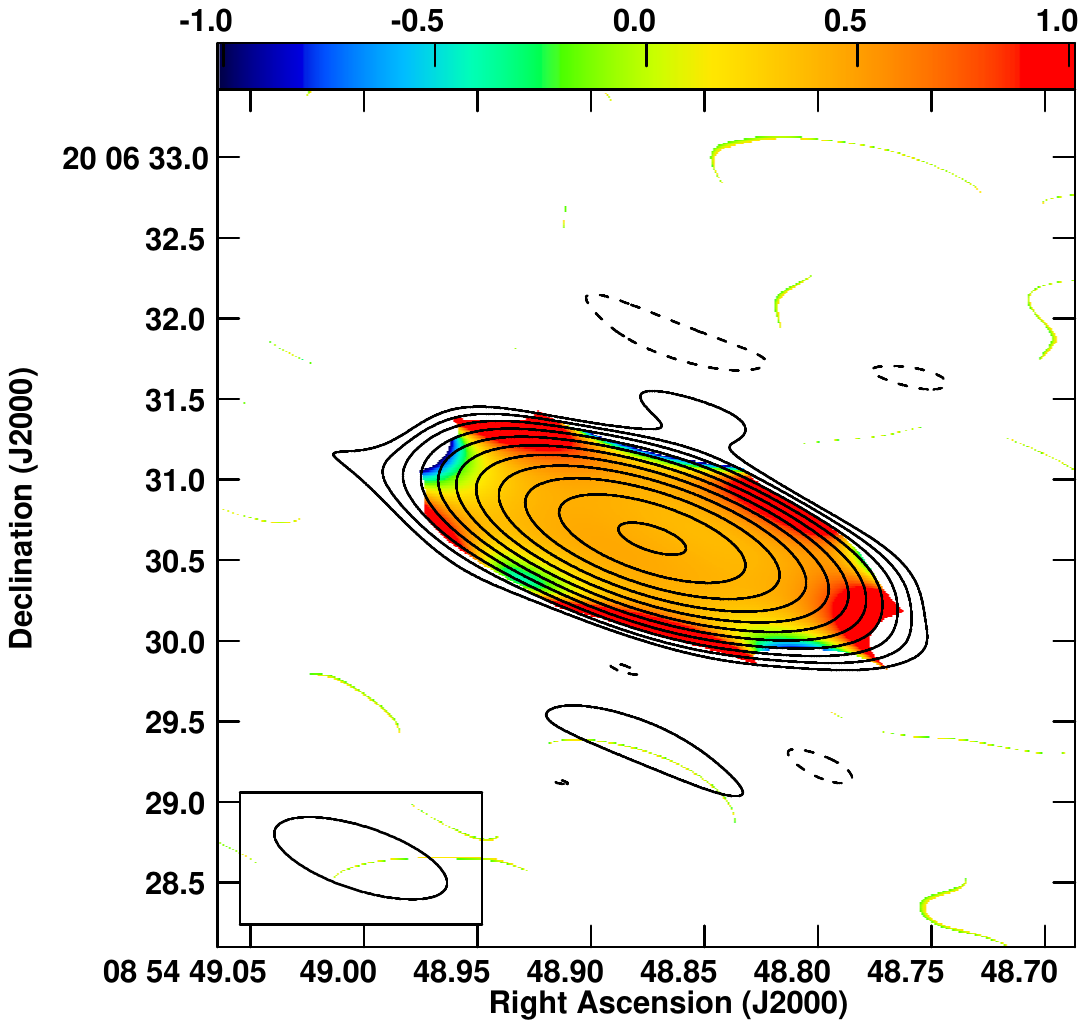}
\caption{\small VLA 6~GHz contour image of BL Lac PG 0851+203 superimposed with (left) red polarized intensity vectors and (right) in-band spectral index image. The VLBI jet direction is shown by the green arrow. The beam is $1.12\arcsec \times 0.39\arcsec$ with a PA of $72\degr$. The peak surface brightness, $I_P$ is 3.833~Jy~beam$^{-1}$ and the contour levels are $I_P \times 10^{-2} \times$  $(-0.09,~0.09,~0.18,~0.35,~0.7,~1.4,~2.8,~5.6,~11.25,~22.5,~45,~90)$~Jy~beam$^{-1}$. 1$\arcsec$ length of the vector corresponds to 0.25~Jy~beam$^{-1}$.}
\label{fig1}
\end{figure*}

\begin{figure*}
\centering
\includegraphics[width=8.7cm,trim=30 270 30 270]{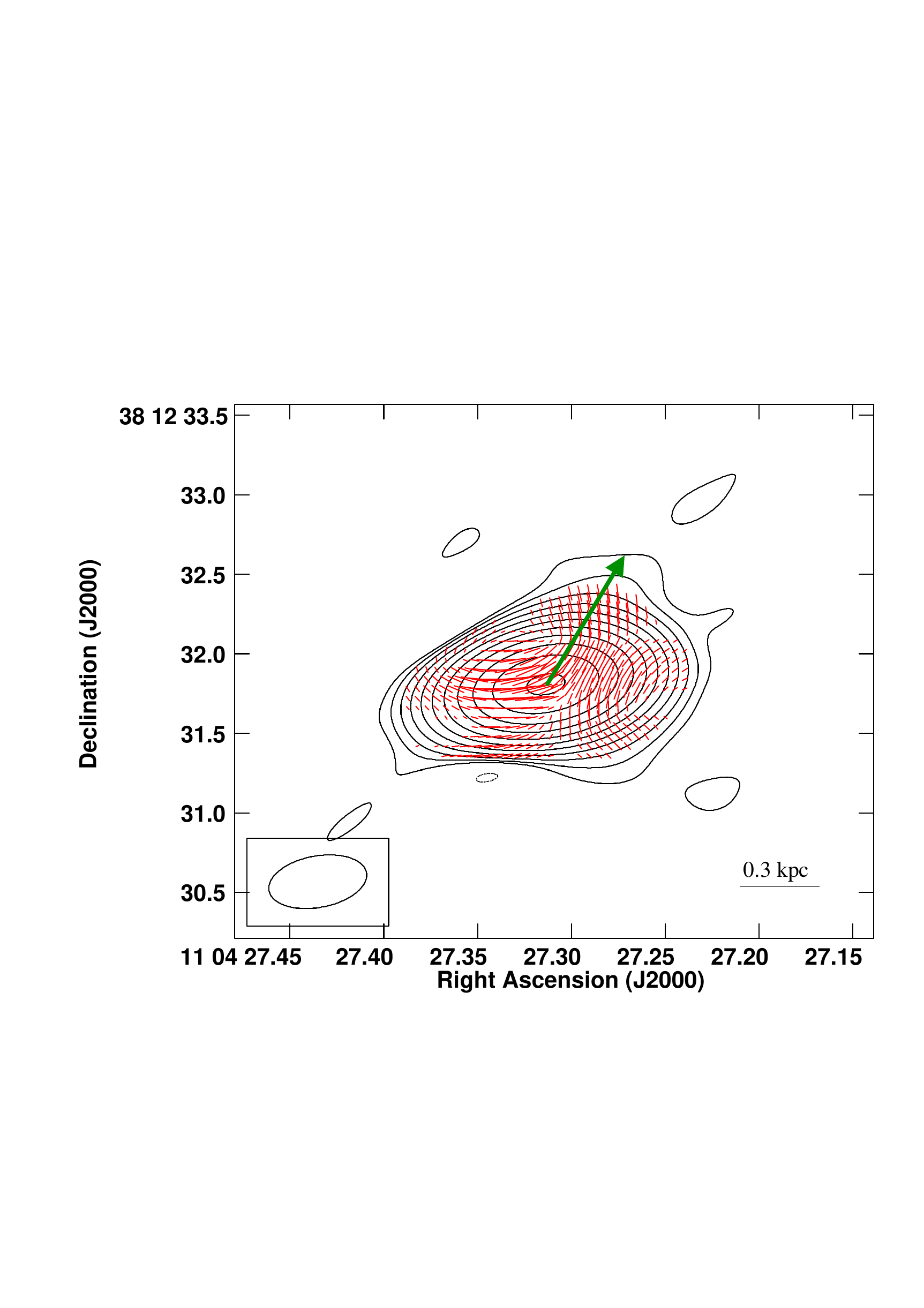}
\includegraphics[width=8.7cm,trim=30 180 30 270]{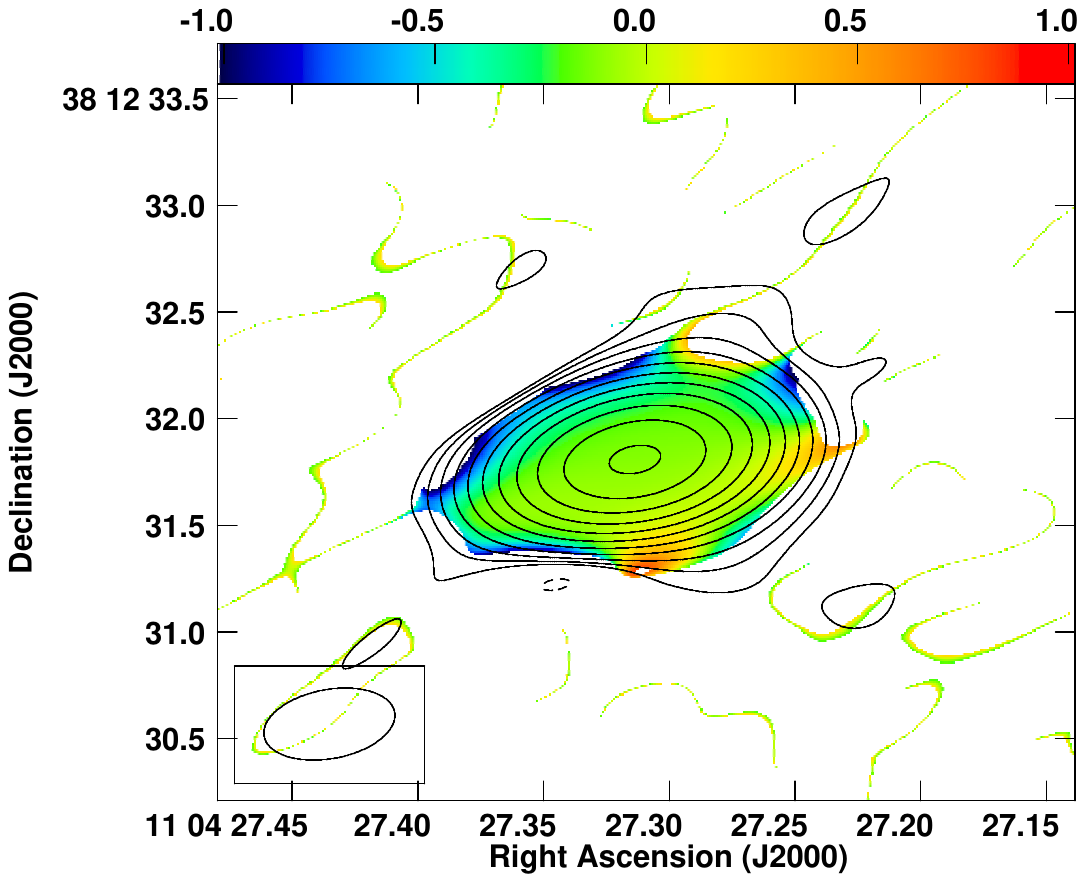}
\caption{\small VLA 6~GHz contour image of BL Lac PG 1101+384 superimposed with (left) red polarized intensity vectors and (right) in-band spectral index image. The VLBI jet direction is shown by the green arrow. The beam is $0.62\arcsec \times 0.32\arcsec$ with a PA of $-80\degr$. The peak surface brightness, $I_P$ is 0.38~Jy~beam$^{-1}$ and the contour levels are $I_P \times 10^{-2} \times$  $(-0.09,~0.09,~0.18,~0.35,~0.7,~1.4,~2.8,~5.6,~11.25,~22.5,~45,~90)$~Jy~beam$^{-1}$. 1$\arcsec$ length of the vector corresponds to 25~mJy~beam$^{-1}$.}
\label{fig2}
\end{figure*}
\begin{figure*}
\centering
\includegraphics[width=8.7cm,trim=30 220 30 220]{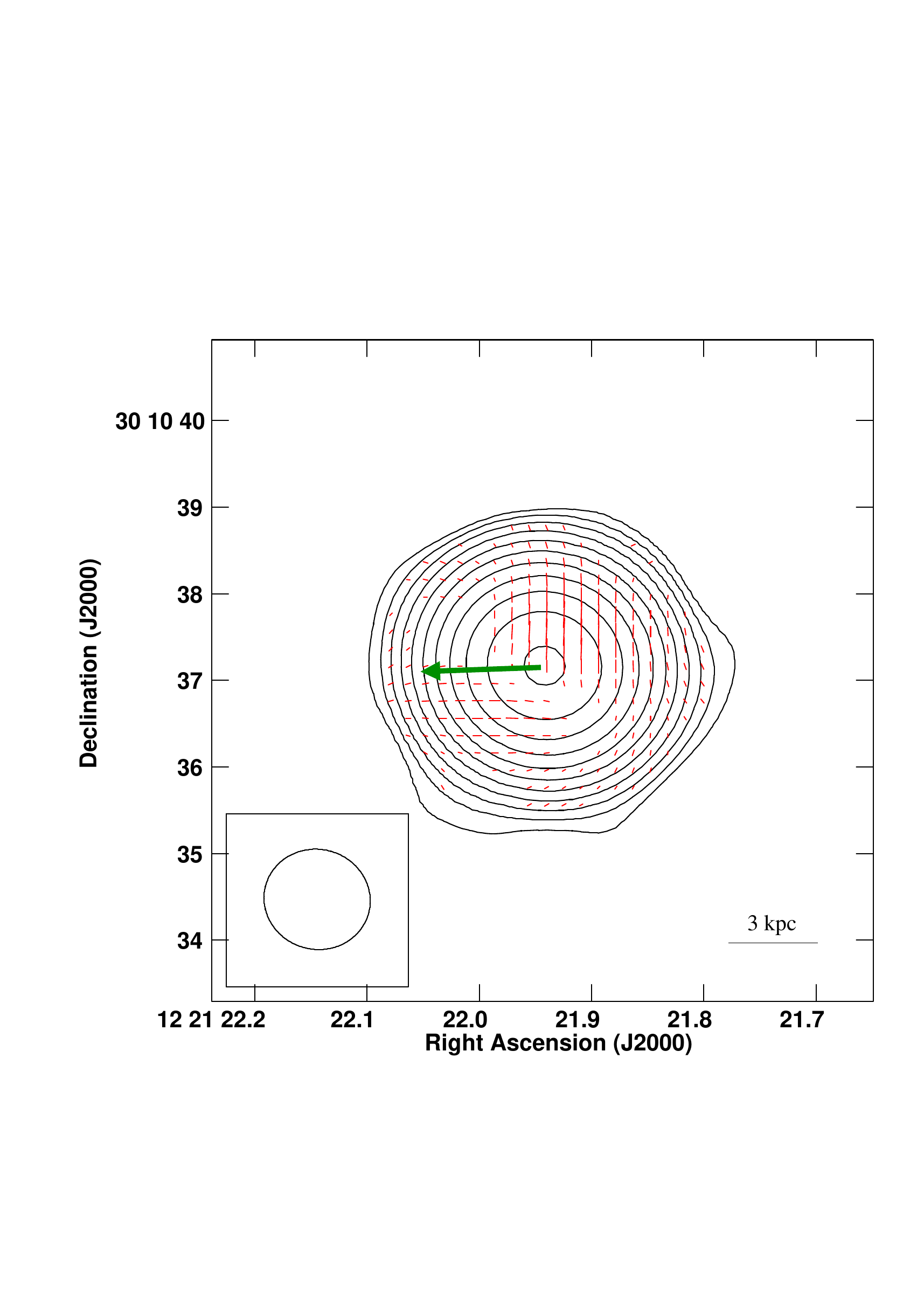}
\includegraphics[width=8.5cm,trim=30 140 30 220]{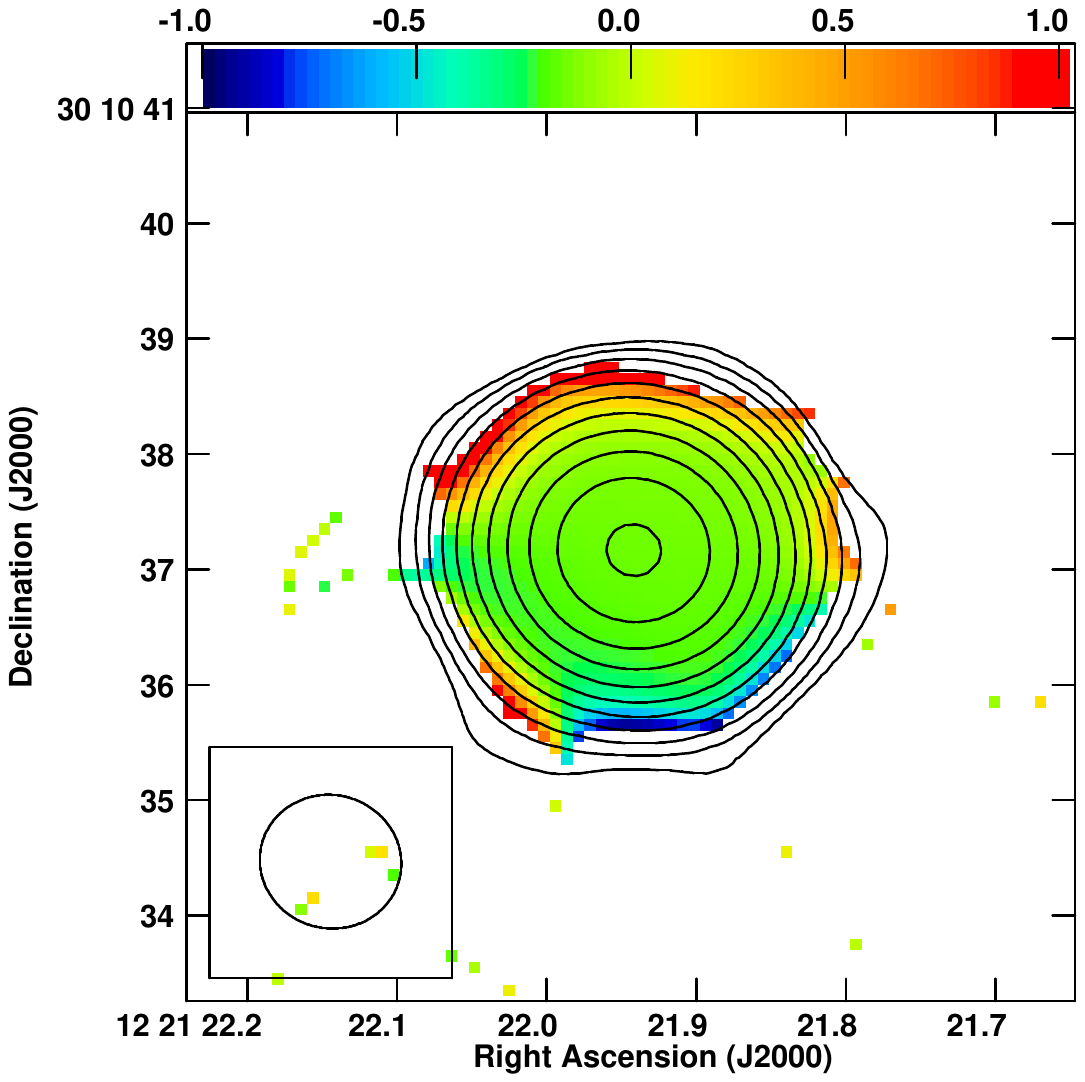}
\caption{\small VLA 6~GHz contour image of BL Lac PG 1218+304 superimposed with (left) red polarized intensity vectors and (right) in-band spectral index image. The VLBI jet direction is shown by the green arrow. The beam is $1.23\arcsec \times 1.15\arcsec$ with a PA of $76\degr$. The peak surface brightness, $I_P$ is 48.3~mJy~beam$^{-1}$ and the contour levels are $I_P \times 10^{-2} \times$  $(-0.09,~0.09,~0.18,~0.35,~0.7,~1.4,~2.8,~5.6,~11.25,~22.5,~45,~90)$~Jy~beam$^{-1}$. 1$\arcsec$ length of the vector corresponds to 2.5~mJy~beam$^{-1}$.}
\label{fig3}
\end{figure*}
\begin{figure*}
\centering
\includegraphics[width=8.7cm,trim=30 220 30 220]{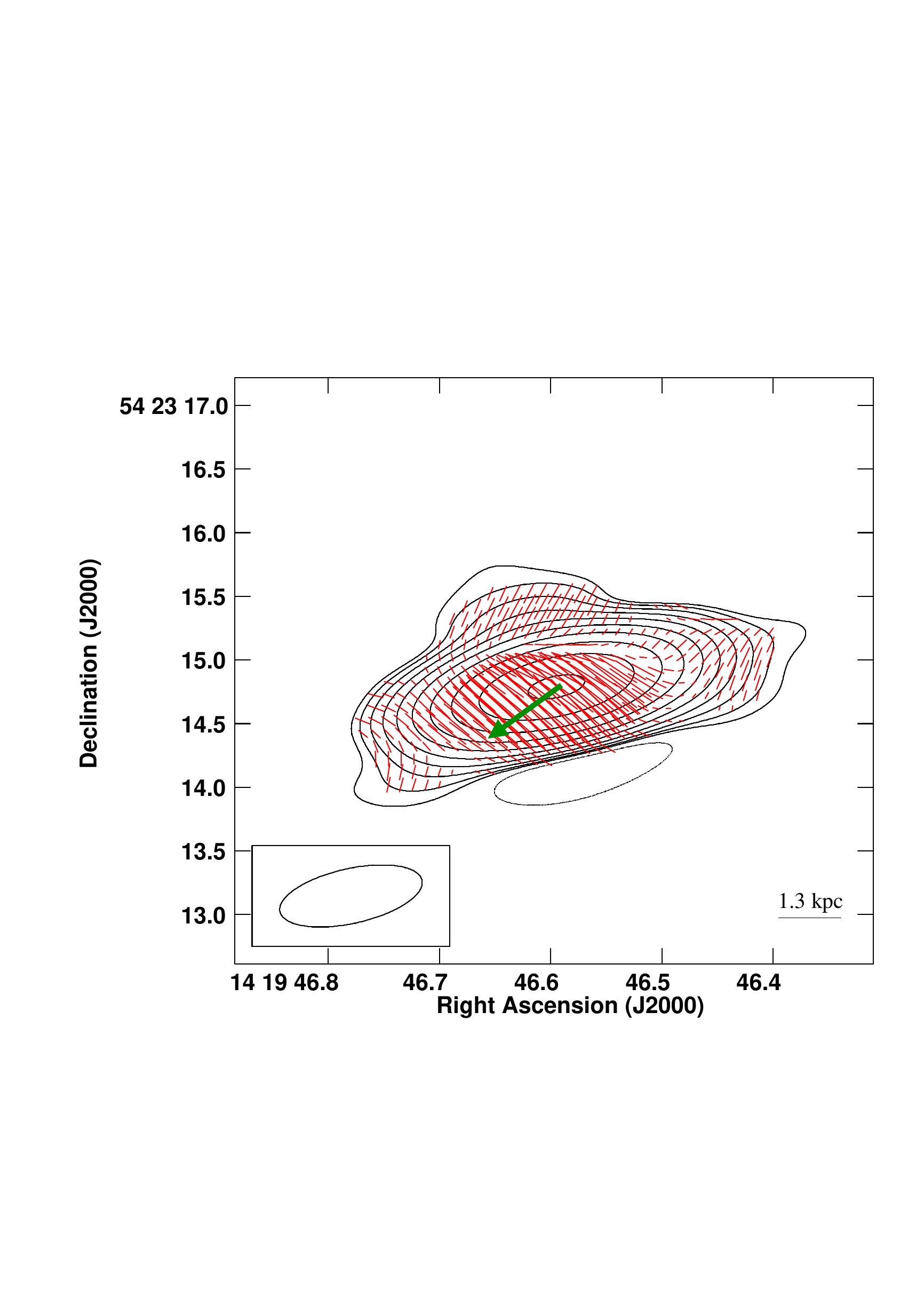}
\includegraphics[width=8.5cm,trim=30 140 30 220]{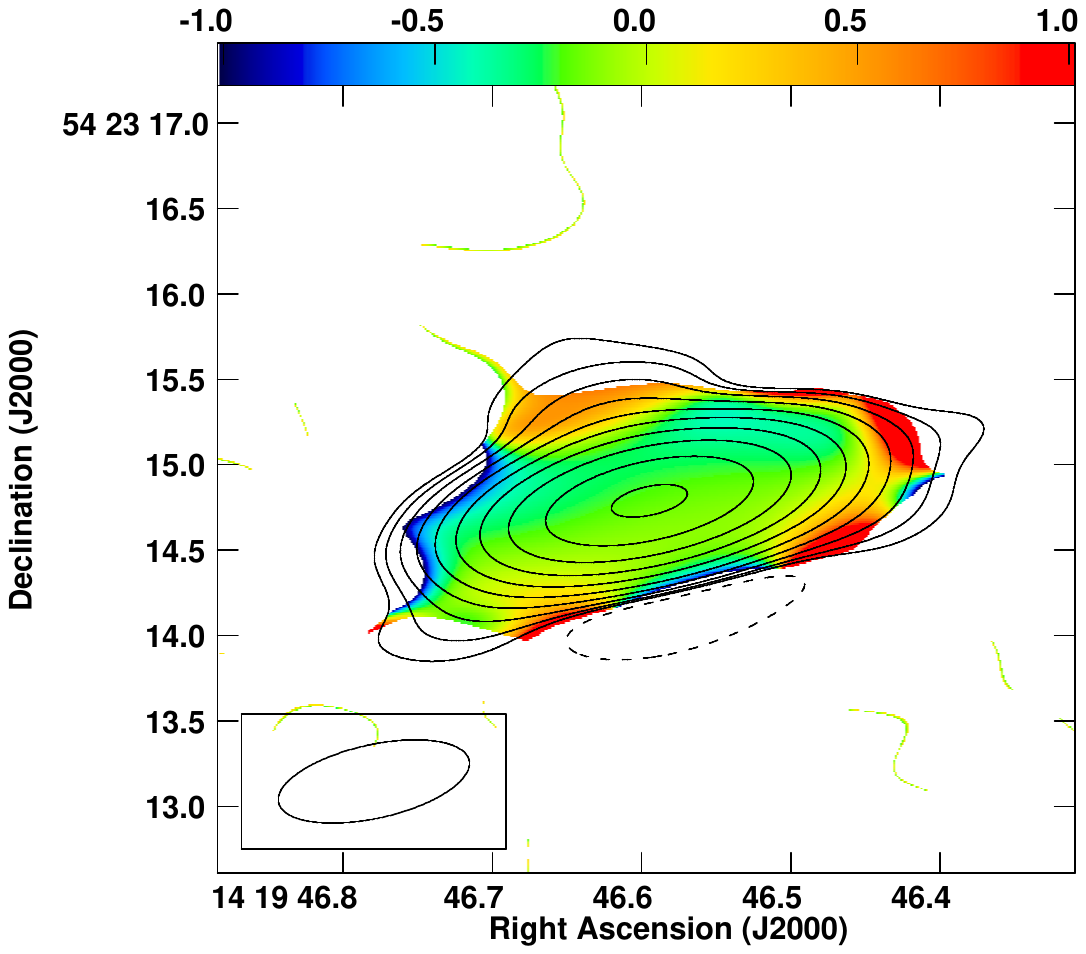}
\caption{\small VLA 6~GHz contour image of BL Lac PG 1418+546 superimposed with (left) red polarized intensity vectors and (right) in-band spectral index image. The VLBI jet direction is shown by the green arrow. The beam is $1.14\arcsec \times 0.43\arcsec$ with a PA of $-78\degr$. The peak surface brightness, $I_P$ is 0.63~Jy~beam$^{-1}$ and the contour levels are $I_P \times 10^{-2} \times$  $(-0.18,~0.18,~0.35,~0.7,~1.4,~2.8,~5.6,~11.25,~22.5,~45,~90)$~Jy~beam$^{-1}$. 1$\arcsec$ length of the vector corresponds to 25~mJy~beam$^{-1}$.}
\label{fig4}
\end{figure*}
\begin{figure*}
\centering
\includegraphics[width=8.2cm,trim=40 140 40 170]{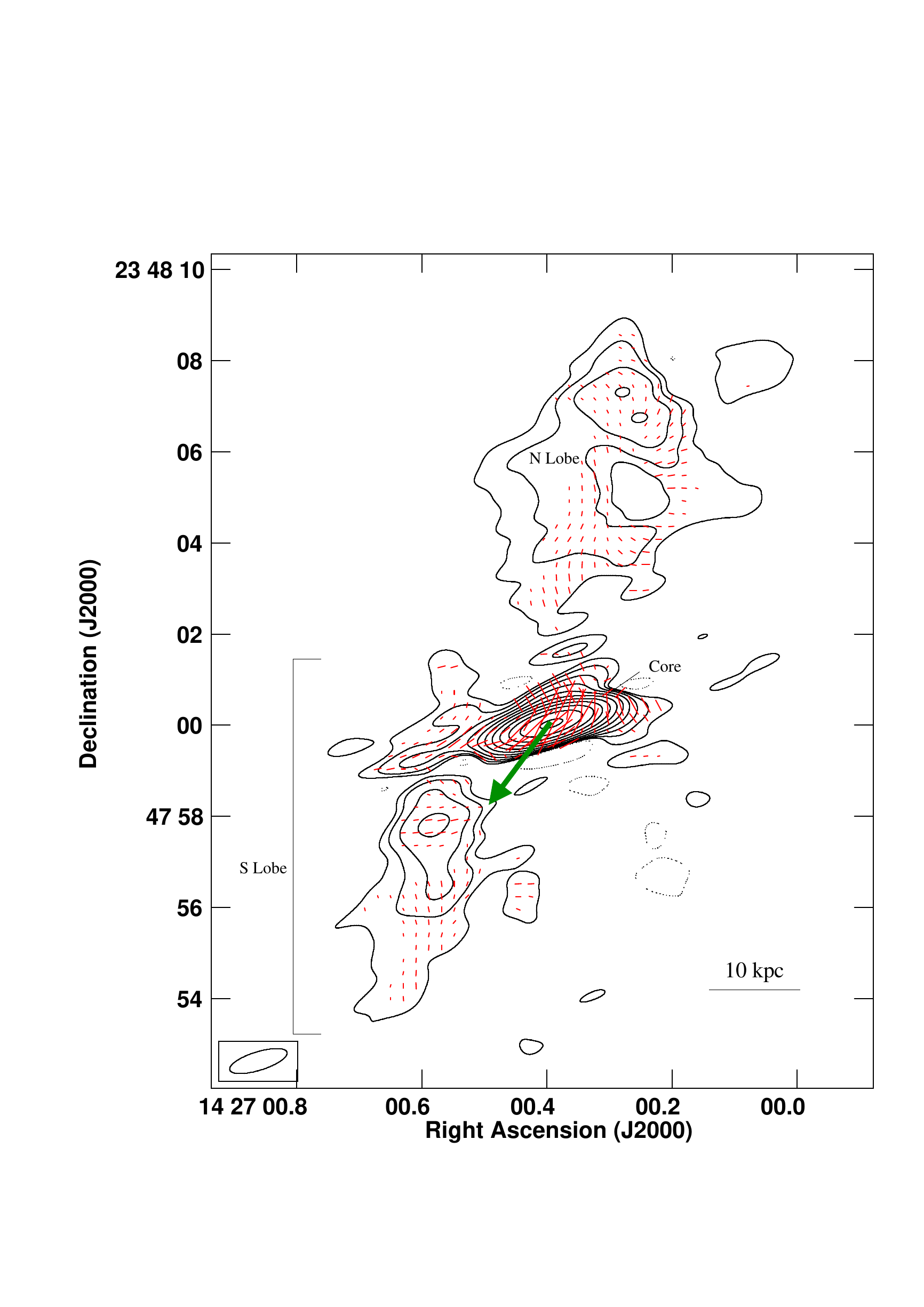}
\includegraphics[width=8.2cm,trim=40 90 40 170]{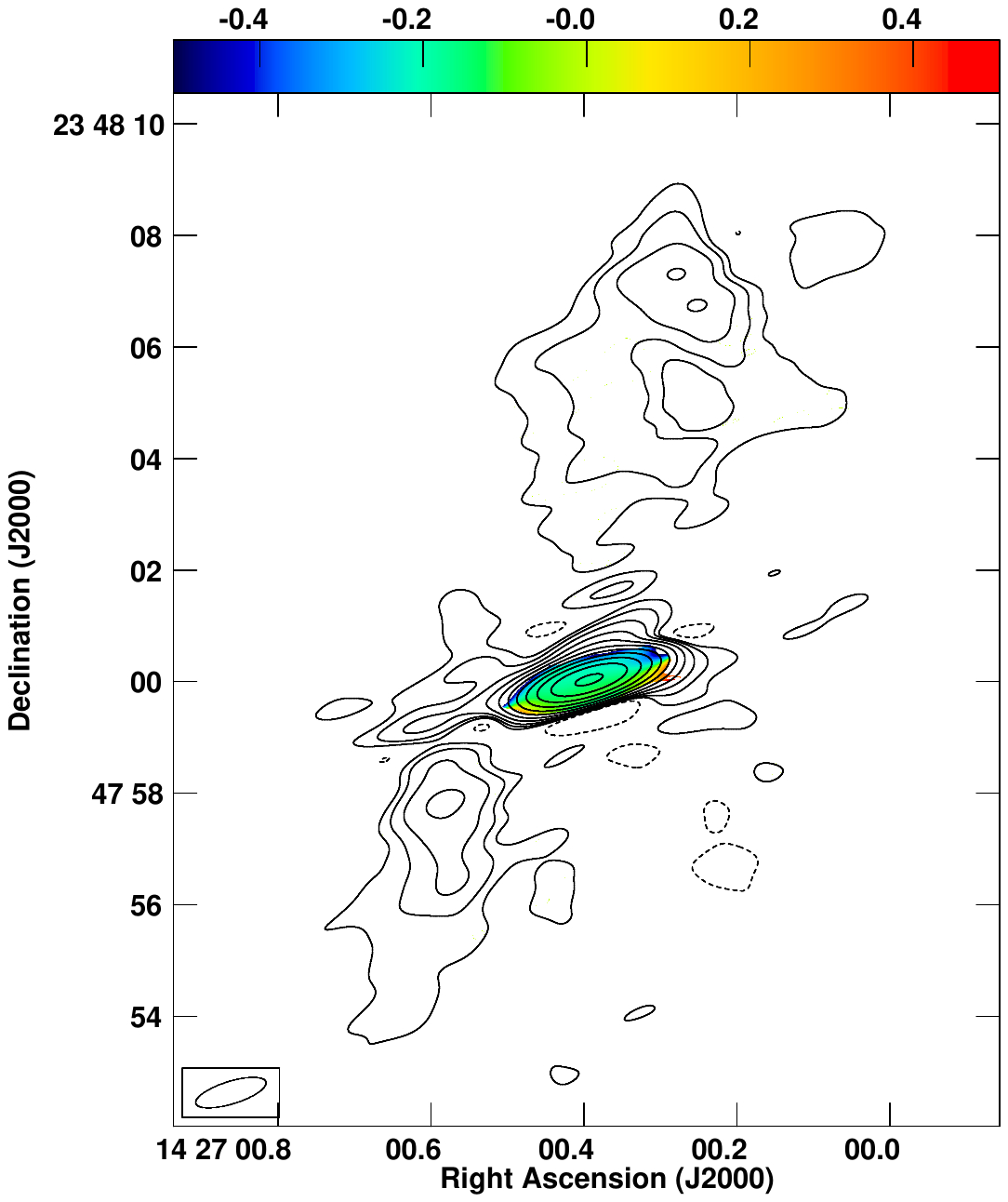}
\caption{\small VLA 6~GHz contour image of BL Lac PG 1424+240 superimposed with (left) red polarized intensity vectors and (right) in-band spectral index image. The VLBI jet direction is shown by the green arrow. The beam is $1.32\arcsec \times 0.40\arcsec$ with a PA of $-73\degr$. The peak surface brightness, $I_P$ is 0.35~Jy~beam$^{-1}$ and the contour levels are $I_P \times 10^{-2} \times$  $(-0.09,~0.09,~0.18,~0.35,~0.7,~1.4,~2.8,~5.6,~11.25,~22.5,~45,~90)$~Jy~beam$^{-1}$. 1$\arcsec$ length of the vector corresponds to 5~mJy~beam$^{-1}$.}
\label{fig5}
\end{figure*}
\begin{figure*}
\includegraphics[width=8.2cm,trim=40 195 40 195]{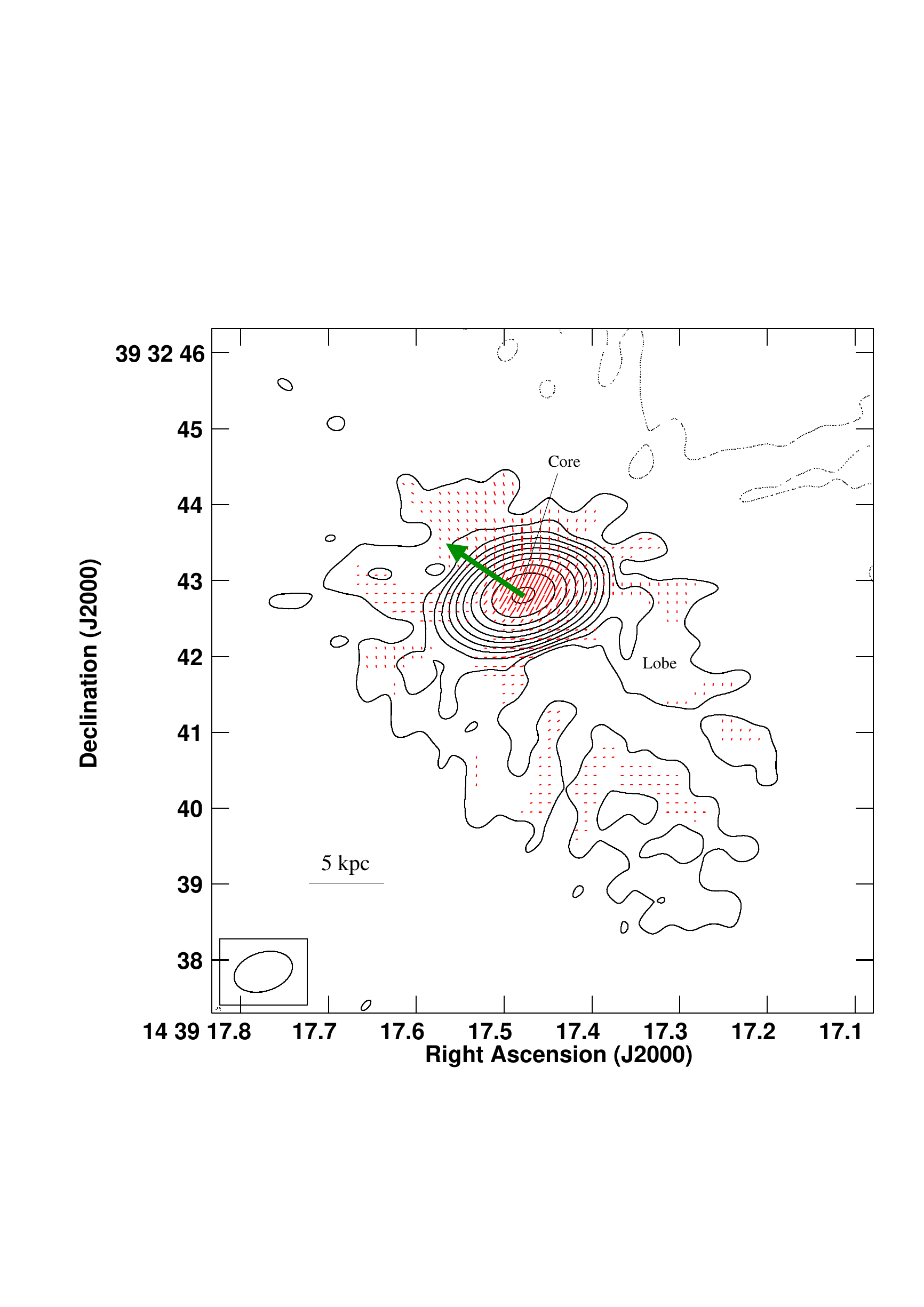}
\includegraphics[width=8.2cm,trim=40 130 40 195]{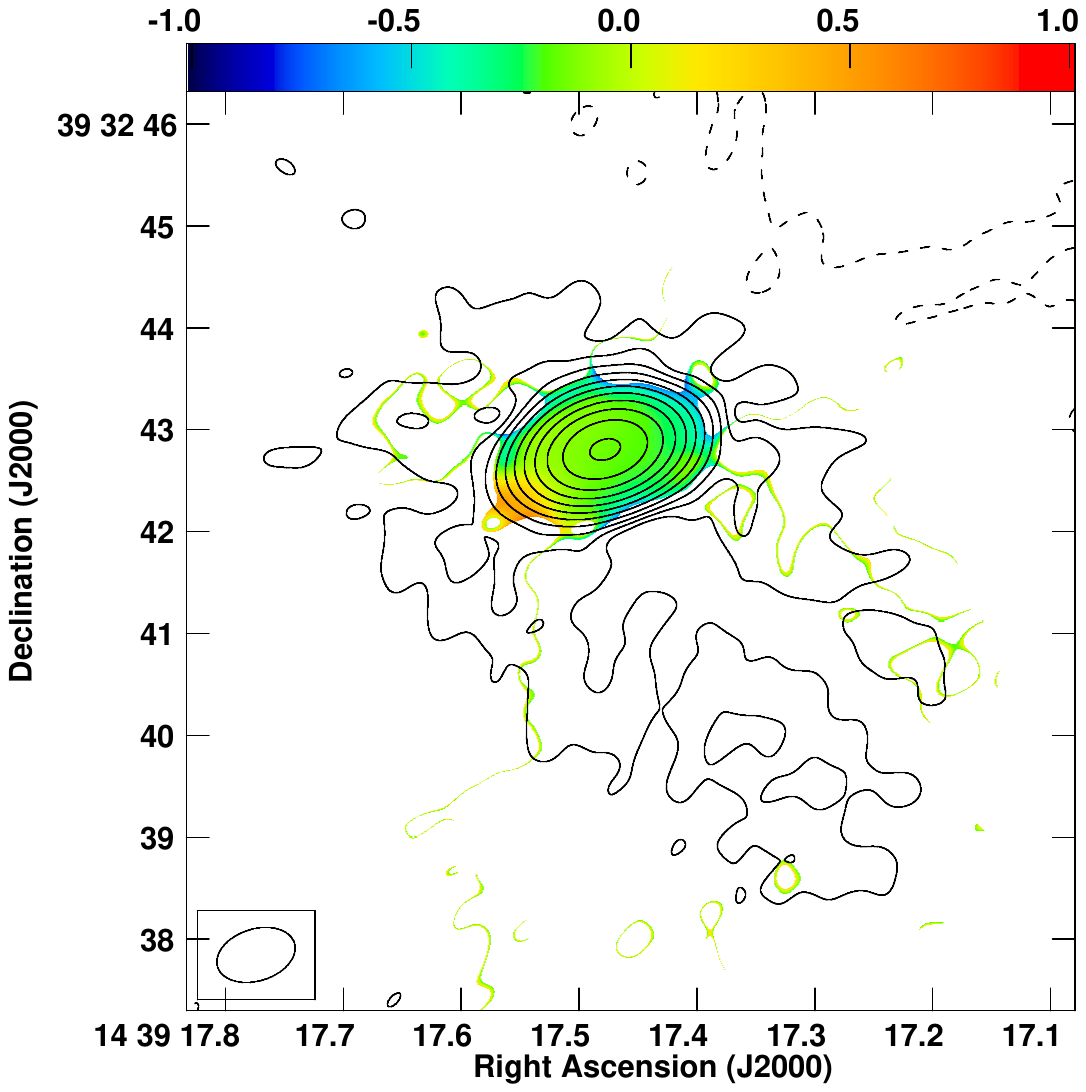}
\caption{\small VLA 6~GHz contour image of BL Lac PG 1437+398 superimposed with (left) red polarized intensity vectors and (right) in-band spectral index image. The beam is $0.79\arcsec \times 0.51\arcsec$ with a PA of $-73\degr$. The peak surface brightness, $I_P$ is 47.5~mJy~beam$^{-1}$ and the contour levels are $I_P \times 10^{-2} \times$  $(-0.09,~0.09,~0.18,~0.35,~0.7,~1.4,~2.8,~5.6,~11.25,~22.5,~45,~90)$~Jy~beam$^{-1}$. 1$\arcsec$ length of the vector corresponds to 50~mJy~beam$^{-1}$.}
\label{fig6}
\end{figure*}
\begin{figure*}
\includegraphics[width=8.7cm,trim=30 200 30 200]{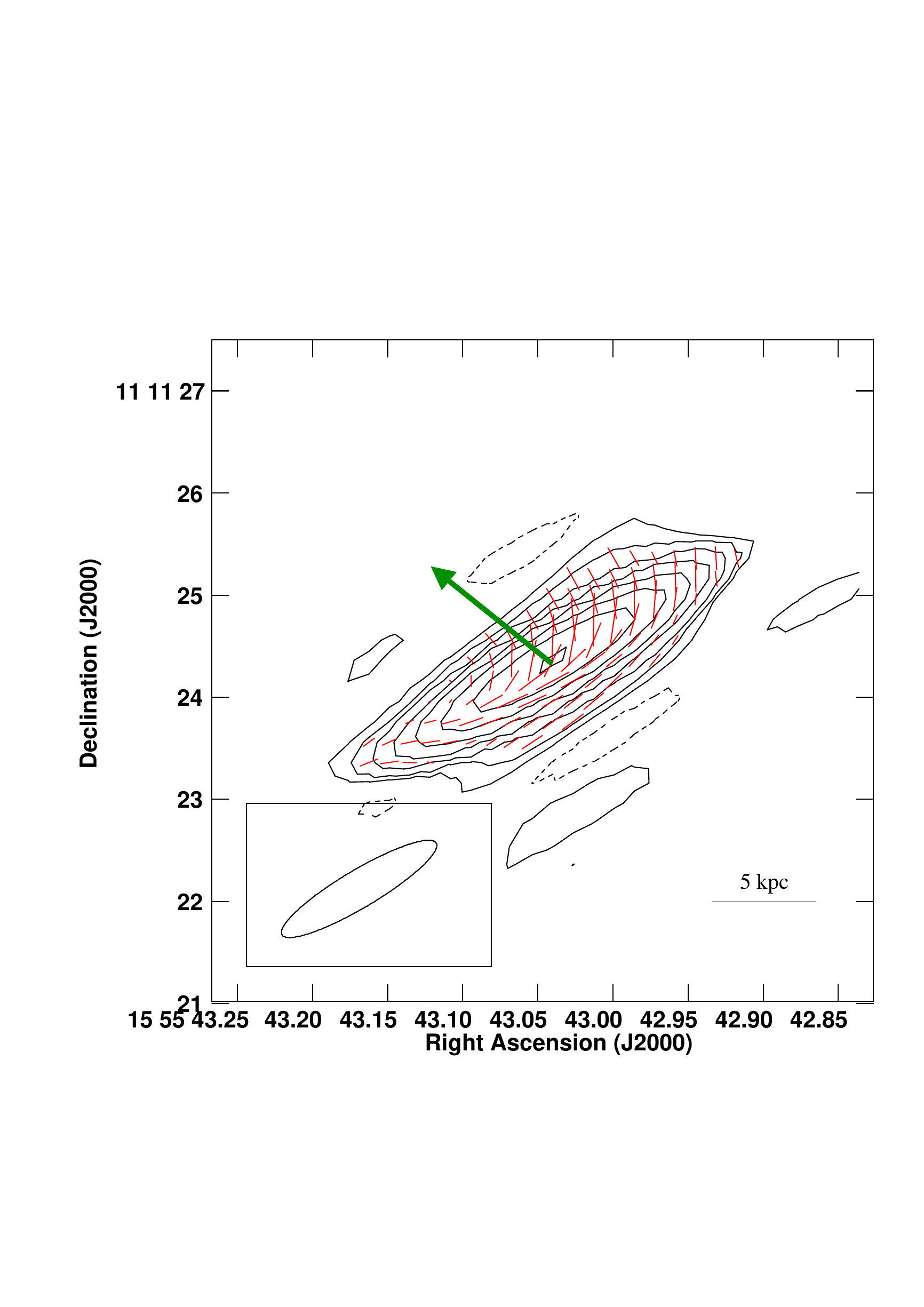}
\includegraphics[width=8.5cm,trim=30 120 30 300]{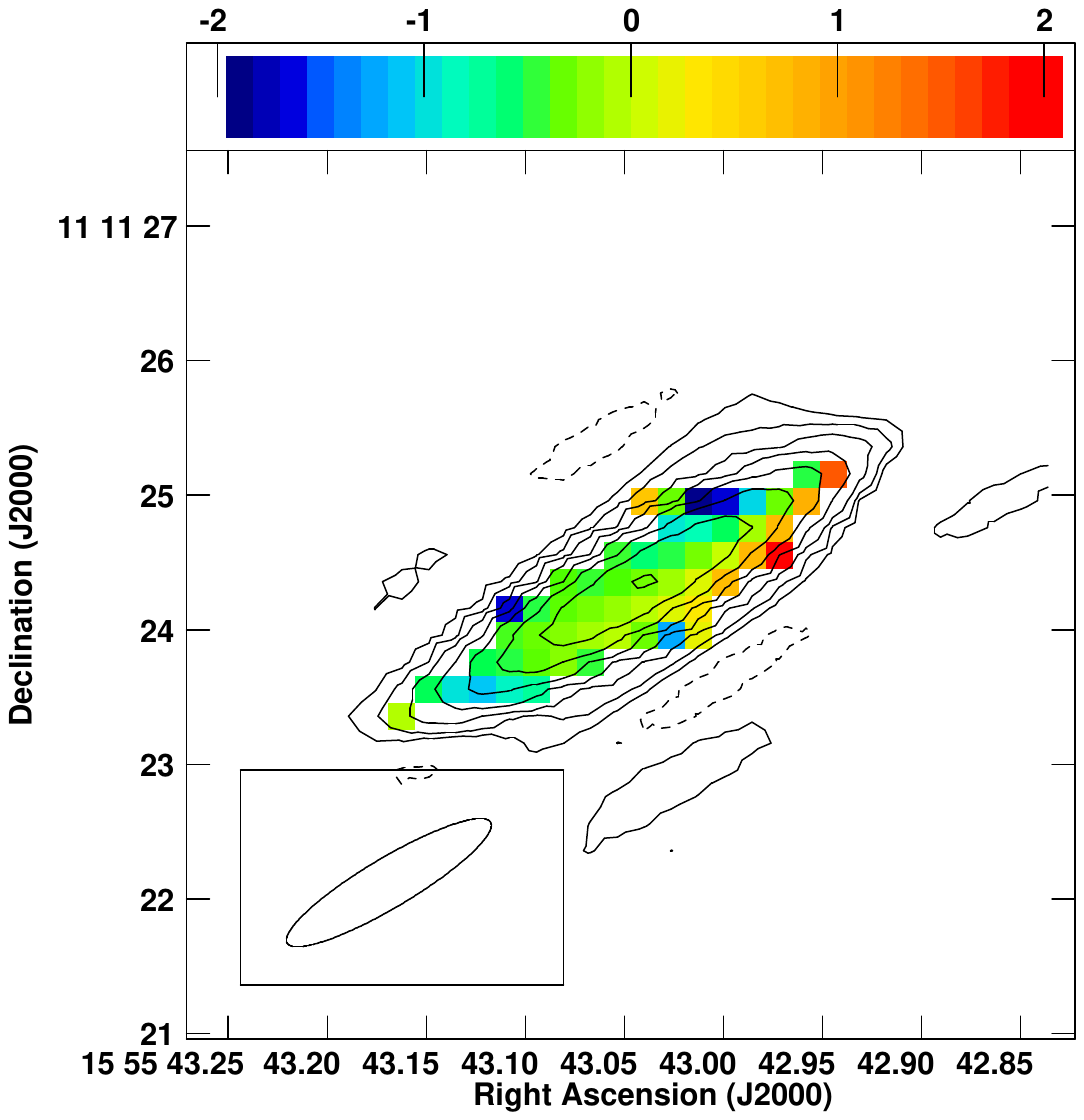}
\caption{\small VLA 6~GHz contour image of BL Lac PG 1553+113 superimposed with (left) red polarized intensity vectors and (right) in-band spectral index image. The VLBI jet direction is shown by the green arrow. The beam is $1.76\arcsec \times 0.37\arcsec$ with a PA of $-59\degr$. The peak surface brightness, $I_P$ is 0.31~Jy~beam$^{-1}$ and the contour levels are $I_P \times 10^{-2} \times$  $(-0.7,~0.7,~1.4,~2.8,~5.6,~11.25,~22.5,~45,~90)$~Jy~beam$^{-1}$. 1$\arcsec$ length of the vector corresponds to 8.33~mJy~beam$^{-1}$.}
\label{fig7}
\end{figure*}
\begin{figure*}
\includegraphics[width=8.7cm,trim=30 330 30 330]{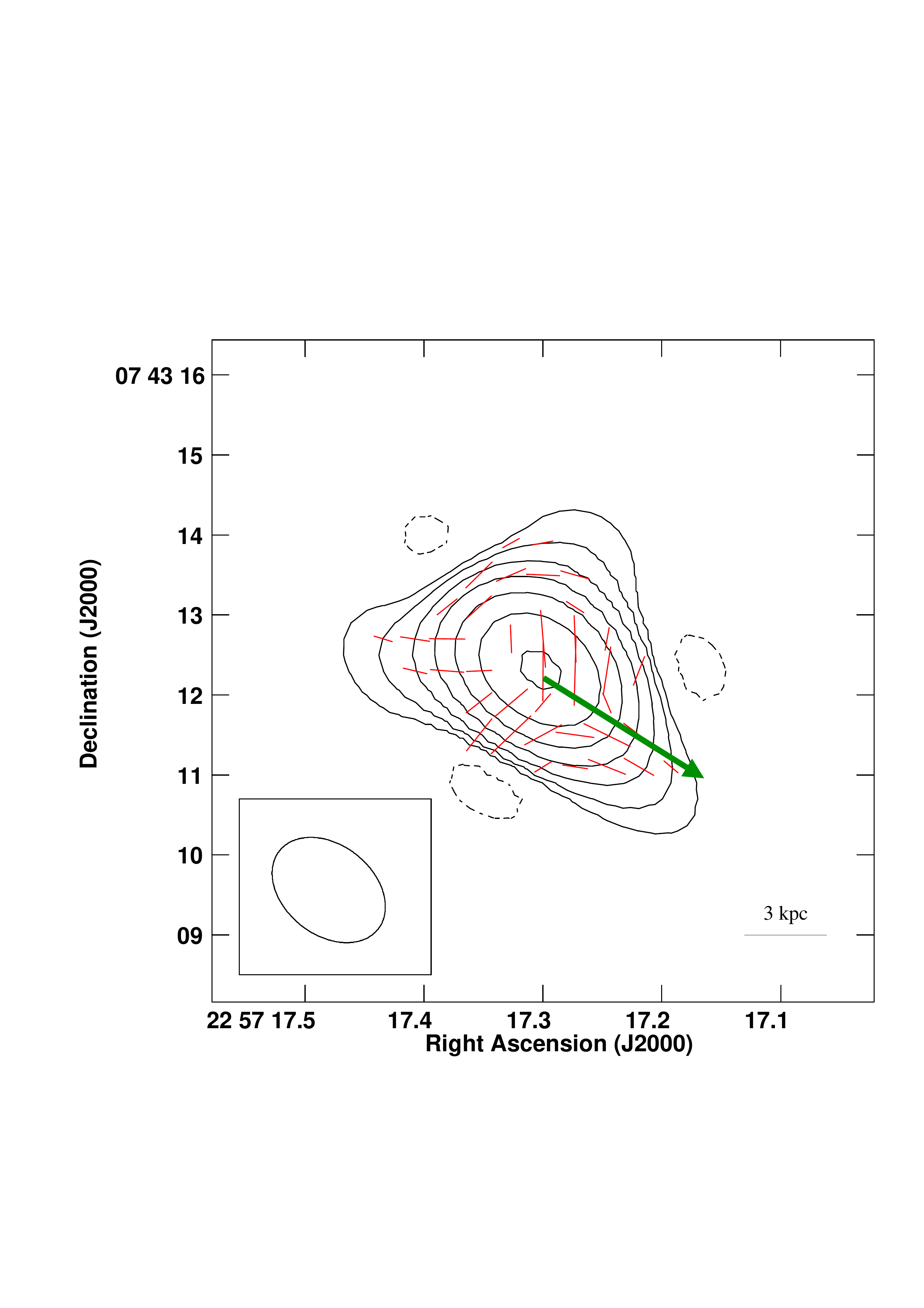}
\includegraphics[width=8.5cm,trim=30 120 30 430]{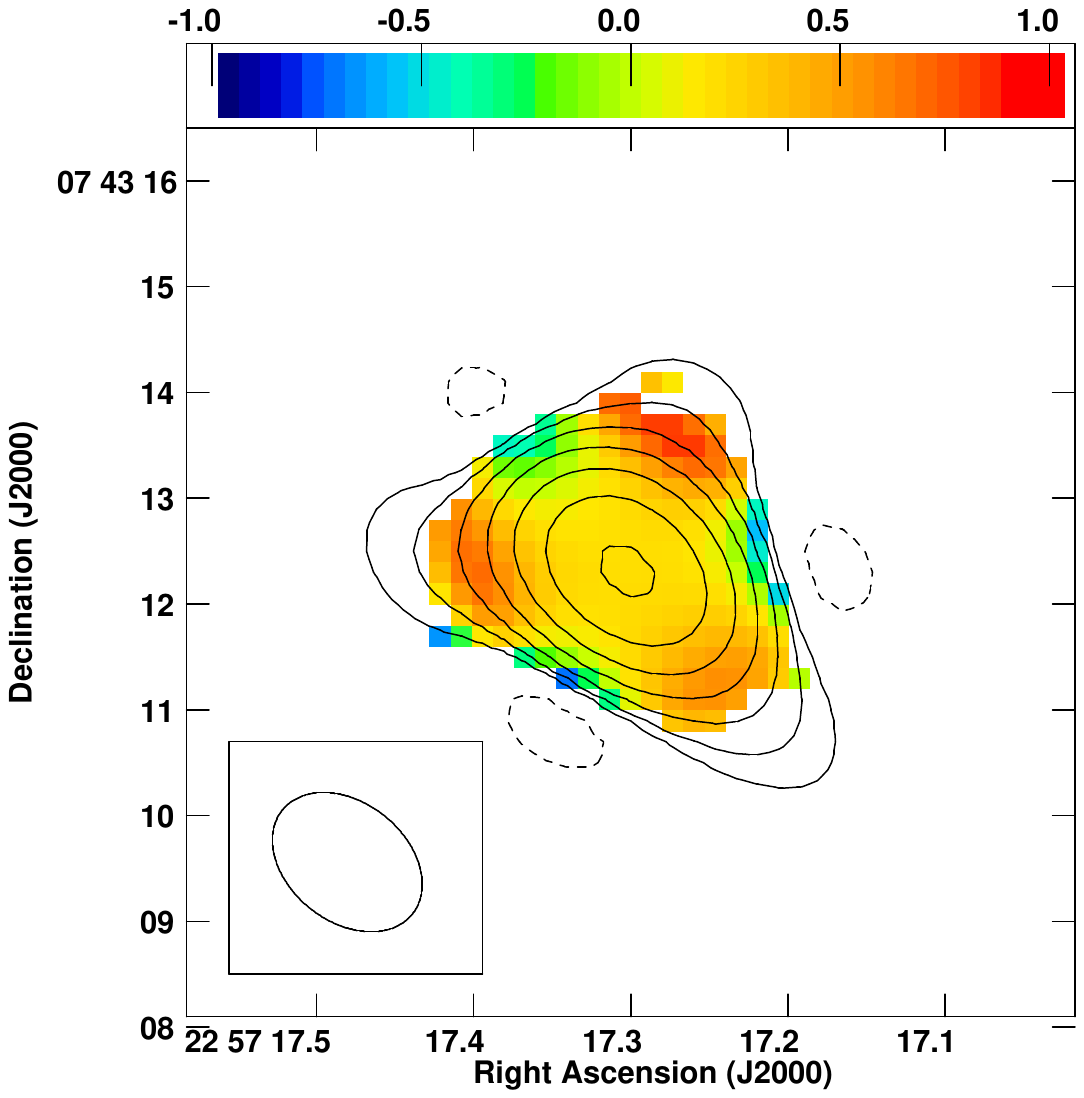}
\caption{\small VLA 6~GHz contour image of BL Lac PG 2254+075 superimposed with (left) red polarized intensity vectors and (right) in-band spectral index image. The VLBI jet direction is shown by the green arrow. The beam is $1.57\arcsec \times 1.12\arcsec$ with a PA of $51\degr$. The peak surface brightness, $I_P$ is 0.587~Jy~beam$^{-1}$ and the contour levels are $I_P \times 10^{-2} \times$  $(-1.4,~1.4,~2.8,~5.6,~11.25,~22.5,~45,~90)$~Jy~beam$^{-1}$. 1$\arcsec$ length of the vector corresponds to 2.50~mJy~beam$^{-1}$.}
\label{fig8}
\end{figure*}


\subsection{PG Quasars}\label{subsec:notesQ}
VLA images of nine of the sixteen PG quasars have been previously presented in \citet{Baghel2023}. Here we discuss the VLA 6~GHz images of the remaining seven quasars. Additional notes on individual sources are provided in the Appendix.

Of the seven quasars presented in this paper, the FSRQ PG 2209+184 displays a compact structure, whereas the other FSRQ PG 1302-102 and all SSRQs { (Figure \ref{figq1}, \ref{figq3}-\ref{figq7})} display extended polarized emission. PG 1302-102 displays a bow-shaped diffuse emission around its core towards the east but without a well-defined hotspot. SSRQs PG 1512+370, PG 1545+210, and PG 2251+113 display typical FRII-like double radio structures with well defined lobes and hotspots (except for PG 2251+113 for which higher resolution imaging than the one presented is needed to see hotspots). PG 2308+098 shows multiple jet knots in its northern jet and no clear hotspot or diffuse lobe to the north. PG 1425+267 displays a large ($\sim1$~Mpc) source with the southwestern approaching jet's hotspot discernable in both total intensity and polarization while no hotspot is visible in the counterjet lobe.

\begin{figure*}
\centering
\includegraphics[width=8.7cm,trim=30 220 30 220]{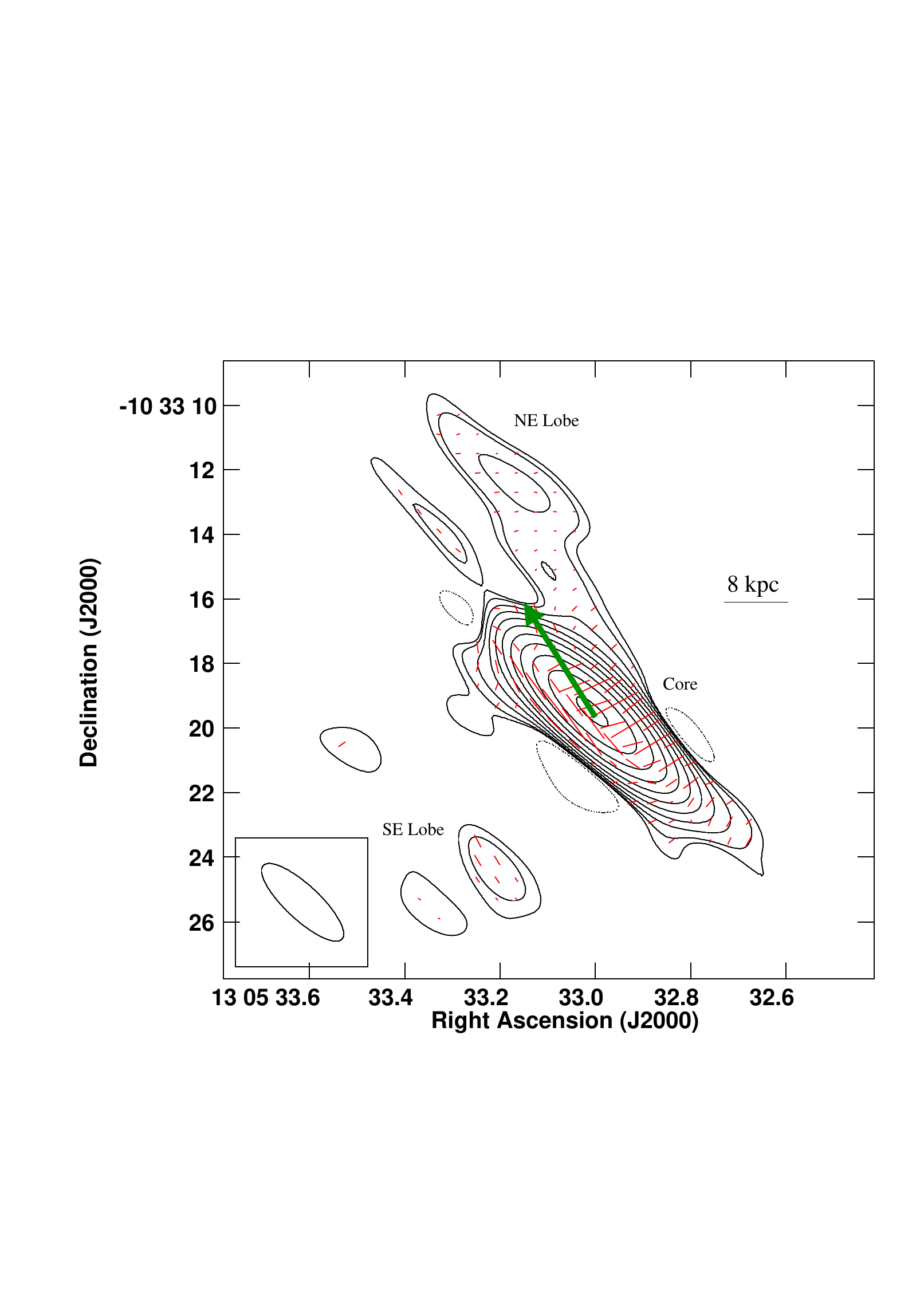}
\includegraphics[width=8.5cm,trim=30 150 30 300]{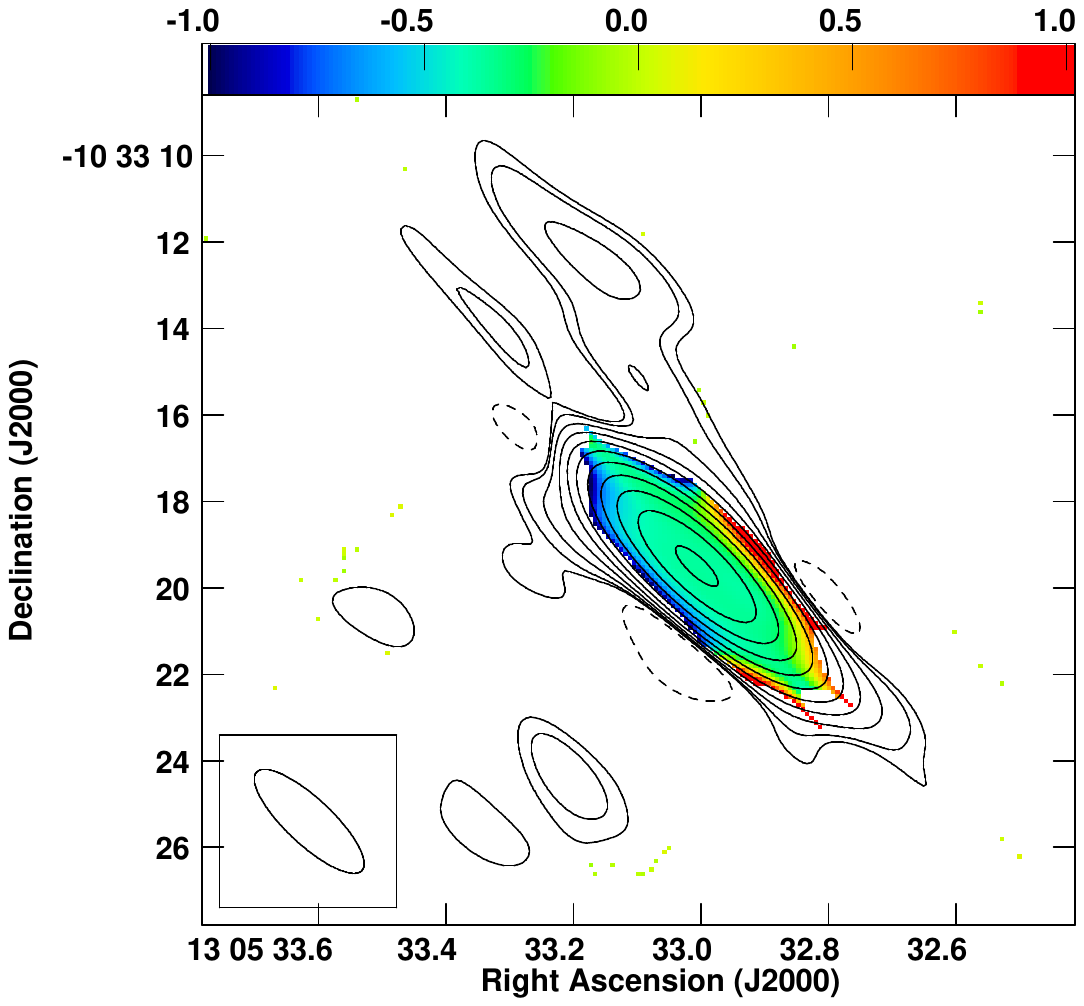}
\caption{\small VLA 6~GHz contour image of quasar PG 1302-102 superimposed with (left) red polarized intensity vectors and (right) in-band spectral index image. VLBI jet direction is shown by the green arrow. The beam is $3.33\arcsec \times 1.06\arcsec$ with a PA of $47\degr$. The peak surface brightness, $I_P$ is 0.68~Jy~beam$^{-1}$ and the contour levels are $I_P \times 10^{-2} \times$  $(-0.12,~0.12,~0.18,~0.35,~0.7,~1.4,~2.8,~5.6,~11.25,~22.5,~45,~90)$~Jy~beam$^{-1}$. 1$\arcsec$ length of the vector corresponds to 6.25~mJy~beam$^{-1}$.}
\label{figq1}
\end{figure*}

\begin{figure*}[hp]
\includegraphics[width=8.5cm,trim=50 400 50 400]{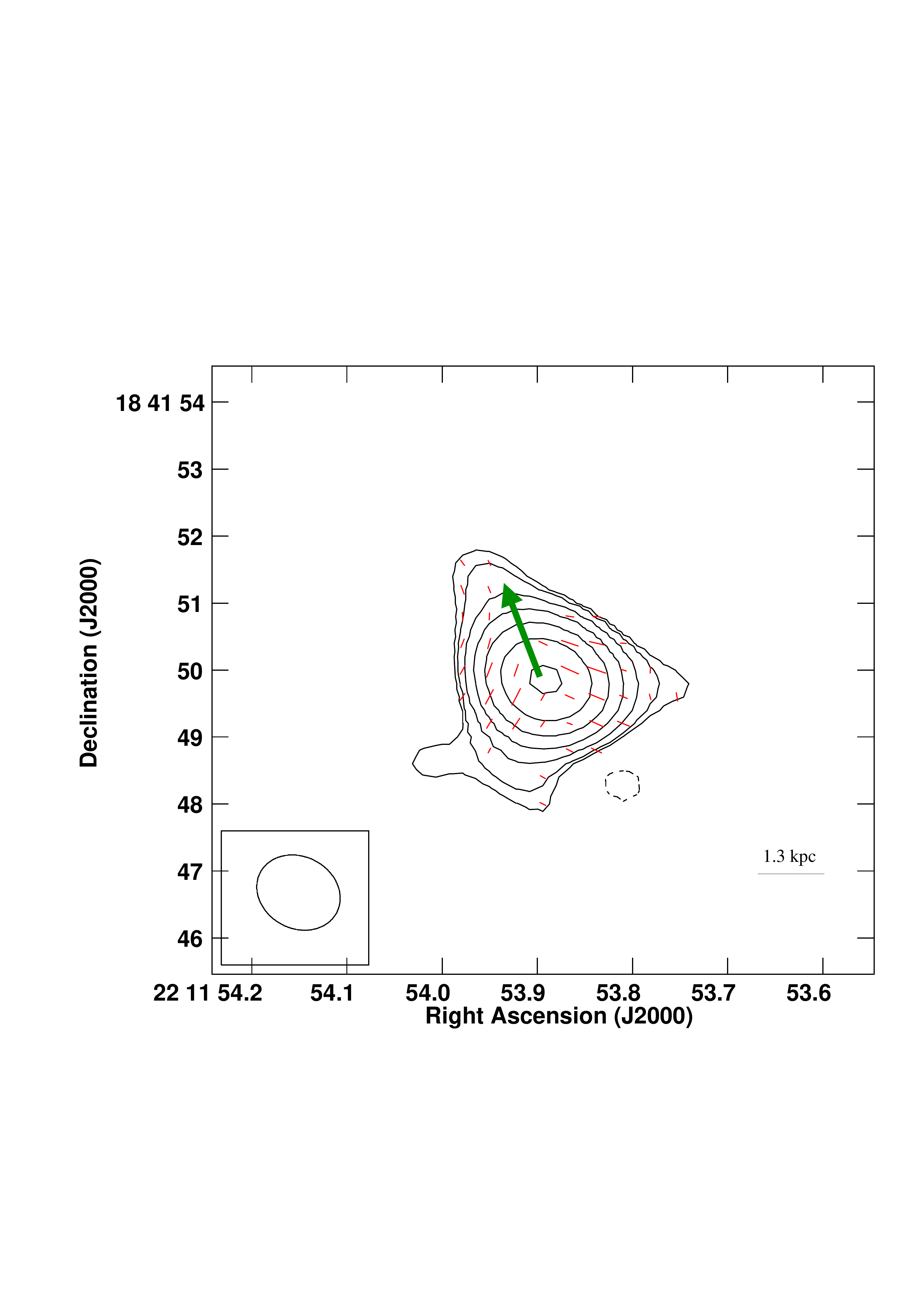}
\includegraphics[width=8.5cm,trim=40 150 40 300]{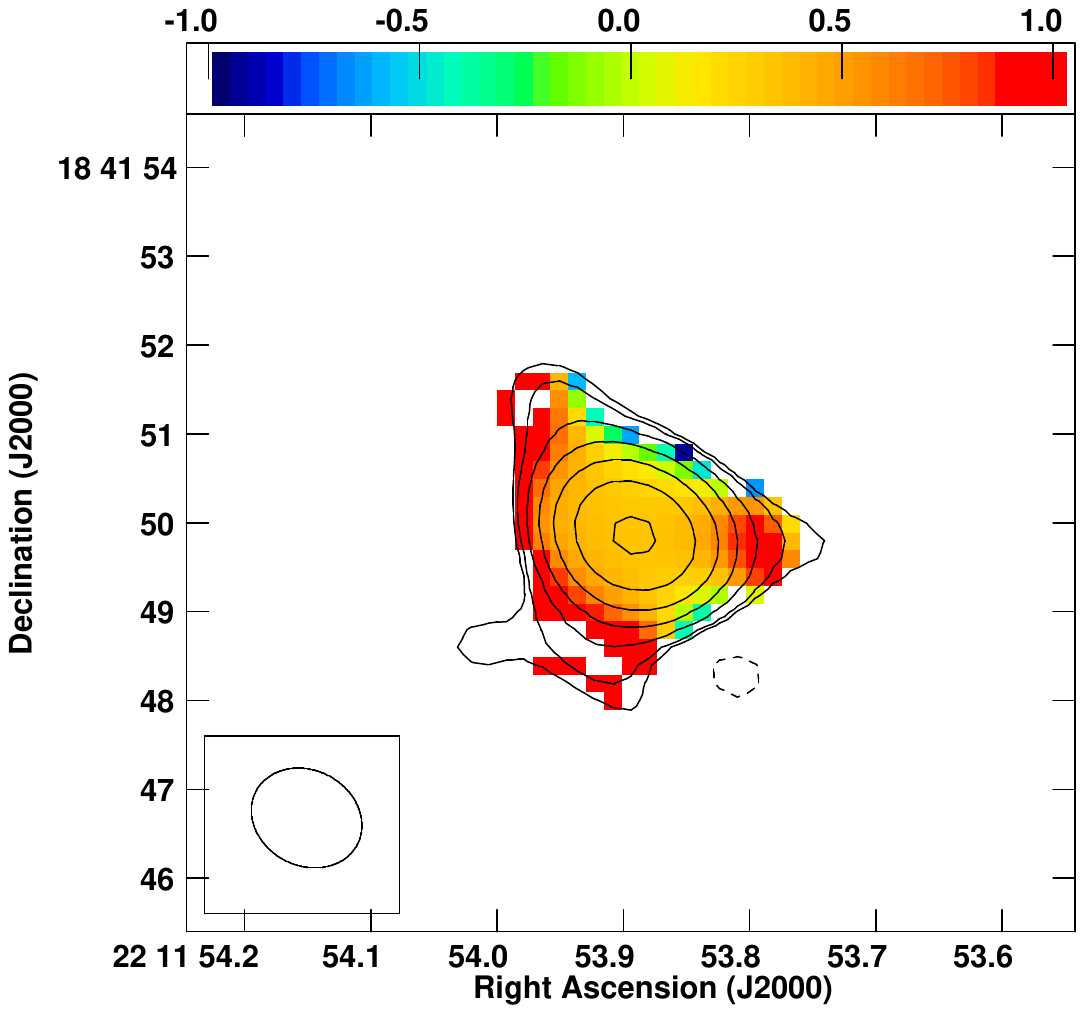}
\caption{\small VLA 6~GHz contour image of quasar PG 2209+184 superimposed with (left) red polarized intensity vectors and (right) in-band spectral index image. The beam is $1.29\arcsec \times 1.07\arcsec$ with a PA of $62\degr$. The peak surface brightness, $I_P$ is 0.103~Jy~beam$^{-1}$ and the contour levels are $I_P \times 10^{-2} \times$  $(-1.9,~1.9, ~2.8,~5.6,~11.25,~22.5,~45,~90)$~Jy~beam$^{-1}$. 1$\arcsec$ length of the vector corresponds to 1.25~mJy~beam$^{-1}$.}
\label{figq2}
\end{figure*}

\begin{figure*}[hp]
\centering
\includegraphics[width=14.6cm,trim=20 250 20 350]{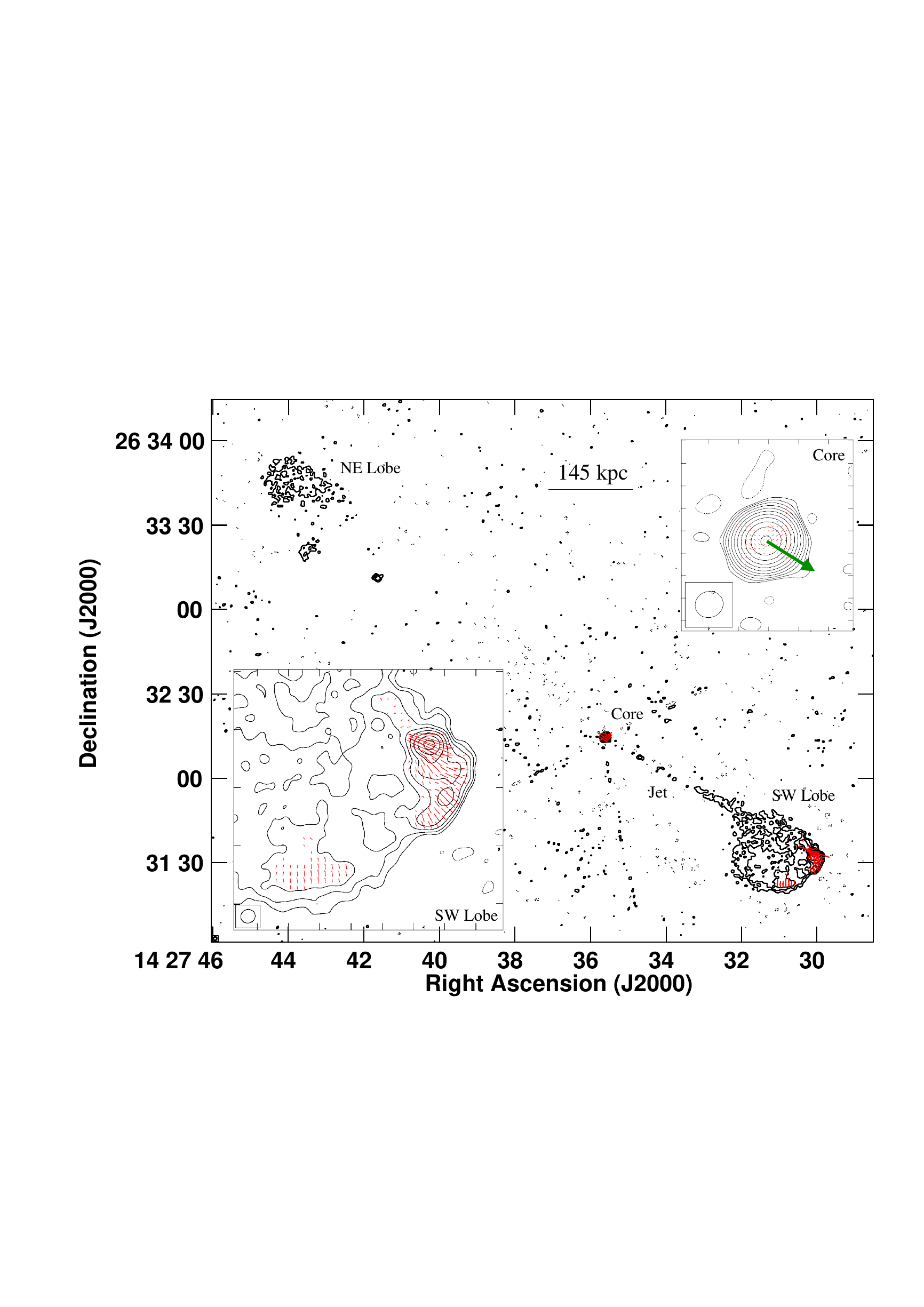}
\includegraphics[width=14.6cm,trim=0 200 22 200]{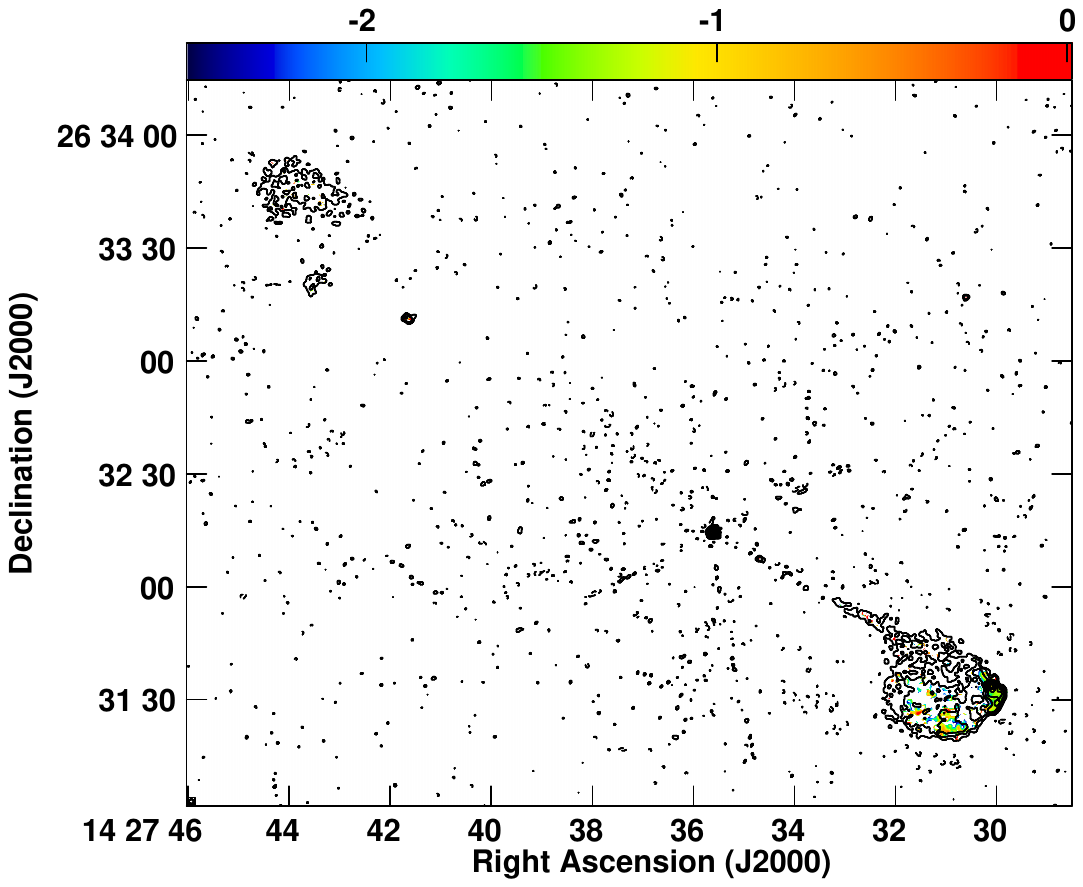}
\caption{\small VLA 6~GHz contour image of quasar PG 1425+267 superimposed with (top) red polarized intensity vectors and (bottom) in-band spectral index image. The insets show the core region and the southern hotspot region. The beam is $1.26\arcsec \times 1.15\arcsec$ with a PA of $-69\degr$. The peak surface brightness, $I_P$ is 25.90~mJy~beam$^{-1}$ and the contour levels are $I_P \times 10^{-2} \times$  $(-0.09,~0.09,~0.18,~0.35,~0.7,~1.4,~2.8,~5.6,~11.25,~22.5,~45,~90)$~Jy~beam$^{-1}$. 20$\arcsec$ length of the vector corresponds to 1~mJy~beam$^{-1}$.}
\label{figq3}
\end{figure*}

\begin{figure*}[hp]
\centering
\includegraphics[width=18.0cm,trim=40 300 40 300]{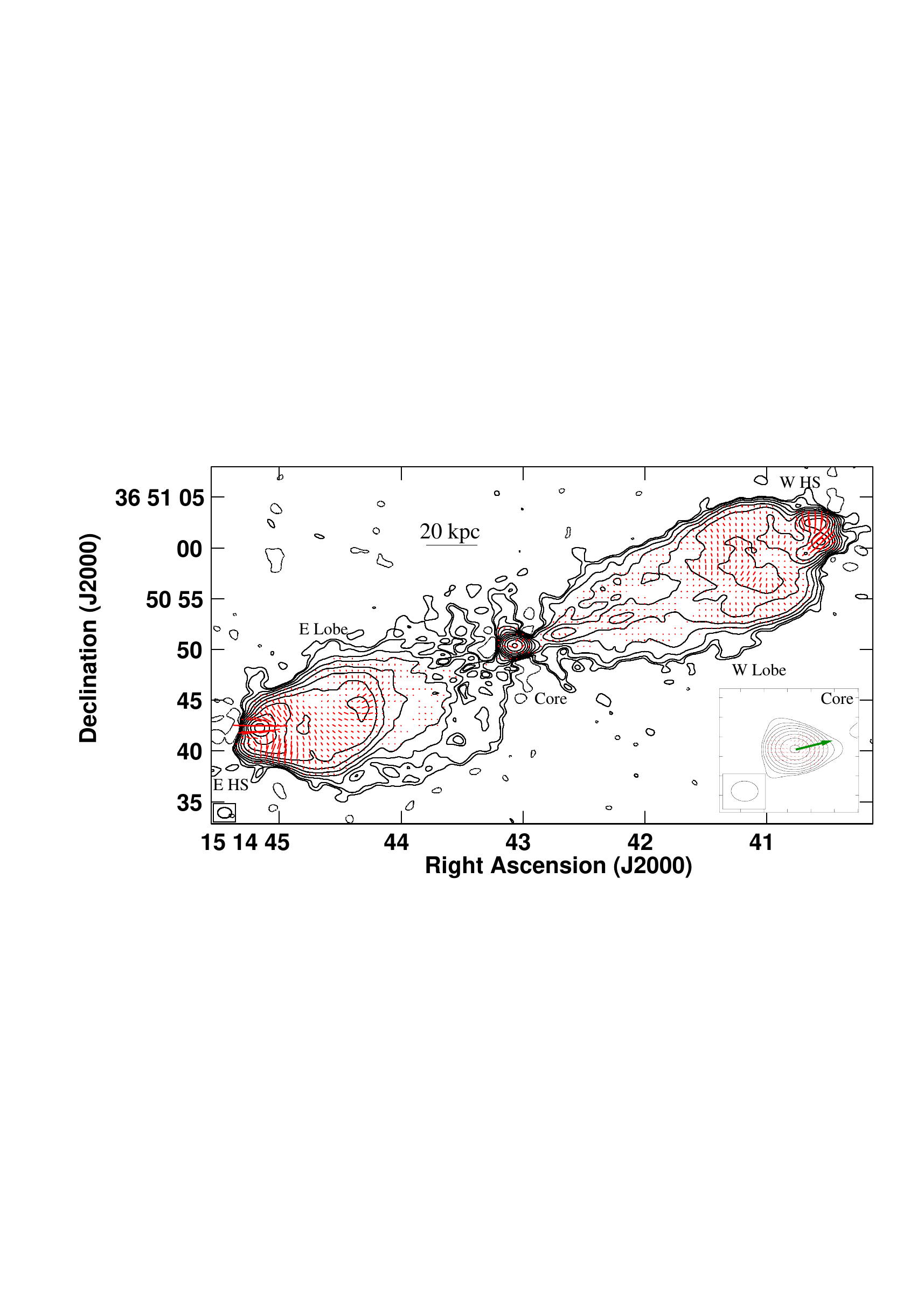}
\includegraphics[width=18.0cm,trim=40 240 40 350]{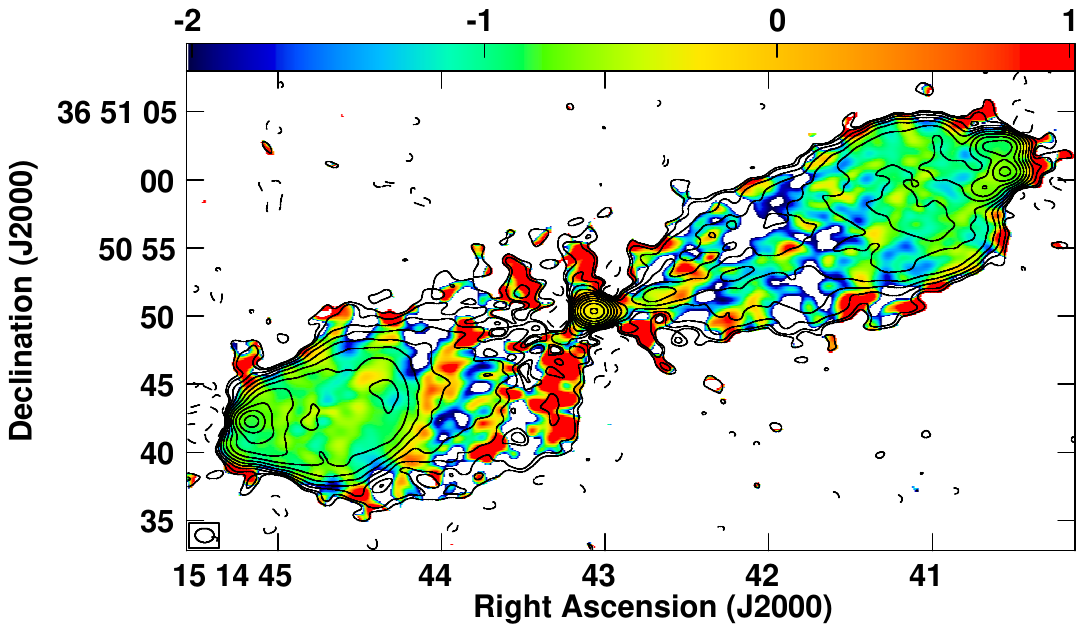}
\caption{\small VLA 6~GHz contour image of quasar PG 1512+370 superimposed with (top) red polarized intensity vectors and (bottom) in-band spectral index image. The inset shows the core region. The beam is $1.38\arcsec \times 1.03\arcsec$ with a PA of $88\degr$. The peak surface brightness, $I_P$ is 42.6~mJy~beam$^{-1}$ and the contour levels are $I_P \times 10^{-2} \times$  $(-0.06,~0.06,~0.09,~0.18,~0.35,~0.7,~1.4,~2.8,~5.6,~11.25,~22.5,~45,~90)$~Jy~beam$^{-1}$. 1$\arcsec$ length of the vector corresponds to 1~mJy~beam$^{-1}$.}
\label{figq4}
\end{figure*}

\begin{figure*}[hp]
\centering
\includegraphics[width=8.8cm,trim=90 50 90 350]{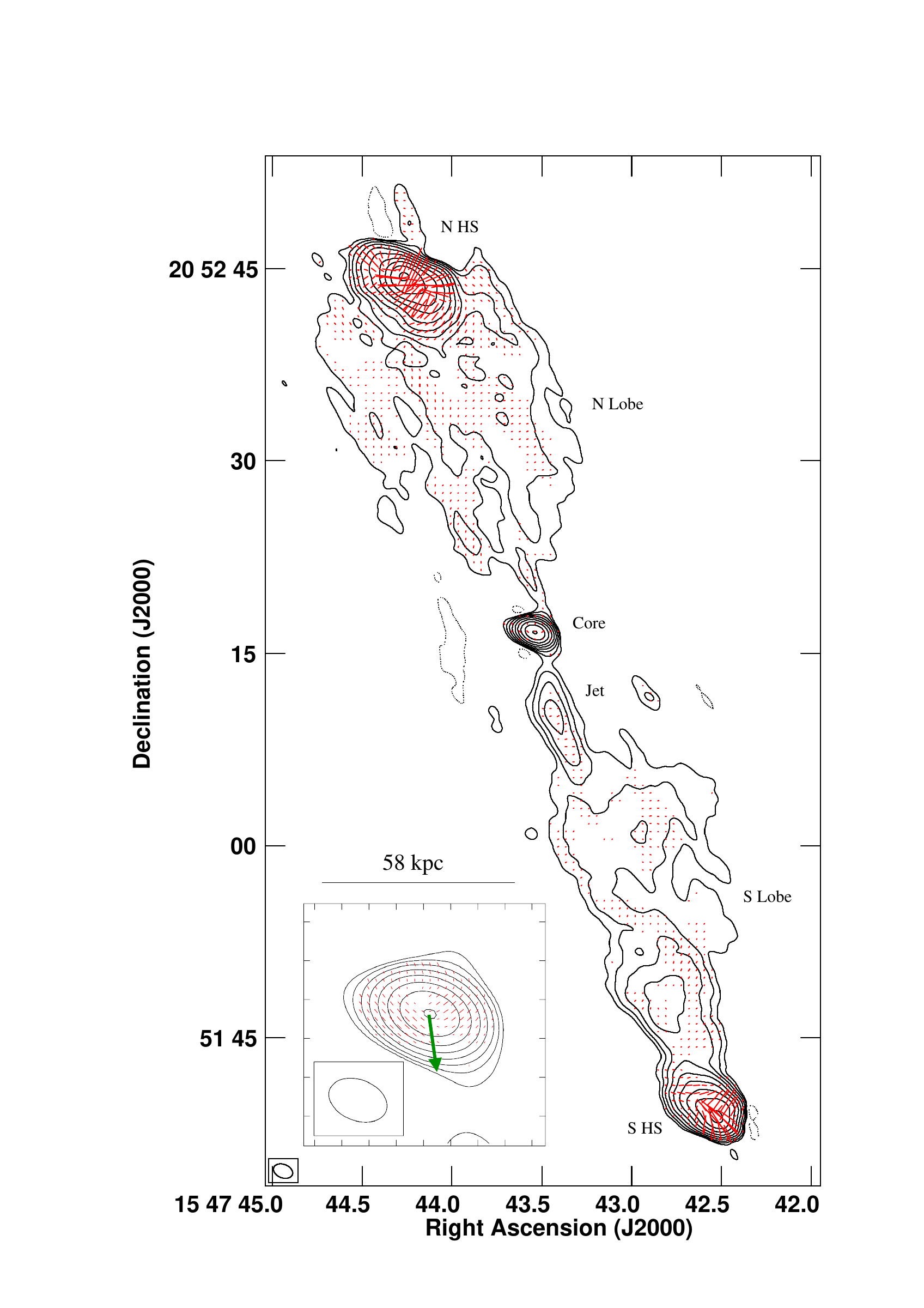}
\includegraphics[width=8.8cm,trim=80 40 80 340]{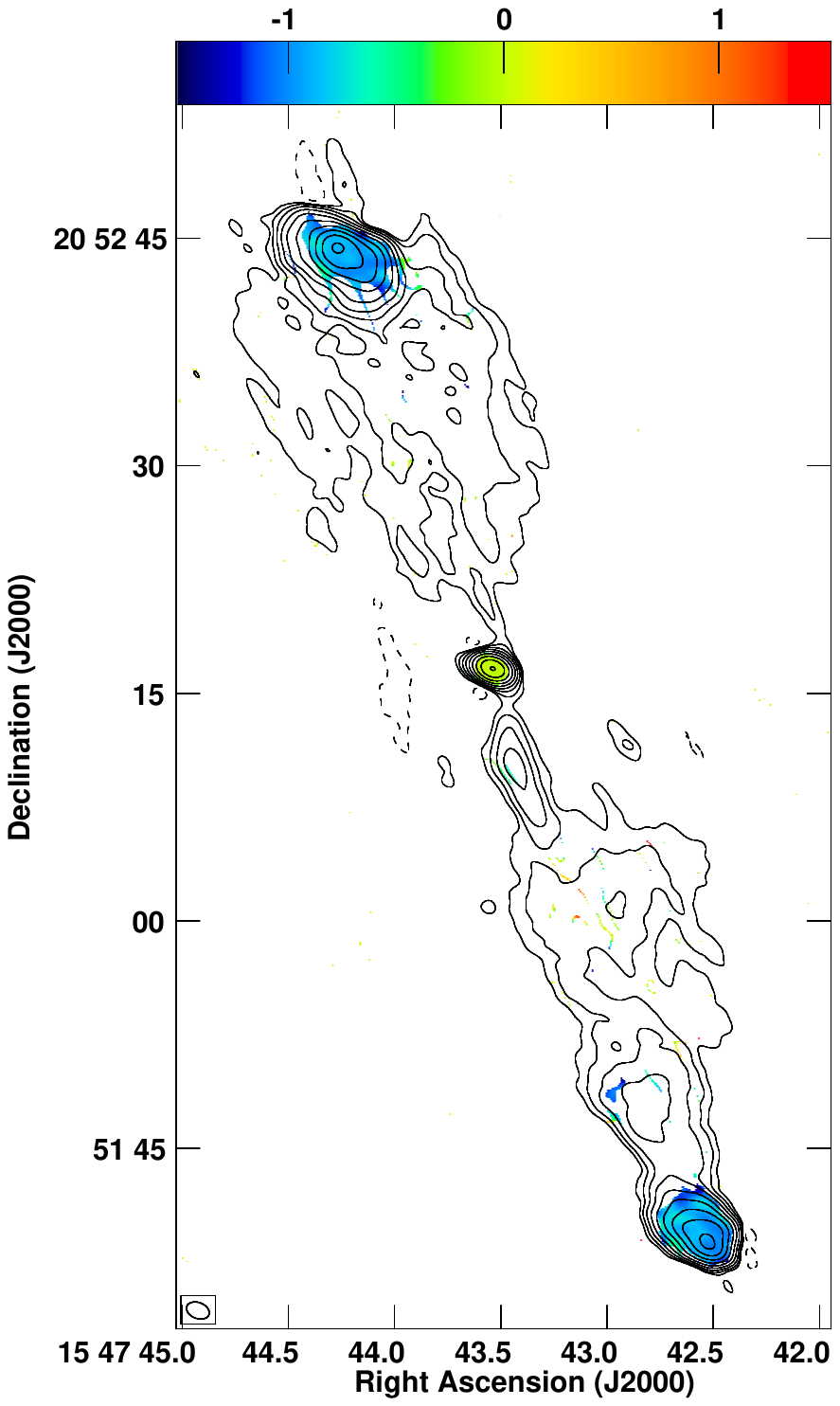}
\caption{\small VLA 6~GHz contour image of quasar PG 1545+210 superimposed with (left) red polarized intensity vectors and (right) in-band spectral index image. The inset shows the core region. The beam is $1.56\arcsec \times 1.05\arcsec$ with a PA of $69\degr$. The peak surface brightness, $I_P$ is 98.25~mJy~beam$^{-1}$ and the contour levels are $I_P \times 10^{-2} \times$  $(-0.18,~0.18,~0.35,~0.7,~1.4,~2.8,~5.6,~11.25,~22.5,~45,~90)$~Jy~beam$^{-1}$. 5$\arcsec$ length of the vector corresponds to 8.33~mJy~beam$^{-1}$.}
\label{figq5}
\end{figure*}

\begin{figure*}[hp]
\centering
\includegraphics[width=8.2cm,trim=40 170 40 170]{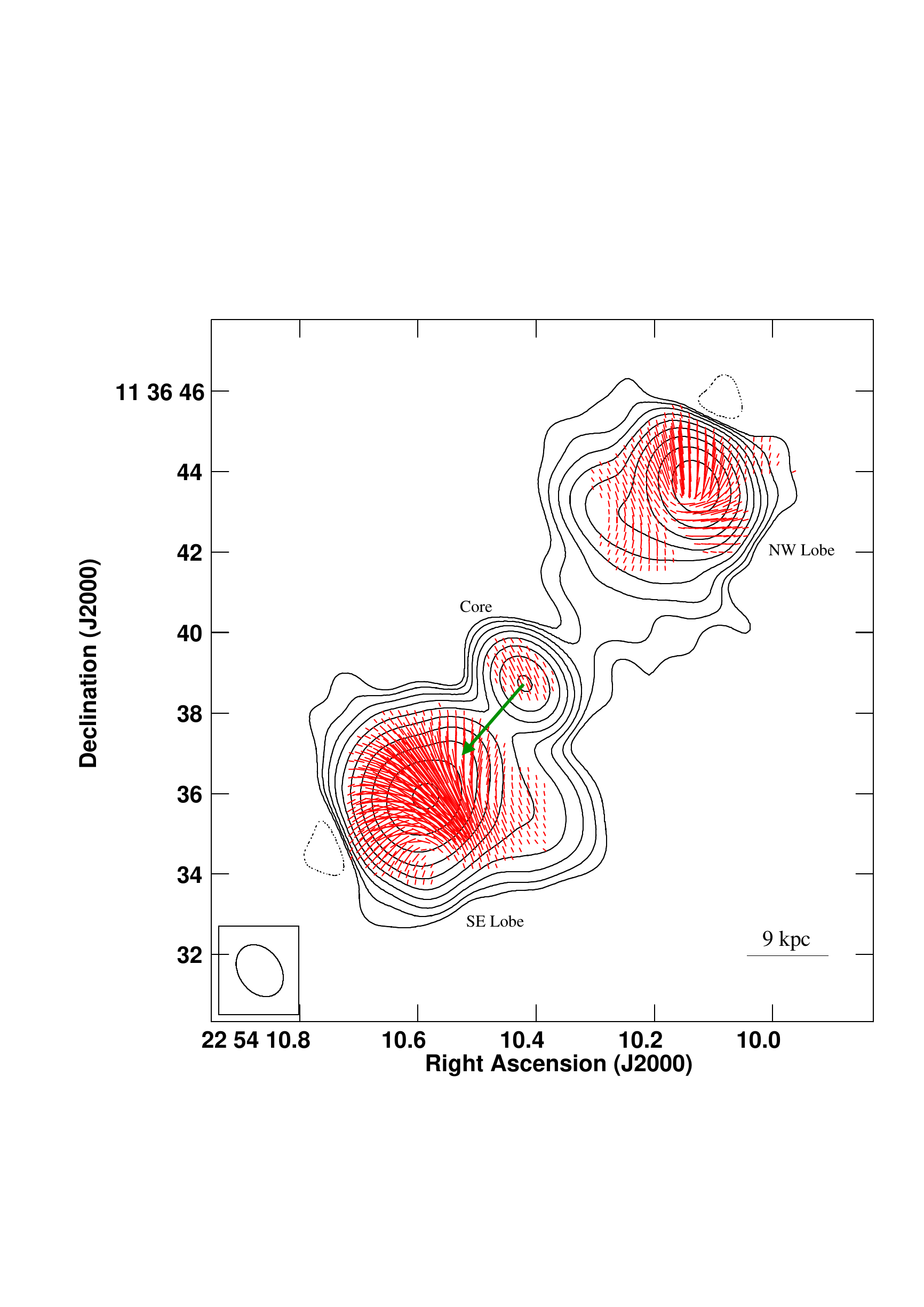}
\includegraphics[width=8.2cm,trim=30 100 30 170]{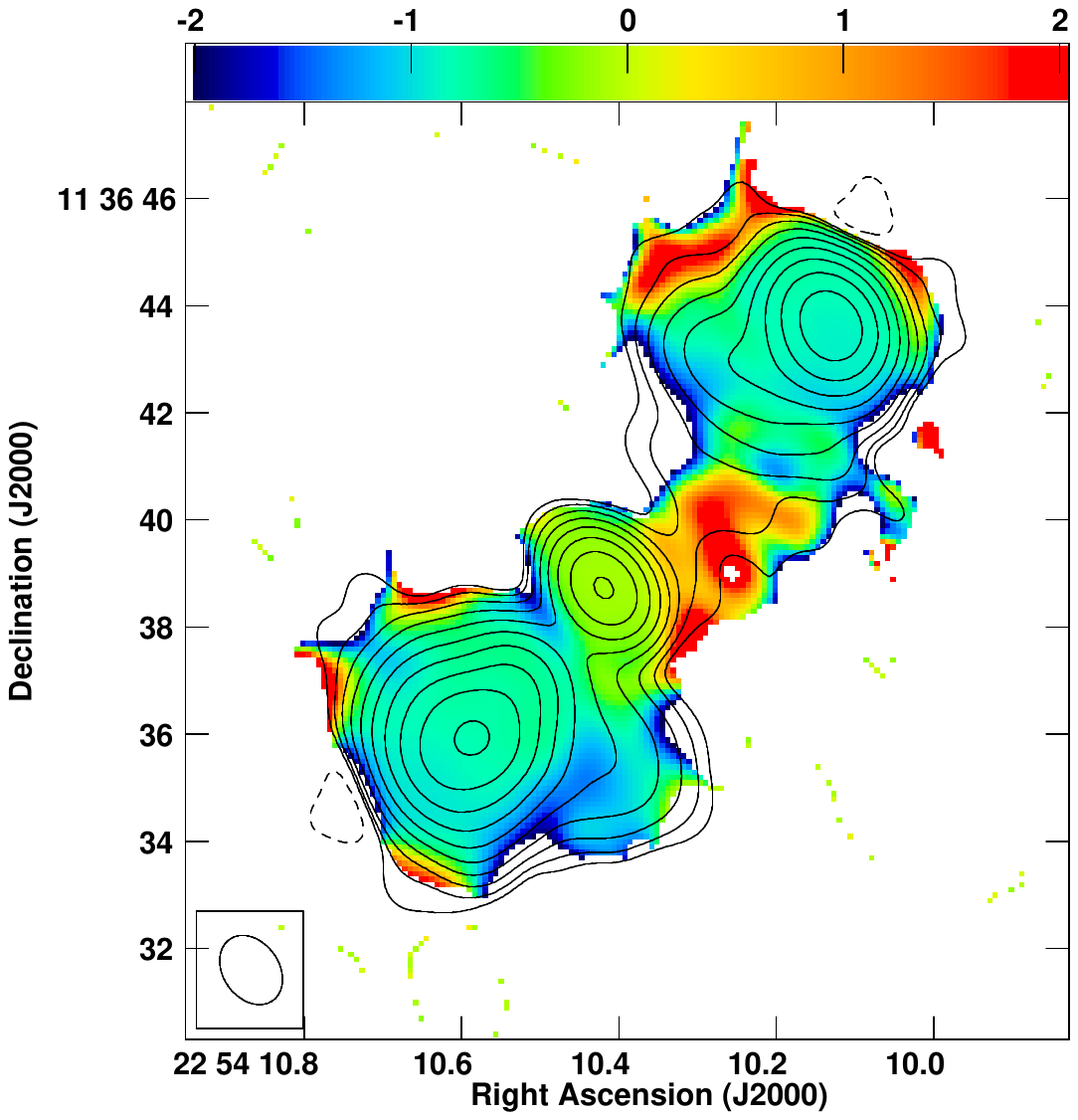}
\caption{\small VLA 6~GHz contour image of quasar PG 2251+113 superimposed with (left) red polarized intensity vectors and (right) in-band spectral index image. The beam is $1.40\arcsec \times 1.04\arcsec$ with a PA of $35\degr$. The peak surface brightness, $I_P$ is 0.134~Jy~beam$^{-1}$ and the contour levels are $I_P \times 10^{-2} \times$  $(-0.18,~0.18,~0.35,~0.7,~1.4,~2.8,~5.6,~11.25,~22.5,~45,~90)$~Jy~beam$^{-1}$. 1$\arcsec$ length of the vector corresponds to 2.5~mJy~beam$^{-1}$.}
\label{figq6}
\end{figure*}

\begin{figure*}[hp]
\centering
\includegraphics[width=8.5cm,trim=50 170 50 170]{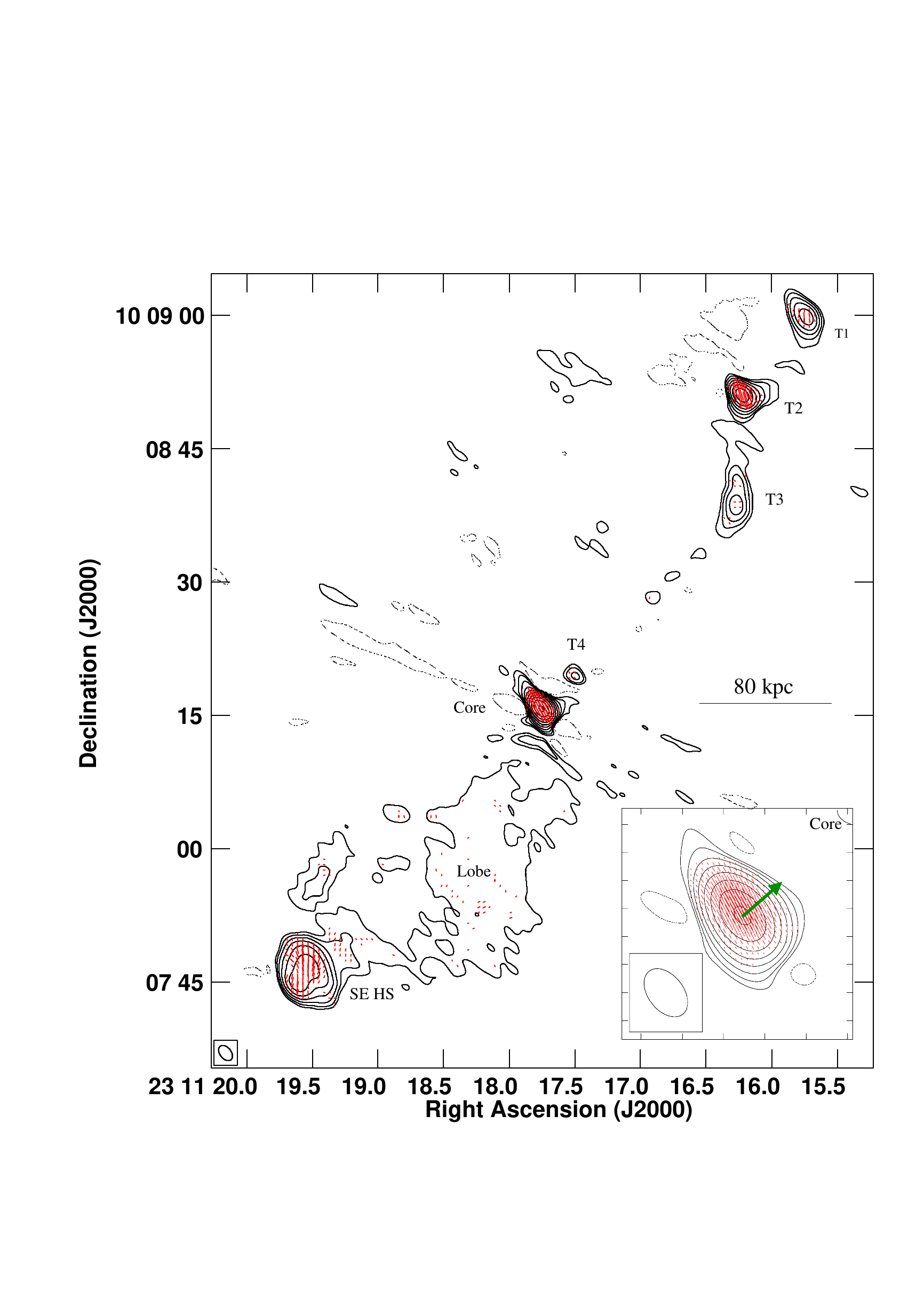}
\includegraphics[width=8.5cm,trim=40 110 40 170]{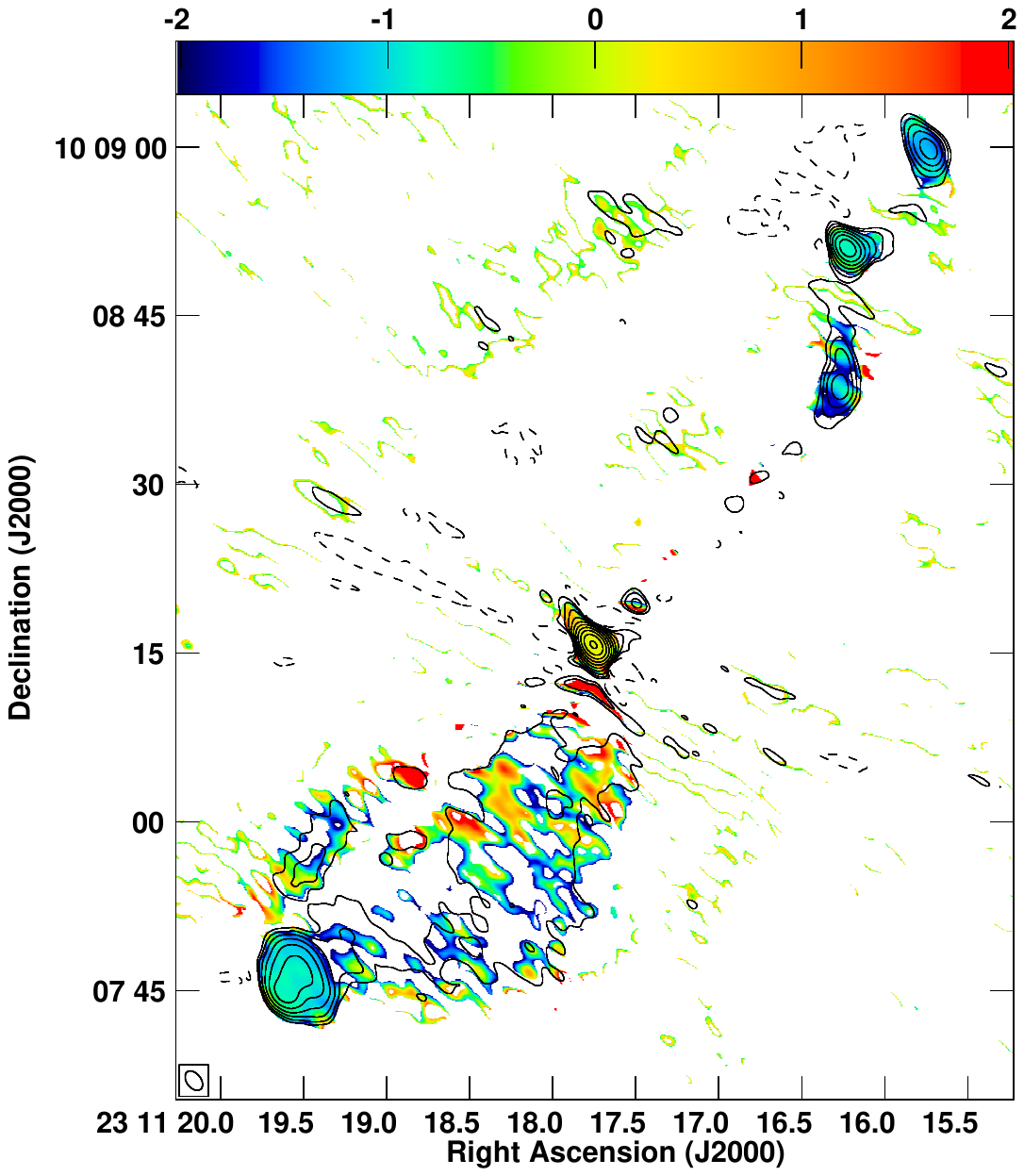}
\caption{\small VLA 6~GHz contour image of quasar PG 2308+098 superimposed with (left) red polarized intensity vectors and (right) in-band spectral index image. The inset shows the core region. The beam is $1.95\arcsec \times 1.26\arcsec$ with a PA of $38\degr$. The peak surface brightness, $I_P$ is 87.05~mJy~beam$^{-1}$ and the contour levels are $I_P \times 10^{-2} \times$  $(-0.18,~0.18,~0.35,~0.7,~1.4,~2.8,~5.6,~11.25,~22.5,~45,~90)$~Jy~beam$^{-1}$. 5$\arcsec$ length of the vector corresponds to 2.5~mJy~beam$^{-1}$.}
\label{figq7}
\end{figure*}

\subsubsection{PG 1302-102}\label{subsubsec:PG1302-102}
This FSRQ (Figure \ref{figq1}) at a redshift of $z=0.279$ was first imaged at 1.4~GHz and 6~GHz by \citet{GowerHutchings1984} who found a $15\arcsec$ two-sided radio structure with radio lobes to the northeast and the southeast and a $570~$mJy core at $1.4~$GHz. The projected bending angle between the two lobes is $~60\degr$ \citep{Rector1995}.  The bent shape of the extended radio emission is over 45~kpc and suggests that the radio source is relatively old and/or not well-confined. 1.4~GHz VLBI image by  \citet{HutchingsNeff1992} and 5~GHz VLBI image by \citet{Wang2023} shows an unresolved core. 15~GHz VLBA observations have provided a mean jet PA of $32\degr$ \citep{Lister2021} which agrees with average jet PA of \citet{Plavin2022}. Our VLA 6~GHz observations for this FSRQ match those of \citet{GowerHutchings1984} with the diffuse emission being patchy with the northeastern lobe picked up better than the southeastern one. The core EVPA is perpendicular to the VLBI jet and the northeastern lobe however no distinct hotspots can be detected. The values of fractional polarization and spectral index are not reliable for the patchy lobe emission.

\subsubsection{PG 2209+184}\label{subsubsec:PG2209+184}
For this FSRQ (Figure \ref{figq2}) the total 5~GHz VLA flux density has been previously measured to lie between 120~mJy \citep{Miller1993} and 290~mJy \citep{Kellermann1989} (claimed by Miller et al. to be an error). It has been found to have a variable flux density at 5~GHz of between 116~mJy to 326~mJy \citep{Machalski1993} and is classified as a radio intermediate quasar \citep{Falcke1995}. The core is the prominent feature with the core flux density accounting for $>95\%$ of the total flux. There is a low brightness knot towards the north-east. It has an inverted spectrum between 5~GHz to 1.4~GHz with an $\alpha = -0.24$ \citep{Falcke1995}. The parsec-scale 5~GHz VLBI image by \citet{Wang2023} shows a jet to the northeast. {\citet{Plavin2022} indicate an average jet PA of $21\degr$.} This FSRQ shows a compact core with some diffuse core halo emission to the northeast in our VLA 6~GHz observations. The core EVPA is complex. However, the EVPA towards the northeast of the core in the direction of the VLBI jet shows EVPA parallel to the jet direction.

\subsubsection{PG 1425+267}\label{subsubsec:PG1425+267}
This SSRQ (Figure \ref{figq3}) has a very large radio extent of about 1.5 Mpc \citep{Kellermann1994}. It has a clear double-lobed structure with a faint jet observed extending towards the southwestern component from the core. The parsec scale 5~GHz VLBI image by \citet{Wang2023} shows a jet to the southwest. We find our VLA 6~GHz observations of this SSRQ match those of \citet{Kellermann1994} and the southwestern jet shows polarized emission in its hotspot with EVPA typical of B-field compression. The jet and the northeastern lobe emission are very diffuse and barely polarized, with high errors in their fractional polarization. The core EVPA is perpendicular to the jet direction. 

\subsubsection{PG 1512+370}\label{subsubsec:PG1512+370}
This SSRQ (Figure \ref{figq4}) has the radio morphology of a typical linear and symmetrical FRII radio galaxy, with the distance between the hotspots being $60\arcsec$ (370~kpc) at a position angle of $110\degr$ \citep{Miller1993}. The parsec scale 5~GHz VLBI image by \citet{Wang2023} shows a tentative jet to the northwest. The VLA 6~GHz observations for this SSRQ (Figure \ref{figq4}) are consistent with those of \cite{Miller1993}, showing hotspots with EVPA characteristic of B-field compression and more of the diffuse lobe emission is detected. The EVPA of the core is orthogonal to the jet direction.

\subsubsection{PG 1545+210}\label{subsubsec:PG1545+210}
This SSRQ (Figure \ref{figq5}) at a redshift of $z=0.264$ is one of the nearest 3C quasars. The triple radio structure was identified in early 5~GHz maps by \citet{Pooley1974}. Even though earlier radio images of this source existed \citet{Bogers1994}, the VLA 1.4~GHz A-array image was the first to identify the jet leading into the southern lobe. The symmetric source is $\sim360$~kpc in extent, straight along its axis. The 5~GHz VLBI image by \citet{Wang2023} shows an unresolved core. 2 and 8~GHz VLBA observations have provided a mean jet PA of $-172\degr$ \citep{Plavin2022}. We find our VLA 6~GHz observations of this SSRQ match those of \citet{Miller1993} with the detection of more diffuse lobe emission and typical FRII-like polarization structures and hotspots with EVPA typical of B-field compression. The core EVPA is perpendicular to the jet direction.

\subsubsection{PG 2251+113}\label{subsubsec:PG2251+113}
This SSRQ (Figure \ref{figq6}) is a relatively compact ($\sim75$~kpc), linear radio source \citep{4CCat1965, Miller1993,Kellermann1994}. The parsec scale 5~GHz VLBI image by  \citet{Wang2023} shows an unresolved core. 8~GHz VLBA observations have provided a mean jet PA of $139\degr$ \citep{Plavin2022}. Our VLA 6~GHz observations of this SSRQ match those of \citet{Miller1993} with more diffuse lobe emission being picked up. The hotspots and other finer features are more clearly resolved in the \citet{Miller1993} image. The core EVPA is perpendicular to the local jet direction.

\subsubsection{PG 2308+098}\label{subsubsec:PG2308+098}
This SSRQ (Figure \ref{figq7}) is a linear radio source with the western component at a slightly different angle than the eastern one \citep{4CCat1965, Swarup1986, Miller1993, Kellermann1994}. The parsec scale 5~GHz VLBI image by \citet{Wang2023} shows a tentative jet to the northwest. We find our VLA 6~GHz observations of this SSRQ match those of \citet{Miller1993} with more diffuse lobe emission being detected in the southeastern lobe and multiple jet knots and no clear hotspot seen in the northwestern jet. The southeastern hotspot shows typical B-field compression. The core EVPA is perpendicular to the jet direction.\\

Overall, we observe that most of the PG quasars have their core EVPAs perpendicular to their jet directions. Of the seven quasars presented in this paper, three have their core EVPAs perpendicular to their jet directions, and the other four show two mutually perpendicular polarization components in their cores. This could be consistent with optical depth effects as highlighted in \citet{Gabuzda2003,Kharb2008}. The kpc-scale jet is also along the VLBI jet direction for all of the seven PG quasars presented here (See Table \ref{tab4rev} and {Figures \ref{figq1}$-$\ref{figq7}}).

\begin{table*}
\centering
\caption{\label{tab:PG2}The PG ``Blazar'' Sample Properties}
\begin{tabular}{ccccccccccc}
\hline \hline
S.No. & 
Name & 
Type & $\mathrm{\log_{10} \frac{M_{BH}}{ M_{\sun}}}$ & Ref & $\bar{Q}$ ($10^{42}$erg/s) & $L_\gamma$ ($\mathrm{erg~s^{-1}}$) & $\mathrm{\log_{10} (\nu_s / Hz)}$ &${\log_{10} \dot{M} \mathrm{\left( \frac{M_{\sun}}{yr}\right)}}$ & Ref 
\\ \hline

1 &
{PG 0851+203} &
BL Lac &
8.5 & 1 & 210 & ($1.62\pm0.04$)$\times10^{46}$ & 13.24 
& ... & ... 
\\

2 &
{PG 1101+384} &
BL Lac &
8.23 & 1 & 6.58 & ($8.4\pm0.1$)$\times10^{44}$ & 16.22 
& ... & ... 
\\

3 &
{PG 1218+304} &
BL Lac &
8.47 & 1 & 9.23 & ($4.0\pm0.1$)$\times10^{45}$ & 16.27 
& ... & ... 
\\

4 &
{PG 1418+546} &
BL Lac &
8.74 & 1 & 61.80 & ($7.3\pm0.4$)$\times10^{44}$ & 13.68 
& ... & ... 
\\

5 &
{PG 1424+240} &
BL Lac &
6.42 & 1 & 988 & ($1.47\pm0.03$)$\times10^{47}$ & 15.29 
& ... & ... 
\\

6 &
{PG 1437+398} &
BL Lac &
8.95 & 1 & 165 & ($2.2\pm0.2$)$\times10^{45}$ & 15.86
& ... & ... 
\\

7 &
{PG 1553+113} &
BL Lac &
7.25 & 1 & 120 & ($6.6\pm0.1$)$\times10^{46}$ & 15.59 
& ... & ... 
\\

8 &
{PG 2254+075} &
BL Lac &
8.85 & 1 & 43  & ($3.4\pm0.5$)$\times10^{44}$ & 12.78 
& ... & ... 
\\

9 &
{PG 0007+106} &
FSRQ &
8.87 & 2 & 7.01 
& ... & ... 
& -0.42 & { 3} 
\\

10 &
{PG 1226+023} &
FSRQ &
9.18 & 2 & 4850 
& ... & ... 
& 1.18 & { 3} 
\\

11 &
{PG 1302$-$102} &
FSRQ &
9.05 & 2 & 161 
& ... & ... 
& 0.92 & { 3} 
\\

12 &
{PG 1309+355} &
FSRQ &
8.48 & 2 & 9.92 
& ... & ... 
& 0.37 & { 3} 
\\

13 &
{PG 2209+184} &
FSRQ &
8.89 & 2 & 1.50 
& ... & ... 
& -0.98 & { 3} 
\\

14 &
{PG 0003+158} &
SSRQ &
9.45 & 2 & 2380 
& ... & ... 
& 0.79 & { 3} 
\\

15 &
{PG 1004+130} &
SSRQ &
9.43 & 2 & 847 
& ... & ... 
& -0.37 & { 4} 
\\

16 &
{PG 1048$-$090} &
SSRQ &
9.37 & 2 & 2690 
& ... & ... 
& 0.3 & { 3} 
\\

17 &
{PG 1100+772} &
SSRQ &
9.44 & 2 & 3540 
& ... & ... 
& 0.29 & { 3} 
\\

18 &
{PG 1103$-$006} &
SSRQ & 
9.49 & 2 & 2250 
& ... & ... 
& 0.21 & { 3} 
\\

19 &
{PG 1425+267} &
SSRQ &
9.9 & 2 & 519 
& ... & ... 
& 0.07 & { 3} 
\\

20 &
{PG 1512+370} &
SSRQ &
9.53 & 2 & 1920 
& ... & ... 
& 0.2 & { 3} 
\\

21 &
{PG 1545+210} &
SSRQ & 
9.47 & 2 & 1930 
& ... & ... 
& 0.01 & { 3}
\\

22 &
{PG 1704+608} &
SSRQ & 
9.55 & 2 & 4640 
& ... & ... 
& 0.38 & { 3}
\\

23 &
{PG 2251+113} &
SSRQ & 
9.15 & 2 & 1760 
& ... & ... 
& 0.66 & { 3}
\\

24 &
{PG 2308+098} &
SSRQ &
9.76 & 2 & 2000 
& ... & ... 
& 0.22 & { 3} 
\\
\hline
\multicolumn{10}{l}{Note. Column (1): Serial Number. Column (2): PG names. Column (3): Blazar type. Column (4): Black hole masses. }\\
\multicolumn{10}{l}{Column (5): References for black hole masses. Column (6): Jet Power. { Column} (7) Gamma-ray luminosity (0.1 - 100 GeV) }\\
\multicolumn{10}{l}{ { from \citet{Abdollahi2022}} Column (8): Synchrotron peak frequency. { from \citet{Abdollahi2022}}}\\
\multicolumn{10}{l}{Column (9): Accretion rates. Column (10): References for accretion rates. }\\
\multicolumn{10}{l}{References- 1: \citet{Wu_2009}, 2: \citet{Shangguan_2018}, { 3}: \citet{Davis2011}, { 4}: \citet{Luo2013}}\\
\end{tabular}
\end{table*}

\begin{table*}
 \caption{Observational Results from \citet{Baghel2023}}
 \label{tab4}
 \begin{tabular}{cccccc}
  \hline
  \hline
  Source & Region & P~(Jy)& I~(Jy) & FP~(\%) & $\alpha$ \\
  \hline
  PG 0003+158 & Core & (3.6 $\pm$ 0.8) E-04 & 1.65E-01 & { 8 $\pm$ 2\%$^\dagger$}  & -0.13$\pm$0.13 \\
  
  PG 0007+106 & Core & (1.2 $\pm$ 0.3) E-04 & 1.26E-01 & 0.8 $\pm$ 0.3\% & 0.12$\pm$0.03\\
  
  PG 1004+130  & Core & (2.7 $\pm$ 0.8) E-05 & 2.90E-02 & {  3 $\pm$ 1\%$^*$} & -0.3 $\pm$ 0.2\\
  
  PG 1048-090 & Core & { (8 $\pm$ 2) E-05 }&{  4.9798E-02 }&{  2.1 $\pm$ 0.4 \%  }&{  0.2 $\pm$ 0.4 }\\
  
  PG 1100+772 & Core & (1.17 $\pm$ 0.05) E-03 & 1.17E-01 & 2.7 $\pm$ 0.4\% & 0.55$\pm$0.006 \\
  
  PG 1103-006  & Core & (2.08 $\pm$ 0.07) E-03 & 1.37E-01 & 5.8 $\pm$ 0.3 \% & -0.319$\pm$0.004\\
  
  PG 1226+023 & Core & (2.128 $\pm$ 0.009)E+00 & 2.16E+01 & 11 $\pm$ 2\% & -0.327 $\pm$ 0.004\\
  
  PG 1309+355 & Core & (1.3 $\pm$ 0.2)E-04 & 4.69E-02 & 0.37 $\pm$ 0.07\% & 0.17 $\pm$ 0.06\\
  
   PG 1704+608 & Core & (1.36 $\pm$ 0.08) E-03 & 2.12E-02 & 8.2 $\pm$ 0.9\% &-0.8$\pm$0.7\\
  \hline
  \multicolumn{6}{l}{Note. Column (1): PG source name. Column (2): Region of the source and location. Column (3): }\\
  \multicolumn{6}{l}{Polarized flux density. Column (4): Total flux density. Column (5): Fractional Polarization.} \\
  \multicolumn{6}{l}{Column (6): Spectral index.}\\
  \multicolumn{6}{l}{{ $^\dagger$ FP at intensity peak position is 0.9 $\pm$ 0.3\%. $^*$ FP at intensity peak position is 0.27 $\pm$ 0.09\%. }}\\
 \end{tabular}
\end{table*}

\section{Global Correlations}\label{sec:corr}
We have attempted to understand the interplay between accretion power, jet power, and the organization of B-fields in kpc-scale jets by examining correlations between various associated quantities discussed ahead. As the gamma-ray luminosity and spectral energy distribution (SED) data are available for all the BL Lac objects, we have attempted to look for relations between them and radio polarization properties. We had previously looked at various global correlations for PG ‘blazar’ sub-samples using our own radio polarization data along with other multi-frequency data from the literature in \citet{Baghel2023,Baghel2024}. Here we look at new correlations using our observational data and other global properties for the entire PG `blazar' sample. { In order to compare the two classes and the findings from our previous studies \citep{Baghel2023,Baghel2024}, we intended to examine and compare RL quasars and BL Lacs independently. We also note that while we did find a marginally significant difference in the polarization distribution of the BL~Lacs and quasars (KS Test p-value= 0.0014). However, given the compact nature of the BL Lac radio emission, there is likely some jet/lobe emission contaminating the core emission, which could increase the `core' fractional polarization.}

High energy luminosity ($0.1-100$~GeV; $L_\gamma$), and synchrotron peak frequency $\nu_s$ for the BL~Lacs have been reported from FERMILPSC - the Fermi LAT 12-Year Point Source Catalog \citep[4FGL-DR3;][]{Abdollahi2022} and the Fourth LAT AGN Catalog \citep[4LAC-DR3;][]{Ajello2022}. { Contrary to the BL~Lacs, Fermi-LAT data was available for only one of the quasars, viz., PG 1226+023. Quasars were therefore not considered for correlations with gamma-ray emission.}

We have estimated the long-term time-averaged bulk radio jet kinetic power $\bar{Q}$ for all blazars using the radio luminosity at 151~MHz as a surrogate for the luminosity of the radio lobes by using the relation given by \citet{Punsly2018}, 

$$\bar{Q} = 3.8 \times 10^{45} \it{f}~L_{151}^{6/7}~\mathrm{erg~s^{-1}}$$ 
where $L_{151}$ is radio luminosity at 151~MHz in units of $10^{28}$~W~Hz$^{-1}$~sr$^{-1}$ and $f\approx15$ \citep{Blundell2000}. We obtained the $L_{151}$ from the TGSS survey flux densities \citep{Intema2017} using the relation, $L_{151} = [ D_{L}^2~F_{151}]/[(1+z)^{(1+\alpha)}]$~W~Hz$^{-1}$~sr$^{-1}$, \footnote{We note that there is no $4\pi$ factor because of the anisotropic jet emission, e.g., see \citet{Peacock1999}.} where we have used the typical average $\alpha = -0.7$ for the extended emission. We report $\bar{Q}$ without errors as systematic uncertainties in the estimation methods dominate the statistical uncertainty in the data. Black hole masses $\mathrm{M_{BH}}$ and accretion rates $\dot{M}$ have been reported from \citet{Shangguan_2018,Wu_2009} and \citet{Davis2011,Luo2013} which do not report errors for the same. { Accretion rates for the quasars have been determined by \citet{Davis2011} using thin accretion disk model spectral fits of optical lines. For the BAL QSO, PG 1004+130, an $\eta = 0.1$ has been adopted by \citet{Luo2013} to compute the accretion rate. For the BL~Lacs, we do not have a similarly determined estimate for accretion rates and hence we have not included them in the correlation tests.} 
However, \citet{Shangguan_2018} use the calibration of \citet{HoKim2015} to calculate $\mathrm{M_{BH}}$ that produces an uncertainty of $\sim0.3-0.4$~dex. The errors in $L_\gamma$ and $\nu_s$ are plotted as reported in the catalog or calculated from reported errors in the flux values.

{ The global parameters for the entire PG `blazar' sample are listed in Table \ref{tab:PG2}). The fractional polarization, spectral index and other observational quantities from our complete set of VLA observations are listed in two tables. Table \ref{tab:BQMSMT} lists the observational parameters from the VLA images of PG BL Lacs and 7 PG Quasars presented in this paper, and Table \ref{tab4} lists the observational parameters of the 9 PG quasars from \citet{Baghel2023}, used in the correlations below.}

With the complete sample of RL PG quasars {(see Tables \ref{tab:BQMSMT}, \ref{tab4} and \ref{tab:PG2})}, we find that the black hole masses $\mathrm{M_{BH}}$ are marginally correlated with the core fractional polarization FP$_C$ (Kendall $\tau$ test probability, p = 0.015) (left panel, Figure \ref{corr12}. This was also observed with the subset of the quasars presented in \citet{Baghel2023} and hence confirms this result for the PG RL quasars. However, we note that this correlation seems to be driven by the values for PG 1309+355 and weakens substantially when it is removed. With the entire quasar sample, we find that jet power $\bar{Q}$ does not correlate with FP$_C$ (Kendall $\tau$ test p = 0.116) (right panel, Figure \ref{corr12}) as opposed to the weak correlation observed with the sub-sample presented in \cite{Baghel2023}. Additionally, we find that the core spectral index $\alpha$ is anti-correlated with FP$_C$ (p value = 0.0189) (left panel, Figure \ref{corr34}). { No such correlation with $\alpha_C$ was found for BL Lac objects (p-value = 0.105).}

{We found a correlation between jet power $\bar{Q}$ and $\dot{M}$ in \citet{Baghel2023} for the PG quasars. In comparing the accretion power, $\dot{M}c^2$ to jet power, we find that in general $\bar{Q} \sim 10^{-4}\dot{M}c^2$ for the FSRQs as in \citet{Sikora2013} and $\bar{Q} \sim 10^{-2}\dot{M}c^2$ for the SSRQs (right panel, Figure \ref{corr34}}{)}.

For the case of the BL Lac objects, we find that compared to the uGMRT 650~MHz FP$_C$ data \citep{Baghel2024}, the correlation of FP$_C$ with $\mathrm{M_{BH}}$ becomes insignificant (p-value = 0.170) (left panel, Figure \ref{corr56}). We found no significant correlation between FP$_C$ and $\bar{Q}$ (p-value = 0.382) (right panel, Figure \ref{corr56}). Correlations with gamma-ray luminosity (p-value = 0.061) (left panel, Figure \ref{corr78}) and synchrotron peak luminosity (p-value = 0.061) (right panel, Figure \ref{corr78}) are weakened compared to those reported in \citet{Baghel2024}. These results are discussed {below}.

\begin{figure*}
\centering
\includegraphics[width=8.7cm,trim=0 0 0 0]{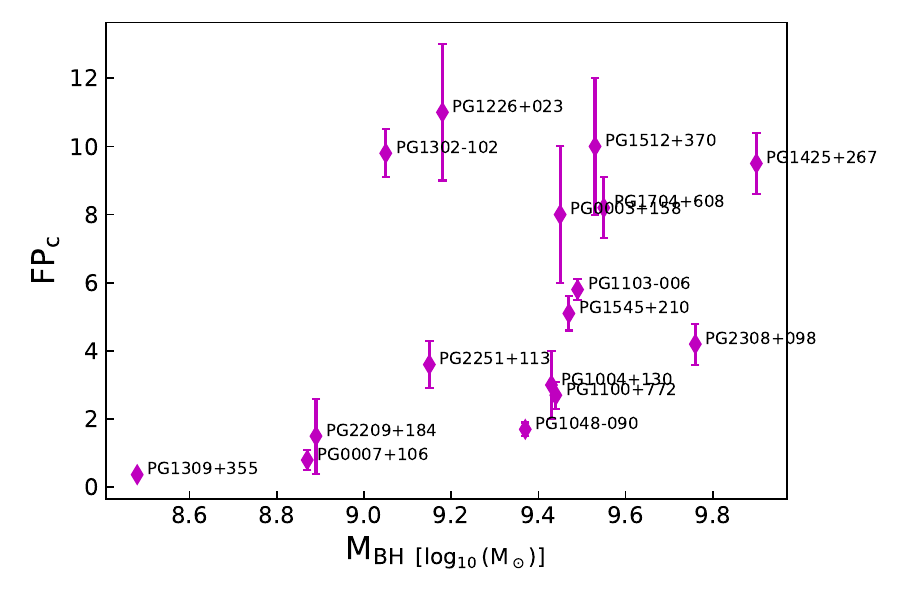}
\includegraphics[width=8.7cm,trim=0 0 0 0]{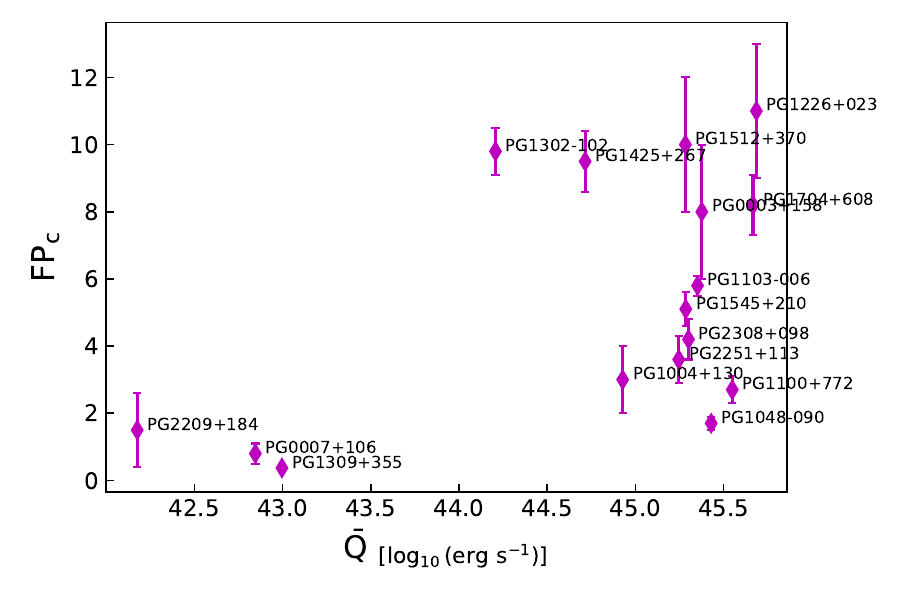}
\caption{\small (Left) Core fractional polarization FP$_C$ (per cent) versus log of BH masses (M$_{BH}$/M$_\odot$), and (Right) FP$_C$ versus log of jet power ($\bar{Q}$ in erg s$^{-1}$) for the PG quasars. }
\label{corr12}
\end{figure*}

\begin{figure*}
\centering
\includegraphics[width=8.7cm,trim=0 0 0 0]{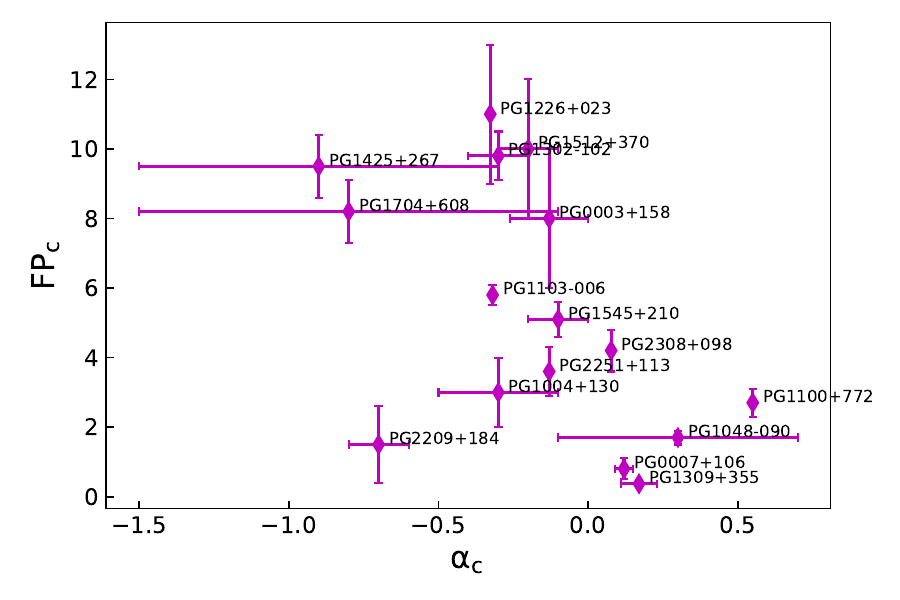}
\includegraphics[width=8.7cm,trim=0 0 0 0]{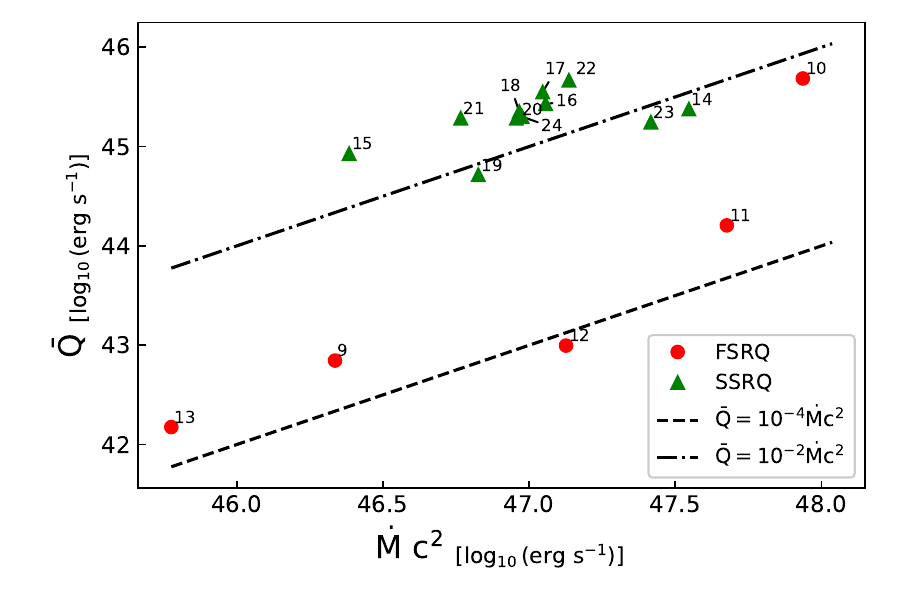}
\caption{\small (Left) Core fractional polarization FP$_C$ (per cent) versus core spectral index $\alpha_c$, and (Right) log of accretion power ($\dot{M}c^2$ in erg s$^{-1}$) versus log of jet power ($\bar{Q}$ in erg s$^{-1}$) for the PG quasars. The dashed line shows $\bar{Q} = 10^{-4} \dot{M}c^2$ and $\bar{Q} = 10^{-2} \dot{M}c^2$. The quasars are numbered by their S.No. in Table \ref{tab:PG2}.}
\label{corr34}
\end{figure*}

\begin{figure*}
\centering
\includegraphics[width=8.7cm,trim=0 0 0 0]{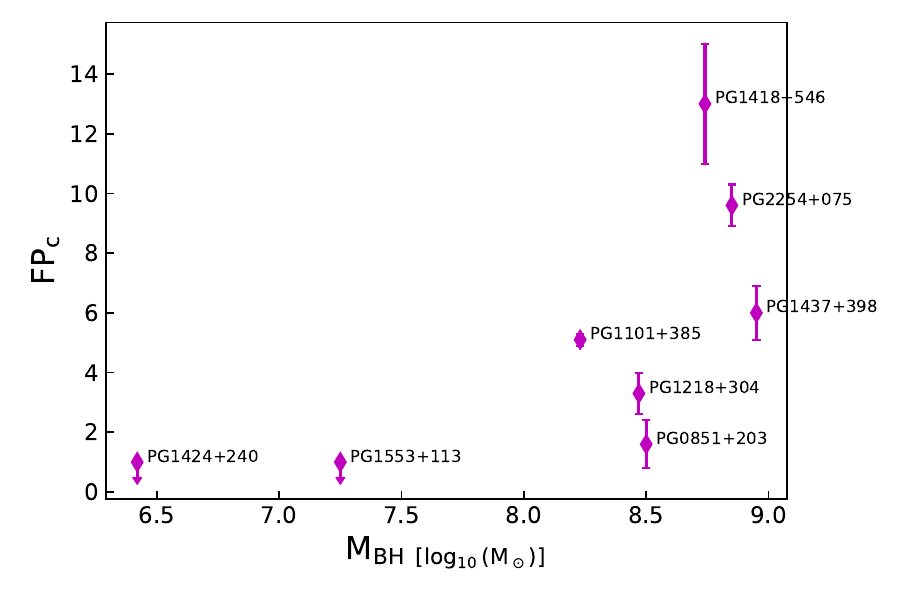}
\includegraphics[width=8.7cm,trim=0 0 0 0]{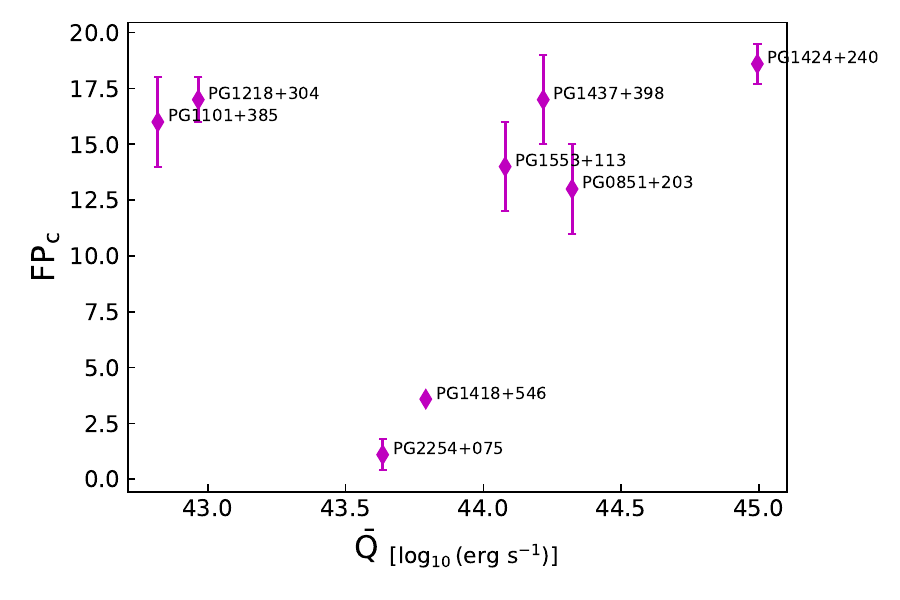}
\caption{\small (Left) Core fractional polarization FP$_C$ (per cent) versus log of BH masses (M$_{BH}$/M$_\odot$), and (Right) FP$_C$ versus log of jet power ($\bar{Q}$ in erg s$^{-1}$) of the PG BL Lac objects.}
\label{corr56}
\end{figure*}

\begin{figure*}
\centering
\includegraphics[width=8.7cm,trim=0 0 0 0]{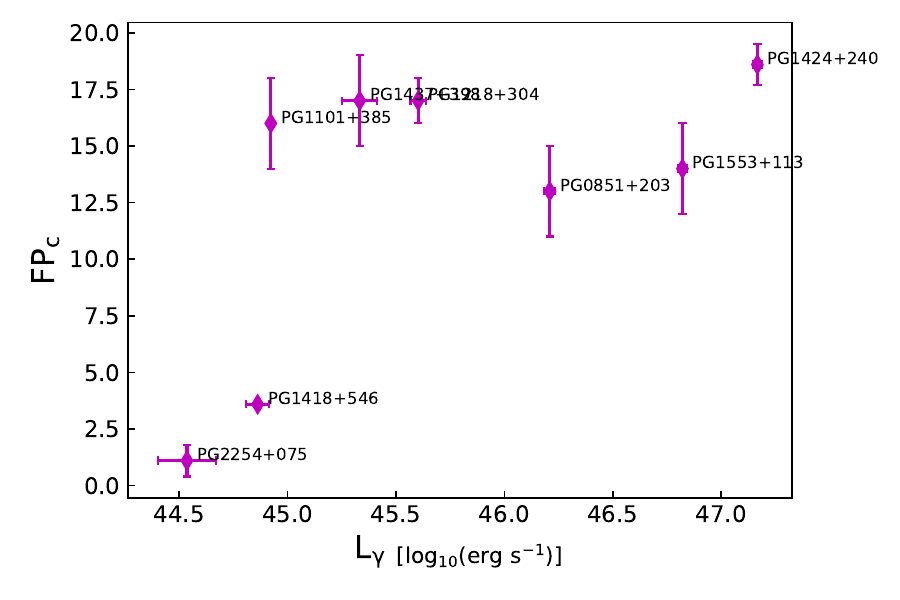}
\includegraphics[width=8.7cm,trim=0 0 0 0]{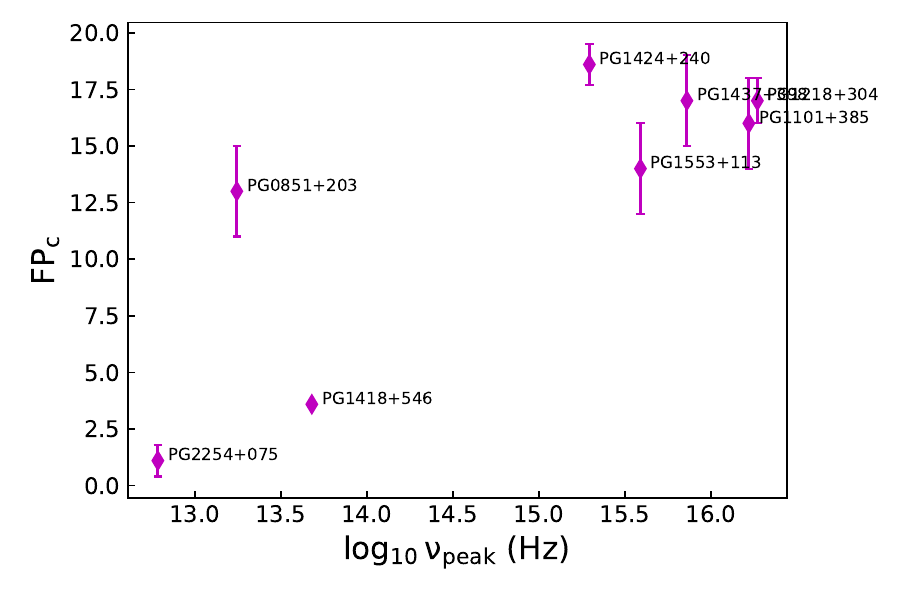}
\caption{\small (Left) Core fractional polarization FP$_C$ (per cent) versus log of gamma-ray luminosity L$_{\gamma}$ {(erg~s$^{-1}$)}, and (Right) FP$_C$ versus log of synchrotron peak frequency $\nu_{peak}$ { (Hz)} of the PG BL Lac objects.}
\label{corr78}
\end{figure*}

\section{Discussion on the complete PG blazar Sample} \label{sec:disc}


Sources in the PG `blazar' sample show a wide variety of radio morphologies not easily classifiable into distinct FRI and FRII categories. We find that seven of the sixteen PG quasars show signatures of episodic or restarted AGN activity { and disturbed morphology. These are PG 0007+106 \citep{Baghel2023,Silpa2021}, PG 1302$-$102, PG 1004+130 \citep{Baghel2023,Ghosh2023}, PG 1100+772 \citep{Baghel2023}, PG 1103$-$006 (this work) \citep{Baghel2023}, PG 1704+608  \citep{Baghel2023}, and PG 2308+098 (this work).} The prevalence of diverse radio morphologies among the RL PG quasars contradicts the picture from the bright 3C sources which typically consist of double radio sources \citep{Mullin2008}. The high number of distorted/hybrid/restarted radio structures in the PG quasars could be a consequence of the optical selection criteria of the PG sample that is biased towards luminous sources with high Eddington ratios \citep{Hooper1996,Laor2000,Jester2005} but remains largely unbiased in terms of their radio properties. A more complex extended source population is also borne out by recent high sensitivity low-frequency surveys with less restrictive selection effects than earlier studies, including candidate hybrid, restarted, and remnant sources \citep[e.g.,][]{Kapinska2017, Mingo2019, Jurlin2021}.

{ The PG quasars appear to show a greater diversity in their morphologies and total radio powers than the PG BL Lac objects. While the number of PG BL~Lac objects is small, this finding is consistent with low-frequency LOFAR studies of larger radio galaxy samples} \citep[e.g.,][]{Mingo2019, Mahatama2023}. Two BL~Lac objects, namely PG 0851+203 and PG 1418+546 show the presence of hotspots/jet bends { in their uGMRT images \citep{Baghel2024}} that are consistent with an FRII morphology which goes against the basic tenets of the simple Unified Scheme \citep{Urry95}. These sources may have been misclassified based on their optical spectra; a boosted optical continuum may have swamped the emission lines in these sources. These sources are also consistent with the blazar envelope scenario discussed ahead. 

\citet{Massaro2020} have suggested that BL Lac objects' environs resemble FR0s, a core-dominated compact class of sources, more than those of FRIs. Two BL Lac objects viz., PG 1218+304 and PG 1553+113, do not exhibit obvious indications of extended emission neither in their VLA images presented here nor in their uGMRT images presented in \citet{Baghel2024} and may resemble FR0s.   

Our polarimetric data indicates overall that magnetic field structures present in the jets of PG quasars are aligned with the jets, while four of the sixteen sources show some perpendicular magnetic field structures in their cores alone. { We note that the inferred B-fields are taken to be perpendicular to the EVPAs even for the core regions in contrast to what we noted earlier in \citet{Baghel2023} (see footnote \ref{footnote1})}. Optical depth effects, as discussed in \citet{Gabuzda2003,Kharb2008} and others, can give rise to parallel/perpendicular EVPAs with respect to local jet directions.

In the PG BL Lac objects, however, {the relationship observed in the magnetic field structures with their local parsec-scale jet directions \citep[e.g.,][]{Pushkarev2023} is not present for the kpc-scale jets} \citep[see also][]{Baghel2024}. We find complex EVPA structures suggesting multiple jet re-orientations in five of the eight PG BL Lacs. This agrees with previous studies suggesting that their weaker jets interact with their environments and change direction \citep[e.g.,][]{Kharb10}. {Complex EVPA structures can also result from turbulent magnetic fields as suggested by \citet{Marscher2017, Zhang2020}.}

However, on the whole, there is a consistency in the observed magnetic field structures in the PG quasars and BL~Lacs with those observed in FRI and FRII radio galaxies by \citet{Bridle1984}; that is, {FRIs have their EVPAs either parallel to the local jet directions or parallel near the centers but perpendicular near the jet edges, while FRIIs have their EVPAs perpendicular to their local jet directions.}

The polarimetric data further suggest that jet-medium interaction is likely to be playing an additional key role in the observed magnetic field structures. Jet-medium interaction can result in the creation of `shear' layers or jet `sheaths' \citep{Laing1996,Kharb2009} and also change the jet propagation direction in BL Lac objects. A clear spine-sheath magnetic field geometry is observed in the BL Lac object, PG 1101+384 and possibly in PG 1424+240, combining the VLA and uGMRT { \citep{Baghel2024}} polarimetric data. Changes in jet direction that are not resolved in our observations could result in a lowering of the degree of polarization on kpc-scales. 

The hotspots of several quasars exhibit a steep spectral index, from $-0.70\pm0.10$ up to $-1.00\pm0.20$, while the cores are typically flat spectrum. This could suggest that the hotspots in some sources are moving through plasma from a previous AGN activity episode \citep[e.g. PG 1004+130, 3C 219,][respectively]{Ghosh2023,Clarke1992}. Alternately, this may be an effect of the total radio power and/or redshift as discussed in \cite{DennettThorpe1999}. Higher-powered, more distant sources have steeper hotspot spectra at a given emitting frequency; i.e., hotspot spectra in more distant objects are detected at a higher emitting frequency. Moreover, the rest-frame spectra of quasar hotspots exhibit greater curvature than those of radio galaxies \citep{DennettThorpe1999}. Greater synchrotron cooling in higher redshift sources can also steepen the hotspot spectra. However, these results have been questioned by \citet{Vaddi2019} who state that the luminosity-spectral index correlation is driven by the luminosity-redshift and the spectral index-redshift correlations and found no difference in the spectral indices of quasars and radio galaxies. Similarly, \citet{Kharb2008b} find steep hotspot spectral indices in several FRII radio galaxies. 

{An association of greater black hole mass with greater jet power has been previously suggested  by \citet{MIGHTEE2022}. Instead, we find a potential correlation between the black hole masses with the core fractional polarization.} While black hole masses are not statistically different between the PG quasars and PG BL Lac objects, we observe {a correlation between} black hole masses and core fractional polarization in the case of quasars alone. This may imply that more massive black holes tend to produce jets with more organized magnetic field structures in the case of quasars \citep[e.g.,][]{Baghel2023}. Alternately, well-organized, large-scale magnetic fields in the environments could promote the formation and early growth of black holes as proposed by \citet{Begelman2023}. {The correlation is, however, marginal and a larger sample is needed to confirm the same.}

Environmental effects can explain the { marginal} anti-correlation observed between core spectral indices and core fractional polarization; the presence of thermal gas emitting free-free emission along with synchrotron plasma at the jet bases could flatten the core spectral indices as well as reduce the core fractional polarization due to the presence of greater Faraday-rotating medium. { Another reason could be that we are sampling different regions in the source. A steep spectrum core could be indicating that the core is observed at a frequency higher than its synchrotron peak, a flat spectrum core would be near its synchrotron peak and an inverted spectrum core would be observed before the synchrotron peak. In the shock-in-jet scenario of \citet{Angelakis2016}, the higher fractional polarization is observed from the shock-compressed compact regions emitting near their synchrotron peaks. We did not find a correlation between core spectral indices and core fractional polarization for the BL~Lacs, which could be resulting from the smaller number of BL~Lacs in our sample, or jet component contamination to the core polarization. However, it could also mean that there is less gas at the jet base of BL Lac objects, or the shock-in-jet scenario, discussed above for the quasars, does not hold.}

The association of total jet power with accretion power is also well known \citep{Rawlings1991, Ghisellini2010b, Punsly2011, Sikora2013} and the PG FSRQs follow the general relation of quasars with $\bar{Q} \sim 10^{-4}\dot{M}c^2$ as noted in \citep{Sikora2013} and the SSRQs closely follow the relation $\bar{Q} \sim 10^{-2}\dot{M}c^2$ with a higher jet production efficiency. This range in jet production efficiency and the tight correlations may be due to the dominance of the Blandford-Znajek mechanism in the formation of these jets \citep{BZ1977}.

The blazar envelope scenario \citep{Meyer2011,Keenan2021} suggests that the FRI and FRII jets have two different structures with the FRIs comprising of structured jets with slow intrinsic speeds and inefficient accretion disks, and FRIIs having single Lorentz factor high-speed jets and efficient accretion disks. This differentiation in the jet velocities would then also explain the differences in the magnetic field structures along the jets in FRI and FRII (consistent with BL Lacs and quasars respectively) even if they have similar black hole spins with either tightly or loosely wound helical magnetic fields resulting from a lower or higher jet velocity \citep{Gabuzda2015}. 

The PG quasar jets also seem to be characterized by perpendicular EVPAs w.r.t. jet direction going from parsec to kpc-scales, possibly suggesting single bulk velocity jets with no substructures. The PG BL~Lacs, however, show a marked discontinuity in their parsec-scale to kpc-scale EVPAs-to-jet-direction behavior \citep{Baghel2024}. This could relate to both internal depolarization occurring inside the structured, possibly spine-sheath-like jets (e.g., PG 1101+384 and PG 1424+240), as well as external depolarization occurring by the jet's interaction with the surrounding medium. We also find that the LSP BL Lacs among the PG BL~Lacs are the ones displaying `strong' jet characteristics in accordance with the blazar envelope scenario. The PG BL Lacs also show a smaller range in their jet powers compared to the PG quasars again hinting at the BL Lacs as a class being `weak' jet sources with inefficient accretion disks. 

Under the magnetic flux paradigm \citep{Sikora2016, Tchekhovskoy2015, Velzen2013}, jets are launched by the magnetically arrested accretion discs (MAD) due to the accumulation of large net magnetic flux. Fluctuating accretion flows can deposit sudden large amounts of flux onto the black hole, leading to intermittent jet production. This mechanism being scale-invariant, is also valid for X-ray binary sources. In such a case, FRI jets can be taken as the counterparts of steady-state X-ray binary jets with weak but persistent radio emission whereas FRII jets would be counterparts to the flaring jet state of X-ray binaries in the hard-to-soft state transition of their accretion disks. In more sensitive recent studies that go down to mJy flux densities, FRIs are found to be more common in the local universe ($z\leq 0.15$) and there is evidence that FRIIs have accretion disks that transition from high to low accretion regimes \citep{Grandi2021}. 

The PG quasars show multiple recurrent patchy knots and disturbed diffuse structures in both their total intensity and polarization structures that might indicate quasi-periodic jet modulation seen in an unstable accretion disk \citep[e.g. as noted by][]{Godfrey2012}. Recurrent activity is also found to be more prevalent in LERG FRIIs than HERG FRIIs where episodic spikes in the accretion rate are expected \citep{Mahatama2023}. Hybrid morphology sources could be the result of restarted activity combined with orientation effects on intrinsic FRII sources \citep{Ghosh2023, Hardwood2020}. \citet{Saripalli2012} show that about 33\% of FRIIs and 13\% of FRIs show evidence of restarted activity in their sample, whereas more recently \citet{Jurlin2020} have found the incidence to be between 13\% and 15\% in their LOFAR study, with restarted double-double radio sources being commonplace. While the greater incidence of such restarted and remnant sources within the PG quasars compared to the BL~Lacs might be attributable to projection effects, these differences persist even when comparing FSRQs and BL~Lacs which both have smaller viewing angles.


The blazar envelope picture \citep{Meyer2011,Keenan2021} can coexist with the scenario suggested by the MAD mechanism of jet launching under the magnetic flux paradigm \citep{Sikora2016, Tchekhovskoy2015, Velzen2013}. The FRII sources will be the ones with the radiatively inefficient geometrically thick accretion flow transitioning in a turbulent manner to a radiatively efficient geometrically thin accretion flow with the launch of flaring jets taking a period of $10^4-10^7$ years. These flaring jets would then also be associated with a perturbed accretion disk which would lead to restarted activity and changes in jet direction and will also have higher bulk velocities \citep{Fender2004, Kylafis2012, Sikora2023}. Our results agree with this combined scenario.

\section{Conclusions} \label{sec: Conclusions}
We present polarization-sensitive 6~GHz VLA images of 7 quasars and 8 BL~Lac objects belonging to the Palomar-Green sample. This paper completes the polarization study of the PG `blazar' sample presented previously in \citet{Baghel2023,Baghel2024}. We summarize below our main conclusions from the entire sample. 
\begin{enumerate}

\item The radio morphology of the PG quasars is more diverse compared to the brighter 3C sources that show double FRII-like radio lobes in the literature. Morphological signatures of re-started activity like changes in jet propagation directions or changes in spectral indices are present in 7 out of 16 PG quasars. The PG BL Lacs in comparison are less diverse in their radio morphologies. The quasi-periodic jet knots, jet precession and restarted activity in quasars might be indicative of an unstable accretion disk agreeing with a possible evolutionary scenario wherein they are undergoing an accretion mode transition. 

\item The radio morphology of the BL Lac objects is typically core-halo type or FR-I like in the case of PG 1424+240 and PG 1437+398, with the main exceptions being the LSP BL Lacs, viz., PG 0851+203 and PG 1418+546. These  display radio morphology and polarization characteristics of `strong' jet sources consistent with the `blazar envelope scenario'. The PG BL Lacs also show a smaller range of total radio power compared to the PG quasars, consistent with them being driven by inefficient accretion disks.

\item We detect kpc-scale polarization with the VLA in all the blazars with fractional polarization ranging from $0.8\pm0.3$\% to $37\pm6$\% in their cores and jets/lobes. In the PG quasars, the polarization structures are consistent with jets primarily displaying aligned magnetic fields along the jets and transverse magnetic fields in the terminal hotspots. The magnetic field structures are more complex in the BL~Lac objects, suggestive of smaller sub-structures with distinct magnetic field orientations that are not resolved in our VLA observations. 

\item The kpc-scale magnetic field orientation in the quasar cores is parallel to their VLBI jet directions, which in most cases is the same as their kpc-scale jet directions. In the case of most BL Lacs, the kpc-scale core magnetic field orientations show no correlation to the VLBI jet directions. This might indicate multiple jet re-orientations for the PG BL~Lac objects.

\item The hotspots of several quasars exhibit a steep spectral index, from  $-0.7 \pm 0.1$ up to $-1.0\pm 0.2$, while the cores are typically flat spectrum. {This could be consistent with episodic AGN activity, spectral aging due to synchrotron cooling, or the observed relations between hotspot spectra and total radio power.} An anti-correlation between the core spectral indices and core fractional polarization could be consistent with the presence of thermal gas along with synchrotron plasma at the jet bases and elsewhere. Environmental effects are also reflected in spine-sheath-like magnetic field structures that are observed in a couple of PG blazars.

\item For the PG quasars, we find that the black hole masses are {marginally} correlated with the kpc-scale core fractional polarization. This may be driven by massive black holes producing jets with highly ordered magnetic fields or environments with highly ordered magnetic fields producing more massive black holes. For the PG quasars, jet power is positively correlated with the accretion power with high jet production efficiencies. This suggests the predominance of the Blandford-Znajek mechanism in the production of these jets.

\end{enumerate}

Overall, we find that the PG `blazar' sample provides a rich set of radio polarimetric and spectral index data that elucidates the nature of radio jets in an optically selected sample. The optical selection criteria bias the sources to have relatively high accretion rates without a corresponding bias in the radio emission produced in these sources. This results in a diverse set of morphologies and magnetic-field structures in jets of radio-loud AGN. 

\section{Acknowledgments}
{We thank the referee for their helpful and insightful suggestions that have improved this paper significantly.}
JB, PK, SG acknowledge the support of the Department of Atomic Energy, Government of India, under the project 12-R\&D-TFR-5.02-0700. LCH was supported by the National Science Foundation of China (11991052, 12233001), the National Key R\&D Program of China (2022YFF0503401), and the China Manned Space Project (CMS-CSST-2021-A04, CMS-CSST-2021-A06). CMH acknowledges funding from a United Kingdom Research and Innovation grant (code: MR/V022830/1). SS acknowledges financial support from Millenium Nucleus NCN23\_002 (TITANs) and Comité Mixto ESO-Chile. The National Radio Astronomy Observatory is a facility of the National Science Foundation operated under cooperative agreement by Associated Universities, Inc. This research has made use of the NASA/IPAC Extragalactic Database (NED), which is operated by the Jet Propulsion Laboratory, California Institute of Technology, under contract with the National Aeronautics and Space Administration.

%

\vspace{5mm}





\appendix
\section{Tables}

\begin{table*}[p]
\centering
\caption{VLBI jet directions of the PG Blazars }
\label{tab4rev}
\begin{tabular}{ccccc}
\hline
\hline
Source & VLBI jet PA & Ref & EVPA core (15~GHz VLBA) 
& EVPA core (6~GHz VLA)\\ 
\hline
\multicolumn{5}{c}{BL Lacs}\\
\hline 
{PG 0851+203} &  $-109\degr$ & 1 & $\perp$ 
& $\perp$\\
{PG 1101+385} &   $-31\degr$ & 1 & oblique/ $\parallel$ & 
oblique/$\perp$\\
{PG 1218+304} &  $+92\degr$ & 1 & ... & 
oblique/$\perp$\\ 
{PG 1418+546} &  $+126\degr$ & 1 & oblique 
& $\perp$\\
{PG 1424+240} &  $+145\degr$ & 1 & $\parallel$ 
& $\parallel$\\
{PG 1437+398} & $+56\degr$ & 1 & ... 
& ...\\
{PG 1553+113} &  $+51\degr$ & 1 & $\parallel$ & 
oblique/$\perp$\\
{PG 2254+075} & $-122\degr$ & 1 & $\parallel$/oblique & 
oblique\\
\hline
\multicolumn{5}{c}{Quasars: FSRQ}\\
\hline 
{PG 0007+106} & $-115\degr$ & 1 & oblique/$\perp$ & $\perp$\\
{PG 1226+023} & $-130\degr$ & 1 & $\perp$ & $\perp$\\
{PG 1302-102} & $+32\degr$ & 1 & $\perp$ & $\perp$\\
{PG 1309+355} & $+130\degr$ & 1 & \nodata  & $\perp$\\
{PG 2209+184} & $+21\degr$ & 1 & \nodata & $\perp$\\
\hline
\multicolumn{5}{c}{Quasars: SSRQ}\\
\hline 
{PG 0003+158} & $+116\degr$ & 1 & \nodata & $\perp$\\
{PG 1004+130} & $\sim +135\degr$ & 2 & \nodata & $\perp$\\
{PG 1048-090} & $-20\degr$ & 1 & \nodata & $\perp$\\
{PG 1100+772} & $+94\degr$ & 1 & \nodata & $\perp$\\
{PG 1103-006} & $-30\degr$ & 1 & \nodata & $\perp$\\
{PG 1425+267} & $-122\degr$ & 1 & \nodata & $\perp$\\
{PG 1512+370} & $-76\degr$ & 1 & \nodata & $\perp$\\
{PG 1545+210} & $-172\degr$ & 1 & \nodata & $\perp$\\
{PG 1704+608} & \nodata$^*$ & 2  & \nodata  & { 
 $~\perp$}$^\dagger$\\
{PG 2251+113} & $+139\degr$ & 1 & \nodata & $\perp$\\
{PG 2308+098} & $-49\degr$ & 1 & \nodata & $\perp$\\
\hline
\multicolumn{5}{l}{Note. Column (1): PG Source names. Column (2): Mean VLBI jet position angle. Column (3):}\\
\multicolumn{5}{l}{References for mean VLBI jet position angle. Column (4): EVPA in core }\\ 
\multicolumn{5}{l}{{\citep[from stacked 15 GHz VLBA data,][]{Pushkarev2023}} versus VLBI jet { PA}}\\ 
\multicolumn{5}{l}{Column (5): EVPA in core { \citep[6 GHz VLA, this work and][]{Baghel2023}} versus VLBI jet { PA}}\\
\multicolumn{5}{l}{References: (1) \citet{Plavin2022} (2) \citet{Wang2023}-VLBA at 5GHz}\\
\multicolumn{5}{l}{$^*$ unresolved in VLBA 5~GHz image by \citet{Wang2023}. $^\dagger$ $\perp$ to kpc-scale jet direction.}\\
\multicolumn{5}{l}{ { ``\nodata'' refers to cases where the 15 GHz VLBI core EVPA was not available.}}\\
\end{tabular}
\end{table*}

\begin{table*}[ht]
\centering
\caption{\label{tab:PGObs}Observational details of PG `blazars' presented in this paper}
\begin{tabular}{ccccccc}
\hline\hline
Name  
& Observation Date & Array Config & Flux Calibrator & Phase Calibrator & Leakage Calibrator & Angle Calibrator \\
\hline

{PG 0851+203} &
19-06-2023 & BnA &
3C 286 &
J0842+1835 &
OQ 208 & 3C 286 
\\

{PG 1101+384} &
22-06-2023 & BnA$\rightarrow$A &
3C 286 &
J1130+3815 &
OQ 208 & 3C 286
\\

{PG 1218+304} &
14-04-2023 & B &
3C 286 &
J1221+2813 &
OQ 208 & 3C 286
\\

{PG 1418+546} &
18-06-2023 & BnA &
3C 286 &
J1349+5341 &
OQ 208 & 3C 286
\\

{PG 1424+240} &
20-06-2023 & BnA &
3C 286 &
J1436+2321 &
OQ 208 & 3C 286
\\

{PG 1437+398} &
21-06-2023 & BnA$\rightarrow$A &
3C 286 &
J1416+3444 &
OQ 208 & 3C 286
\\

{PG 1553+113} &
20-06-2023 & BnA$\rightarrow$A &
3C 286 &
J1608+1029 &
OQ 208 & 3C 286
\\

{PG 2254+075} &
16-04-2023 & B &
3C 138 &
J2241+0953 &
3C 84 & 3C 138 
\\

{PG 1302$-$102} &
13-05-2023 & B &
3C 286 &
J1246$-$0730 &
OQ 208 & 3C 286
\\

{PG 2209+184} &
24-03-2023 & B &
3C 138 &
J2139+1423 &
3C 84 & 3C 138 
\\

{PG 1425+267} &
15-01-2023 & B &
3C 286 &
J1407+2827 &
OQ 208 & 3C 286
\\

{PG 1512+370} &
28-04-2023 & B &
3C 286 &
J1602+3326 &
OQ 208 & 3C 286
\\

{PG 1545+210} &
28-04-2023 & B &
3C 286 &
J1513+2338 &
OQ 208 & 3C 286
\\

{PG 2251+113} &
10-03-2023 & B &
3C 138 &
J2241+0953 &
3C 84 & 3C 138 
\\

{PG 2308+098} &
10-03-2023 & B &
3C 138 &
J2241+0953 &
3C 84 & 3C 138 
\\

\hline
\multicolumn{7}{l}{Note. Column (1): PG names of sources. Column (2): Observation date. Column (3): Array Configuration. Column (4): Flux }\\
\multicolumn{7}{l}{Calibrator. Column (5): Phase Calibrator. Column (6): Polarization Leakage Calibrator. Column (7): Polarization Angle Calibrator}\\
\end{tabular}
\end{table*}

\newpage

\section{Notes on Quasars}
\subsection{PG 1302-102}\label{subsubsec:AP_PG1302-102}

\citet{HutchingsNeff1992} and \citet{Bahcall1995} presented optical images that show a smooth elliptical host galaxy with a small eccentricity for PG 1302-102. These optical images also show a $0.5\arcsec$ wide object 2$\arcsec$ to the north, and also a bright feature 0.8$\arcsec$ to the west of the nucleus. \citet{Veilleux2009} and \citet{Kim2017} use the Spitzer and HST images respectively to argue that the host morphology is not strictly elliptical but ambiguous. There is no correlation between the optical and radio structures. Instead, the radio lobes appear to be pushed back from the inner optical knot. The hybrid optical luminosity profile of the mildly disturbed host elliptical, along with the linear shape of the nearest optical feature at $0.8\arcsec$ (possible small merging companion) and the relatively undisturbed extended radio emission suggests this to be a minor merger with the companion spiraling into the primary elliptical galaxy for some time \citep{HutchingsNeff1992}. HST images of \citet{Zhao2021} also detect possible remnant from the merger as extended emission to the north. There is marginal evidence for PG 1302-102 being in a galaxy cluster \citep{Yates1989, Yee1993}. 

\subsection{PG 2209+184}\label{subsubsec:AP_PG2209+184}

PG 2209+184 is very compact and resides in a spherical galaxy \citep{Zwicky1971}. 
It also shows night-to-night optical variability \citep{Jang1995}. \citet{Villafana2022} have modeled the velocity-resolved reverberation response of the H$\beta$ broad emission line for PG 2209+184 and found a mean BLR mean radius of 15.2 light days and BLR opening and inclination angles of 29.1$\degr$ and 30.1$\degr$ respectively indicating BLR is a thick disk slightly inclined to the line of sight.

\subsection{PG 1425+267}\label{subsubsec:AP_PG1425+267}

\citet{Laor1997} found PG 1425+267 to be weaker in soft X-rays compared to other radio-loud quasars. It is hosted by a bright elliptical galaxy that has three companions within $10\arcsec$ hinting at a possible interaction \citep{Marquez2001}. It has associated absorption at z = 0.3605 and 0.3643 \citep{Bechtold2002}. It shows intranight optical variability of $\sim$ 4 hours \citep{Stalin2005}. An emission line cloud up to 3 arcsec from the nucleus has been detected \citep{Boroson1984, StocktonMacKenty1987}. The [O III] emission extends into two opposing extensions to the east and west \citep{Stockton1983}.

\subsection{PG 1512+370}\label{subsubsec:AP_PG1512+370}
PG 1512+370 was first identified with an optical faint galaxy by \citet{Olsen1970}. \citet{Bahcall1993} noted a possible association with C$_{IV}$ absorption. Optical observations by \citet{Stockton1978} and others show another galaxy with the same redshift in the field, with \citet{Bergeron1987} finding a tentative continuum bridge linking the galaxy and the quasar. \citet{Ellingson1994} note that there are two other galaxies lying close to the quasar ($<60$~kpc) but it is uncertain if they are co-spatial with the quasar or foreground/background sources. Hence, the quasar may lie in a small group of galaxies \citep{YeeGreen1987, Ellingson1994}. PG 1512+370 also has an extremely luminous extended emission line region (EELR) extending up to 200~kpc \citep{StocktonMacKenty1987, Block1991}, exceptionally so among quasars at $z\lesssim0.5$. \citet{Crawford2000} carried out optical integral field spectroscopy of the EELR around PG 1512+370 using the ARGUS instrument on the Canada-France-Hawaii Telescope (CFHT) and found two off-nuclear [O III] clouds to the east and north-west of the quasar. The distribution of the optical continuum and line-emission were uncorrelated but both seem to be embedded in a low-level diffuse emission. Additionally, the clouds had their most blueshifted regions spatially coinciding with the radio source axis, though any relationship between the regions is not obvious given the difference in the extent of the two emissions with the radio lying on a much larger scale. This [O III] region also reveals filamentary structure in its HST image \citep{Stockton2002}. In hard X-ray, \citet{Akiyama2003} found a source coincident with PG 1512+370 using the High-Resolution Imager (HRI) of ROSAT.

\subsection{PG 1545+210}\label{subsubsec:AP_PG1545+210}

The optical counterpart for PG 1545+210 was found by \citet{Wyndham1966} and there have been several investigations into the luminous elliptical host galaxy and environment of PG 1545+210 \citep{Neugebauer1995,Bahcall1997}. Optical continuum and emission line images by \citet{VanHeerde1988,Hes1996} show two companion objects $19\arcsec$ to the south-east (PA$=117\degr$) and $12\arcsec$ towards the west (PA$=276\degr$) with faint trails of [O III] emission linking them to the quasar core. The [O III] line emission is mostly concentrated towards the western companion, extending from the southeast of the quasar, making it asymmetrical. \citet{Crawford2000} have carried out optical integral field spectroscopy of the EELR in PG 1545+210 using the ARGUS instrument on the CFHT. \citet{Stockton1982} had earlier identified a closer companion $2.7\arcsec$ to the north-west in their continuum image, with line emission confirming that it is at the same redshift and associated with the quasar. \citet{Hutchings1988} optical contour plots, after subtracting the quasar light, identified several compacts objects or knots, specifically one at $1.3\arcsec$ towards the north-west and another at $3\arcsec$ towards the south in the asymmetrical host galaxy, further indicating this to be a merging source. HST imaging by \citet{Bahcall1995, Bahcall1997} and by \citet{Kim2017} further confirms the companion at $2.7\arcsec$ to be another elliptical galaxy. PG 1545+210 is located towards the edges of a rich galaxy cluster as it lies $7\arcmin$ away from the center and at the same redshift as a compact Zwicky cluster 1545.1+2104 \citep{Oemler1972, YeeGreen1984}.

\subsection{PG 2251+113}\label{subsubsec:AP_PG2251+113}
PG 2251+113 is bright in infrared \citep{HutchingsNeff1992} and has an extended [O III] emission region similar in size and spatially coincident with the radio source \citep{StocktonMacKenty1987, Hutchings1990, Crawford2000}. The gas exhibiting the strongest blueshift is found at the same location as the southern radio hotspot along with being slightly more ionized, suggesting an interaction between the radio and optical plasma in this region \citep{Crawford2000}. The source also shows associated absorption of C IV, N V, Si IV, and Ly$\alpha$ possibly due to the group of galaxies that this quasar lies in, with galaxies within 60~kpc of the quasar at the same redshift \citep{Gunn1971,Ellingson1994, Bahcall1993}. It has a large (tens of kpc) irregular disturbed gas velocity field (over several hundred km~s$^{-1}$ \citep{Hutchings1990}. The AGN resides in a $10\arcsec$ undisturbed round host galaxy with a $R^{1/4}$ luminosity profile \citep{HutchingsNeff1992, Zhao2021}. The undisturbed nature of the host galaxy, the smooth optical profile, and the large size of the radio structure suggest that any AGN-triggering nuclear activity happened long ago given the typical major merger periods of $10^8$-$10^9$ years \citep{Lotz2011}. The source is moderately weak in the soft X-ray band \citep{Brandt2000}. 

\subsection{PG 2308+098}\label{subsubsec:AP_PG2308+098}
PG 2308+098 has an elliptical host galaxy with faint emission filaments extending to two nearby galaxies lying $\sim$10$\arcsec$ to the northeast and northwest \citep{Gehren1984}. It also has a weak associated absorber at z=0.434 \citep{Bechtold2002}. The optical–near-infrared spectral index of the polarized flux spectrum of PG 2308+098 suggests that the disk temperature profile of PG 2308+098 is consistent with the standard thin disk model \citep{Kishimoto2008}. 


\bibliography{main}{}
\bibliographystyle{aasjournal}



\end{document}